\documentclass{aa}

\usepackage{graphicx}
\usepackage{xcolor}
\usepackage{float}
\usepackage{color}
\usepackage{amsmath}    
\usepackage{txfonts}
\usepackage{hyperref}
\usepackage{comment}
\usepackage{tablefootnote}
\usepackage{booktabs, multirow}
\usepackage{notes2bib}
\usepackage{fontawesome}
\hypersetup{colorlinks=true,linkcolor=[rgb]{1.,0.2,0.2},citecolor=[rgb]{0.1,0.4,1.},filecolor=[rgb]{0.7,0.2,0.2},urlcolor=[rgb]{0.7,0.2,0.2}}
\definecolor{darkspringgreen}{rgb}{0.09, 0.45, 0.27}
\definecolor{amber}{rgb}{0.85, 0.39, 0.0}

\newcommand{\stars}{\texttt{STAR}~}
\newcommand{\galaxy}{\texttt{GALAXY}~}
\newcommand{\qso}{\texttt{QSO}~}

\usepackage[normalem]{ulem}

\begin{document} 

   \title{VEXAS:  VISTA EXtension to Auxiliary Surveys} 
   \subtitle{Data Release 2. Machine-learning based classification of sources \\ in the Southern Hemisphere\thanks{VEXAS is publicly available through the ESO Phase 3, \url{https://archive.eso.org/scienceportal/home?data_collection=VEXAS}}\fnmsep \thanks{The DR2  catalogues are also available  from the CDS via anonymous ftp to \url{cdsarc.u-strasbg.fr} (\url{130.79.128.5}) or via \url{http://cdsarc.u-strasbg.fr/viz-bin/cat/J/A+A/}}}
   \titlerunning{VEXAS DR2, object classification}
   \author{V.~Khramtsov\inst{1,2}, C.~Spiniello\inst{3,4}, A.~Agnello\inst{5} \and
   A.~Sergeyev\inst{1,6}
   }   \offprints{V. Khramtsov,\\\email{vld.khramtsov@gmail.com}}
    \institute{Institute of Astronomy, V. N. Karazin Kharkiv  National University, 35 Sumska Str., Kharkiv, Ukraine 
    \and
    Department of Data Science, Quantum, 20, Otakara Yarosha lane, Kharkiv, Ukraine 
    \and
    Sub-Dep. of Astrophysics, Dep. of Physics, University of Oxford, Denys Wilkinson Building, Keble Road, Oxford OX1 3RH, UK 
    \and
    INAF, Osservatorio Astronomico di Capodimonte, Via Moiariello  16, 80131, Naples, Italy 
    \and
    DARK, Niels Bohr Institute, University of Copenhagen,
    Jagtvej 128, 2200 Copenhagen \O, Denmark 
    \and
    Institute of Radio Astronomy of the National Academy of Sciences of Ukraine, 4, Mystetstv St., Kharkiv, 61002, Ukraine
    }

   \date{Submitted, February, 2021}
 \abstract
  {We present the second public data release of the VISTA EXtension to Auxiliary Surveys (VEXAS), where we classify objects into stars, galaxies and quasars based on an ensemble of machine learning algorithms.}
  {The aim of VEXAS is to build the widest 
  multi-wavelength catalogue, providing reference magnitudes, colours and morphological information for a large number of scientific uses. }
  {We apply an ensemble of thirty-two different machine learning models, based on three different algorithms and on different magnitude sets, training samples and classification problems (two or three classes) on the three VEXAS Data Release 1 (DR1) optical+infrared (IR) tables. The tables were created in DR1 cross-matching VISTA near-infrared data with Wide-field Infrared Survey Explorer  far-infrared data and with optical magnitudes from the Dark~Energy~Survey (VEXAS-DESW), the Sky~Mapper Survey (VEXAS-SMW), and the  Panoramic Survey Telescope and Rapid Response System Survey (VEXAS-PSW).
  We assemble a large table of spectroscopically confirmed objects (VEXAS-SPEC-GOOD,  415\,628 unique objects), based on the combination of six different spectroscopic surveys that we use for training.
  We develop feature imputation to classify also objects for which magnitudes in one or more bands are missing. }
{We classify in total $\approx90\times 10^{6}$ objects in the Southern Hemisphere. Among these, $\approx62.9\times 10^{6}$ ($\approx52.6\times 10^{6}$) are classified as 'high confidence' ('secure') stars,  $\approx920\,000$ ($\approx750\,000$) as 'high confidence' ('secure') quasars and $\approx34.8$ ($\approx34.1$) millions as 'high confidence' ('secure') galaxies, with $p_{\rm class}\ge 0.7$ ($p_{\rm class}\ge0.9$). The DR2 tables update the DR1 with the addition of imputed magnitudes and membership probabilities to each of the three classes.  }
{The density of high-confidence extragalactic objects varies strongly with the survey depth: at $p_{\rm class}>0.7,$ there are 111/deg$^2$ quasars in the VEXAS-DESW footprint and 103/deg$^2$ in the VEXAS-PSW footprint, while only 10.7/deg$^2$ in the VEXAS-SM footprint. Improved depth in the midIR and coverage in the optical and nearIR are needed for the SM footprint that is not already covered by DESW and PSW.} 
  

   \maketitle
%

\begin{figure*}
    \centering
    \includegraphics[width=\textwidth]{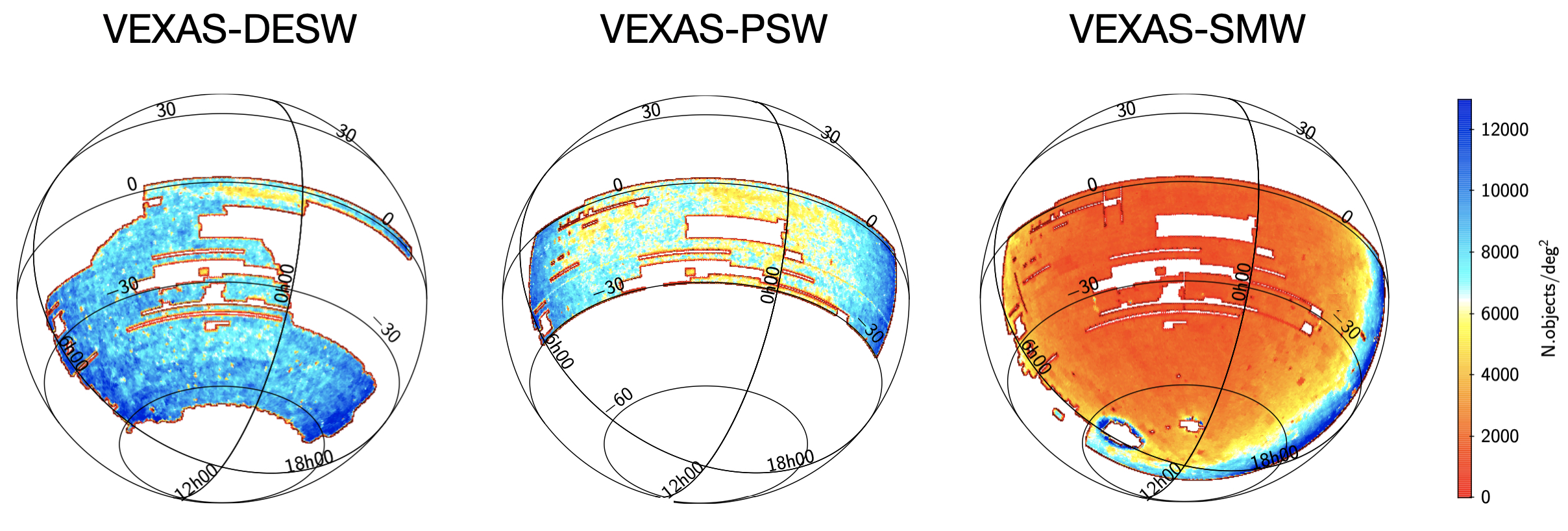}
    \caption{Sky coverage view of the three input VEXAS optical+IR tables. The colours indicate the number of objects per deg$^2$, as shown by the side bar, obtained using a Hierarchical Equal Area isoLatitude Pixelation of a sphere (\textsc{HealPix}) 
    with resolution equal $9$.}
    \label{fig:cover}
\end{figure*}

\section{Introduction}
\label{sec:intro}

Wide-field, digital, astronomical surveys have the potential to advance our knowledge in observational astrophysics. Footprint, depth and above all accessibility are paramount in the search for rare objects (e.g. extremely-metal-poor or ultra-cool-dwarf stars, high-redshift quasars, strong gravitational lenses), and for the statistical properties of astronomical objects at large, such as the clustering properties of galaxies or the assembly history of our own Milky Way (MW, see e.g. the review of \citealt{Helmi2020} and references therein).

Over the last decade, different facilities have surveyed wide areas of the Sky, over various frequencies in the optical and near-infrared (NIR). They have extended the pioneering work of the Sloan Digital Sky Survey (SDSS, DR1, \citealt{Abazajian03};
DR14, \citealt{SDSS_DR14})  to wider and homogeneous footprints, and in some cases with better depth and image quality. The Panoramic Survey Telescope \& Rapid Response System DR1 (PanSTARRS1, \citealt{Chambers16}) has covered 30'000 deg$^2$ contiguously, to the North of declinations $\delta>-30$ deg, in $grizy$ magnitudes, down to $i\approx23.1$ (5$\sigma$),  the Dark Energy Survey (DES, \citealt{des16}) has reached deeper ($i\approx23.50$, 10$\sigma$) over a 5'000 deg$^2$ contiguous footprint in the South in $grizY$, and the SkyMapper Southern Sky Survey (SM, \citealt{Wolf18}) has covered the whole Southern Hemisphere in $uvgriz$, albeit to shallower limits and coarser image quality. To complement those optical surveys, different initiatives have mapped wide areas in the NIR, mostly in the South (e.g. with the Visible and Infrared Survey Telescope for Astronomy, VISTA, \citealt{Emerson06}). 

In principle, these existing surveys can already provide a wealth of information on multiple fields of observational astrophysics, especially in preparation for more ambitious endeavours such as the \textit{Vera Rubin Observatory LSST} \citep{Ivezic2019} and the  ESA-\textit{Euclid} mission. In practice, matters are complicated by their heterogeneous and separate coverage. The SDSS was transformational also because it provided homogeneised information and additional tables with higher-level data, such as e.g. object classifications, photometric redshifts, and stellar masses and sizes of galaxies. 
All in all, calibrated and cross-matched multi-wavelength catalogues are crucial. 

This is indeed the purpose of the VISTA EXtension to Auxiliary Surveys (VEXAS) project (\citealt{VEXASdr1}, hereafter Paper~\textsc{I}), which extends infrared data of VISTA 
from the X-ray (ROSAT All Sky Survey, \citealt{Rosat_I, Rosat_II}; The XMM-Newton Serendipitous Survey, \citealt{XMM}) up to the radio domain (SUMSS, \citealt{Sumss_I}), in a fully homogenised database. 

The VEXAS DR1 catalogues, publicly available via the ESO Phase 3\footnote{The VEXAS tables are available both from the  \href{https://archive.eso.org/scienceportal/home?data_collection=VEXAS}{Archive Science Portal} or through the  \href{https://www.eso.org/qi/}{Catalog Facility} (active links in this footnote in the online version).}, include multi-bands photometric data, 
as well as objects that have been followed up spectroscopically (from the Sloan Digital Sky Survey, SDSS, DR14, \citealt{SDSS_DR14} and/or from the 6dF Galaxy Survey, 6dFGS, \citealt{6dfgs_I}).   
The core requirement of VEXAS is a reliable photometry in more than one band. This condition, together with the detection in at least two surveys (via cross-match), should minimise, if not completely eliminate, the number of spurious detections in the final catalogues.

For the VEXAS Data Release 2 (DR2), described in this paper and released to ESO Phase3, we continue the effort of providing the scientific community with useful multi-wavelength photometric catalogues, but we also add, for each catalogue source, a probability of it being a galaxy, a star or a quasar. This macro-classification is obtained through a machine learning (ML) ensemble of algorithms 
initially developed for a similar effort on a smaller footprint \citep{Khramtsov19} but improved and re-trained on purpose to match the VEXAS catalogues composition and varying depth and coverage.

\medskip
The paper is organised as follows. In Section~\ref{sec:VEXAS_table} and Section~\ref{sec:training} we provide details on the input tables and training tables that we used for the classification. In Section~\ref{sec:classification} we present an exhaustive description of our classification pipeline, including details on the feature imputation technique and on the ensemble learning. This latter is based on three different classifiers which were used to build 32 single models solving different problems, with different features and training datasets. In Section~\ref{sec:quality_results} we assess the performance and quality of such pipeline 
on a test sample. 
In Section~\ref{sec:results} we present the main results of the classification process, including the validation with external data. Finally, in  Section~\ref{sec:conclusions}, we draw our conclusions and discuss possible future developments of the VEXAS Project.


\section{The input data: the VEXAS optical+infrared tables}
\label{sec:VEXAS_table}

In Paper~\textsc{I}, we covered the Southern Galactic Hemisphere (SGH) below the Galactic plane (i.e. to $b<-20^{\circ}$). 
As a first step, we assembled a table of 198\,608\,360 detections from the VISTA-Kilo-Degree Infrared Galaxy Survey (VIKING, \citealt{Sutherland12,Edge13}) and the VISTA Hemisphere Survey (VHS, \citealt{McMahon12}), with reliable Petrosian magnitudes and small uncertainties in at least one band (see eq.~1 in Paper~\textsc{I}). 
We then cross-matched the table with the AllWISE Source Catalogue (\citealt{Cutri13}), which includes $\approx$747 million objects over the whole sky, with median angular resolution of 6.1, 6.4, 6.5, and 12.0 arcseconds respectively for the four bands ($W1$, $W2$, $W3$, $W4$).   
Setting a matching radius of 3 arcsec, we built the VEXAS-AllWISE table of 126\,372\,293 objects with near and far infrared photometry. 
Subsequently, we matched the VEXAS-Allwise table with three different multi-band optical wide-sky photometric surveys: DES, PanSTARRS1  and SkyMapper, resulting in the tables VEXAS-DESW, VEXAS-PSW and VEXAS-SMW.  
In each table, each entry has reliable photometry measured in at least one of the VISTA infrared bands ($Z$, $Y$, $J$, $H$, $K_s$), two of the WISE bands ($W1$, $W2$, $W3$, $W4$) and three optical bands  ($u$, $g$, $r$, $i$, $z$, $y$)\footnote{Throughout the paper we always indicate with uppercase $Y$ the infrared band from VISTA and with lowercase $y$ the optical one from DES or PS.}.  
These magnitudes are always provided in their native system of reference (AB for optical, Vega for infrared) and corrected for extinction. 

In this VEXAS-DR2, we used a filtered version of the DR1 tables: we removed all sources fainter than $25^m$ in each considered band since below this value the extrapolation of the training sample cannot be tested and thus trusted (see Sec.~\ref{sec:saferange} and App~\ref{app:DESW_faint}). In the case of VEXAS-PSW, we also applied a more severe cut on the optical magnitudes and associated uncertainties: we restricted ourselves to magnitudes brigther than the 
mean $5\sigma$ point-source limiting sensitivity values given in \citet{Chambers16} and we filtered out all sources with uncertainties $>1^m$. 
For the VEXAS-SMW table, in DR1 we limited ourselves to $\delta<-30^{\circ}$, since above this declination the PSW coverage is uniform and at least two magnitudes deeper. Here in DR2 we consider the whole coverage of Sky Mapper in the SGH, extending the VEXAS-SMW table to $\sim32$M unique  objects. This allowed us to compare the results obtained from training the pipeline on SMW with that obtained training it on DESW and PSW (see Sec.~\ref{sec:int_validation}).

These three tables constitute the photometric datasets used for this VEXAS-DR2. Their sky coverage is shown in Figure~\ref{fig:cover} and the numbers of objects are listed in Table~\ref{tab:numb_objects}.

Throughout the paper, these three tables are treated separately. As we describe in Section~\ref{sec:results}, we used our classification pipeline on each of them and obtained three tables of classified objects in output, which update and replace the three optical+NIR VEXAS DR1 tables. 

\begin{table}
\begin{center}
\begin{tabular}{c|c|c|c}
\hline\hline
{\bf Table}	& {\bf Sky area} & {\bf Declination} & 	{\bf Num (\#) of} \\
& {\bf (deg$^2$)} &  & {\bf Objects}	\\
\hline\hline
VEXAS-DESW & $4900$ & $-70^{\circ} < \delta < 0^{\circ}$ &    35\,381\,482 \\
\hline
VEXAS-PSW &  $3800$ &  $-30^{\circ} < \delta < 0^{\circ}$  &    21\,891\,609 \\ 
\hline
VEXAS-SMW  &  $9300$ & $\delta < 0^{\circ}$  & 31\,999\,995 \\
\hline\hline
\end{tabular}
\end{center}
\caption{Number of objects and sky coverage of each of the three optical cross-matched VEXAS tables we gave as input to our classification pipeline. These tables were assembled in Paper \textsc{I} but have been slightly changed here (see text for more details). Both VEXAS DR1 and DR2 include only objects with Galactic latitude $b<-20^{\circ}$.} 
\label{tab:numb_objects}

\end{table}

\section{The training samples}
\label{sec:training}
In this section we describe the different training datasets that we used for the ML process. 
We combined data from six different spectroscopic surveys (SDSS DR16, \citealt{SDSS16}; WiggleZ, \citealt{Drinkwater2018}; GAMA DR3 \citealt{GAMADR3}; OzDES, \citealt{Childress2017}, 2QZ \citealt{2QZ10k} and 6dFFS DR3, \citealt{6dfgs_DR3}), in order to build a training sample as large as possible and as complete as possible in all the three classes of objects  (\texttt{STAR}, \galaxy and \qso). 

In the following sub-sections, we describe how we selected and combined these different sets into the training, validation and testing samples.  
The data always include spectroscopic classification of the sources and redshift information for extra-galactic objects. We applied some selection criteria on the spectra to select only sources passing the quality level thresholds recommended directly by the data providers. These criteria are described in the following, for each of the six spectroscopic surveys separately. 

\subsection{SDSS DR16}
The largest spectroscopic sample we used for training is 
the sixteenth data release of Sloan Digital Sky Survey \citep[SDSS DR16, ][]{SDSS16}. 
SDSS DR16 is the forth release of the current SDSS phase IV \citep{sdssiv}, and it collects:
\begin{itemize}
    \item the most recent imaging, photometric and spectroscopic data;
    \item optical spectra from the extended Baryon Oscillation Spectroscopic Survey \citep[eBOSS,][]{eboss}, including data from the eBOSS sub-projects: the SPectroscopic IDentfication of ERosita Sources \citep[SPIDERS, ][]{spiders1,spiders2} survey and the Time-Domain Spectroscopic Survey \citep[TDSS, ][]{tdss};
    \item infrared stellar spectra from the Apache Point Observatory Galaxy Evolution Experiment 2 \citep[APOGEE-2,][]{apogee2};
    \item spectroscopic observations of nearby galaxies from the Mapping Nearby Galaxies at Apache Point Observatory \citep[MaNGA, ][]{manga}.
\end{itemize}
SDSS DR16 covers $\sim15\,000$ deg$^2$ and includes 5\,789\,200 spectra in total, among which 5\,107\,045 are unique\footnote{Selected from the `SpecObj' table via the CasJobs platform.}. 
From these, we selected only sources with 
$\texttt{zWarning}=0$, that resulted in a sample of 890\,374 stars, 751\,741 quasars and 2\,638\,083 galaxies.

\subsection{WiggleZ Final DR}
The WiggleZ Dark Energy Survey \citep{WIGGLEZ} is a survey of $\sim10^5$ objects 
with ultraviolet photometry from the \textit{Galaxy Evolution Explorer} \citep[GALEX, ][]{GALEX} survey, specifically selected in order to limit the sample to $z\sim0.5$ emission-line galaxies ($z_{median}\approx0.6$).

The final data release of the WiggleZ survey \citep{Drinkwater2018} contains a spectroscopic classification for $225\,415$ sources (mostly blue galaxies) across a $1000$ deg$^2$ area split into seven fields. 
We used here only galaxies with well-defined redshifts (with the quality parameter $Q>3$), and limited the sample to sources with $z>0.0024,$ as recommended by \citet{WIGGLEZ} to remove possible stellar contamination from the catalogue. This results in a set of 144\,040 galaxies.

\subsection{GAMA}
The Galaxy And Mass Assembly \citep[GAMA\footnote{\url{http://www.gama-survey.org/}},][]{GAMA} survey is mainly aimed at redshift measurements of galaxies.  We employed here the GAMA Data Release 3 \citep{GAMADR3}, that consists of $\sim$ 214\,000 galaxies, some of which were also observed with other surveys (SDSS, WiggleZ, etc.) and added to the catalogue. This survey is split into five sky fields (with area of $\approx60$ deg$^2$ each), three of which are near the celestial equator, and two are in the southern celestial hemisphere. 

We cleaned the GAMA DR3 sample of galaxies with the following criteria: $0.05<z<0.9$, $nQ>1$, where $nQ$ is the 'normalized quality scale' and measures the probability that the redshift estimate for a given spectrum is correct\footnote{For spectra with $nQ<1$ it is not possible to measure a redshift}. This allowed us to collect a sample of 192\,422 galaxies with precisely measured redshifts. 

\subsection{OzDES DR1}
The Australian Dark Energy Survey \citep[OzDES, ][]{Childress2017,Yuan2015} is a spectroscopic survey to measure redshifts of $\sim2500$ Type Ia supernovae host-galaxies over the redshift range $0.1 < z < 1.2$, and derive reverberation-mapped black hole masses for $\sim500$ quasars over $0.3 < z < 4.5$. The OzDES First Data Release (OzDES DR1\footnote{\url{http://www.mso.anu.edu.au/ozdes/DR1}}) contains the redshifts of $\sim15\,000$ sources that were observed during the first three years of observations. We used only the 14\,693 sources with redshift-quality flag $>3$. 

\subsection{2QZ}
The 2dF QSO Redshift Survey \citep[2QZ, ][]{2QZ10k} is a spectroscopic survey of $\sim40\,000$ quasars, observed over two $75^{\circ}\times5^{\circ}$ declination strips.

We retrieved the final catalogue, which combines 2QZ with the 6dF QSO Redshift Survey \citep{2QZ10k}, and comprises 49\,424 sources. Since 2QZ includes observations over two different epochs, we filtered out sources that have been classified in different classes or that have different redshift in one epoch with respect to the other. 
Moreover, we also excluded all sources for which redshift estimation is missing in both epochs (see  Table~2 in \citealt{2QZ}). For sources with more than one redshift estimates, we simply computed the average.

The labels \texttt{STAR}, \texttt{QSO}, or \galaxy for the selected objects were retrieved directly from the catalogue. The final sample, after cleaning, consisted of 39\,639 sources. 

\subsection{6dFGS}
The 6dF Galaxy Survey \citep[6dFGS, ][]{6DF} aims at spectroscopic studying galaxies in the Southern Sky. The final data release of 6dFGS contains 136\,304 spectra of mostly extragalactic objects. We selected only objects with $q_z\geq3$ and labelled as stars those with  $q_z=6$, as described on the 6dFGS description pages. 

\subsection{The final VEXAS spectroscopic table }
\label{sec:final_spec}
Combining the six `cleaned' spectroscopic tables described above, and cross-matching with the three input tables (using a matching radius of $1.5\arcsec$ since the resolution in the optical is better than that in the IR), we assembled a final spectroscopic table for each of the VEXAS table. In particular, for VEXAS-DESW, VEXAS-PSW, and VEXAS-SMW we find 293\,584, 328\,821, and 211\,092 unique spectroscopic sources, respectively. More detailed information about how many objects per each class are found in each survey is available in Table~\ref{tab:spectra}. 

\begin{figure}
    \centering
    \includegraphics[width=0.48\textwidth]{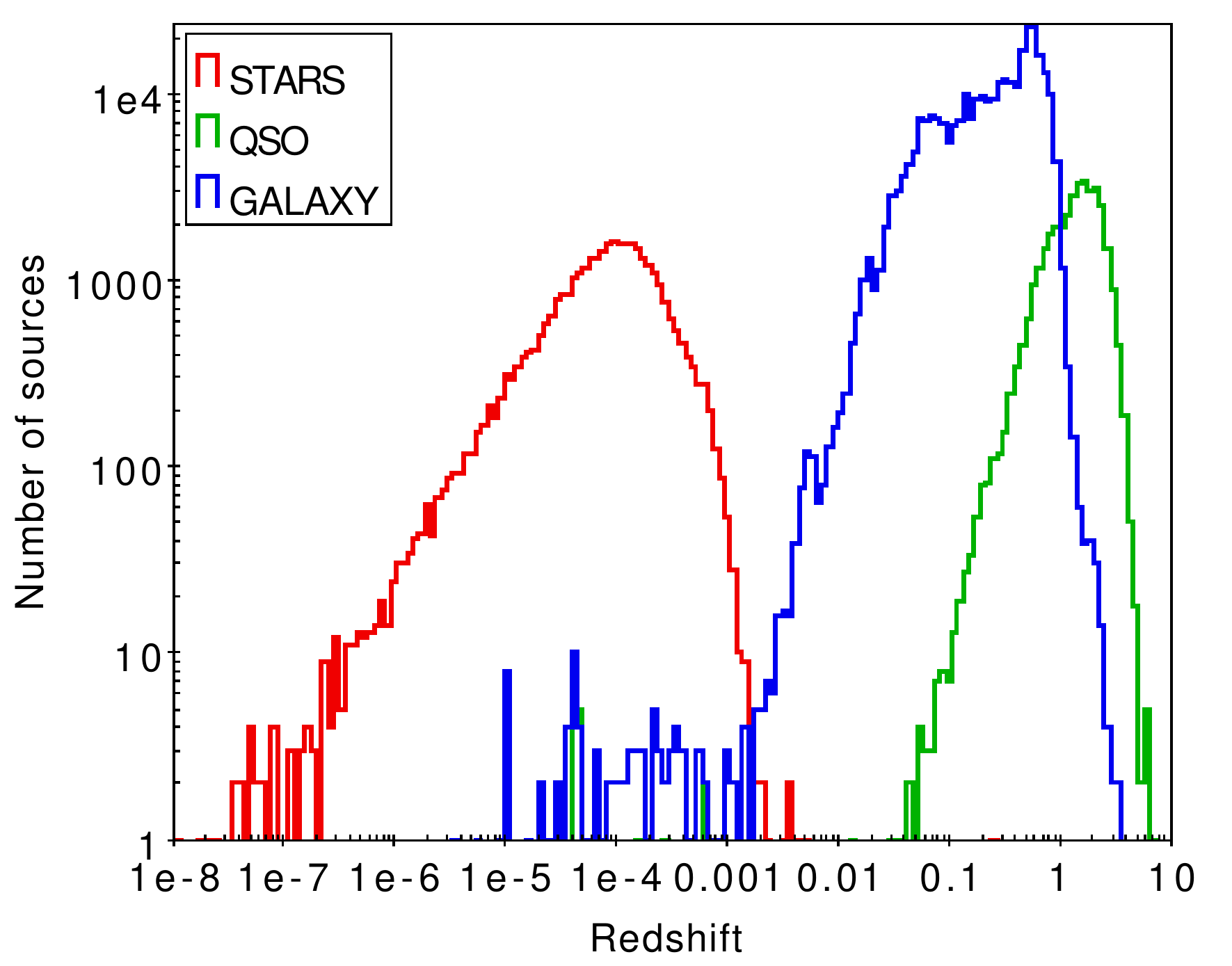}
    \caption{Redshift distribution of the sources in the VEXAS-SPEC-GOOD table, color coded by object class.}
 \label{fig:redshift_training}
 \end{figure}

\begin{figure}
    \centering
    \includegraphics[width=0.48\textwidth]{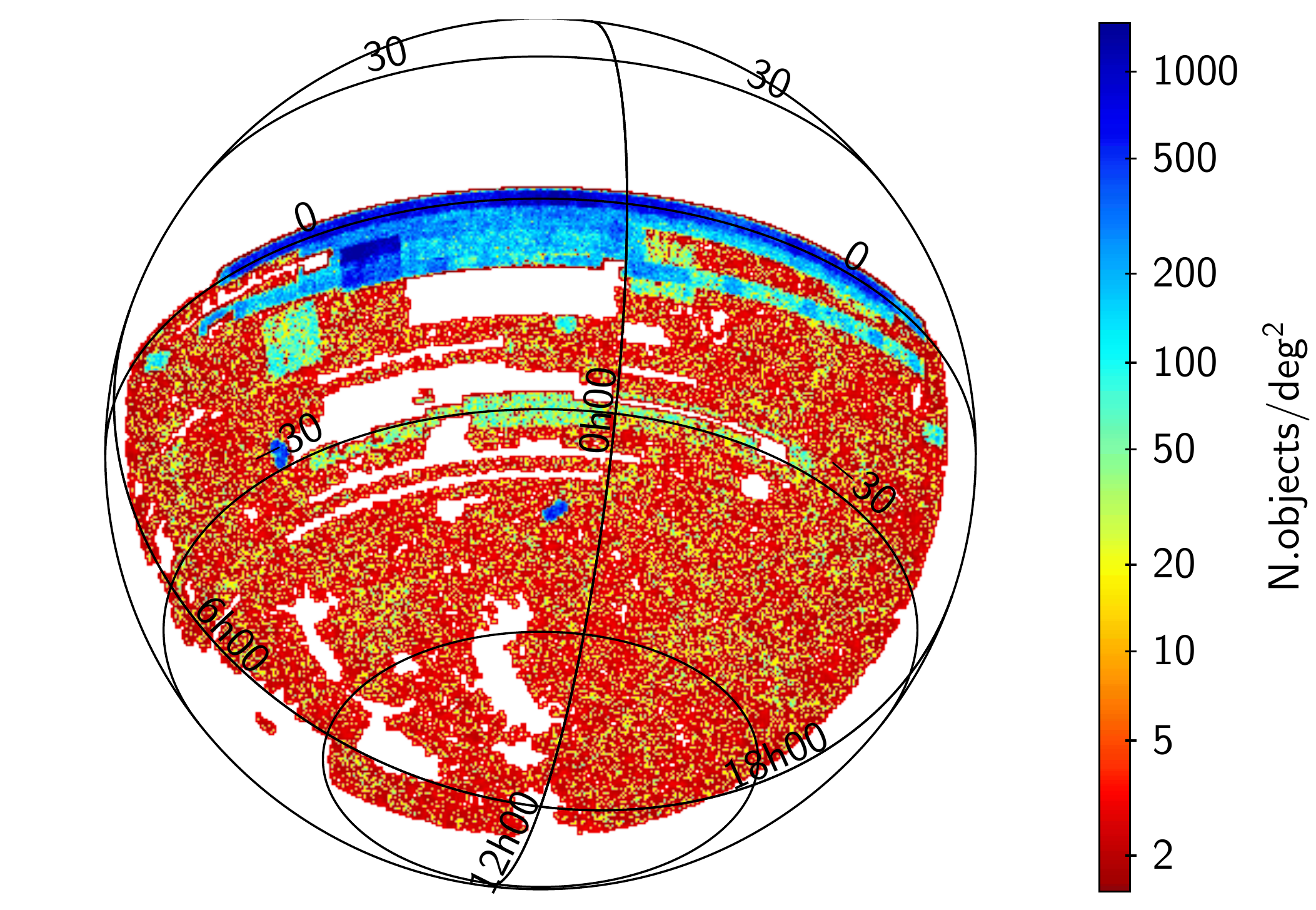}
    \caption{Sky coverage view of the VEXAS-SPEC-GOOD final table. The colour indicates the number of objects per deg$^2$, in logarithmic scale, as shown by the side bar, obtained as in Fig.~\ref{fig:cover}.}
 \label{fig:cover_spec}
 \end{figure}

Despite the fact that we used these three separate spectroscopic samples for the three VEXAS input tables, we also release the VEXAS-SPEC-GOOD, comprising 415\,628 unique VEXAS objects with photometry in the optical and infrared and a secure and clean spectroscopic classification. 
In total there are 89\,222 unique \stars, 35\,179 unique \qso and 291\,227 unique \galaxy in the released spectroscopic table. 
The redshift distribution of the spectroscopic objects in plotted in Figure~\ref{fig:redshift_training} where the sources are color coded by their class, as indicated by the caption.
The footprint of the table, colour-coded by object density (number of objects per deg$^2$) is instead plotted in Figure~\ref{fig:cover_spec}. 

In this paper, the three final VEXAS spectroscopic tables are used in Section~\ref{sec:classification} to train our classification models. They are further split into main and auxiliary and training, testing and validating samples, as described below.

\begin{table}[]
    \centering
\begin{tabular}{c|c|c|c|c}
 \hline \hline
\multirow{2}{*}{\textbf{Survey}} &\multirow{2}{*}{\textbf{Class}} &\textbf{VEXAS} &\textbf{VEXAS} &\textbf{VEXAS} \\   &  &\textbf{DESW} &\textbf{PSW} &\textbf{SMW} \\ 
\hline
\hline
\multirow{3}{*}{SDSS} 
&\stars &56\,187 &74\,794 &77\,100 \\
&\qso &23\,820 &26\,872 &4\,660 \\
&\galaxy &144\,495 &162\,717 &65\,937 \\ \hline
\multirow{3}{*}{2QZ} 
&\stars &537 &151 &701 \\
&\qso &3499 &657 &699 \\
&\galaxy &368 &114 &46 \\ 
\hline
\multirow{3}{*}{OzDES} 
&\stars &1\,614 &1\,216 &1\,569 \\
&\qso &1\,191 &998 &220 \\
&\galaxy &7\,246 &4\,438 &1\,778 \\ 
\hline
\multirow{2}{*}{6dFGS} &\stars &506 &649 &1\,544 \\
&\galaxy &25\,274 &21\,792 &54\,093 \\ \hline
GAMA &\galaxy &33\,050 &34\,551 &11\,578 \\ 
\hline
WiggleZ &\galaxy &7\,851 &12\,759 &723 \\
\hline
\hline
\end{tabular}
    \caption{Number of objects with a spectroscopic match from one or more spectroscopic surveys we used in this paper as training sample.}
    \label{tab:spectra}
\end{table}

\subsection{Splitting spectroscopic datasets}
\label{sec:splittingdata}
Each final spectroscopic table described above is separated into \textit{main} and \textit{auxiliary} samples for the workflow described in Section~\ref{sec:classification}. 
These two parts are used at different stages during the training of the algorithms to mitigate the effect of the presence of wrong classification labels.
In fact, despite the selection criteria we applied on the spectroscopic tables, the final labeling and classification of the spectroscopic datasets is not ideal (as shown in Appendix~\ref{app:noisy_spec}). Many sources have spectra that are too noisy to provide a trustworthy classification and some objects have been also misclassified by the survey pipelines. 
Therefore, we undertook an iterative process, alternating the training and sample composition. At training, the classifier(s) learns on the main training sample, and then it predicts the classes of all the sources in the auxiliary sample. At this point, all the sources for which the predicted class results consistent with the original spectroscopic-based  class are added to the main training sample, and the classifier is run again, learning each time on a larger data-set. 
As initial main sample we adopted the SDSS DR16 catalogue, as it comprises the largest number of objects over the three macro-classes.  
We used the remaining five spectroscopic datasets as auxiliary instead.

The main (SDSS) sample was divided into training (60\%), validation (20\%) and testing (20\%). 
The training sample  
was then used to train the classification algorithm, the  validation sample to control and tune the classifier during its learning process, and finally, the testing sample was used to check the quality of the classifier, which was run on previously unseen data, as described in Section~\ref{sec:quality_results}. 
The auxiliary sample was added only to the main training sample.

\section{Classification Pipeline}
\label{sec:classification}
In this section, we highlight the main characteristics of our classification pipeline in more detail. We make the code available via Github\footnote{ \faGithub\href{https://github.com/VEXAS-team/VEXAS-DR2}{VEXAS Github repository} (active link in the online version)}.
 
To classify the sources of the VEXAS input tables into \texttt{STAR}, \qso and \texttt{GALAXY}, we set up a pipeline able to resolve, or at least to take into account, the major limitation of the VEXAS DR1 dataset, in particular regarding the fact that some magnitudes are missing for a not-negligible number of sources in one or more bands. This can represent an issue, because not all machine learning algorithms can handle missing data.
We used the technique of \textit{feature imputation} to fill the missing magnitudes, and trained 
 a separate ML algorithm on purpose, which we describe below (Section~\ref{subsec:featureimput}). 

We finally give details on the feature sets (Sec~\ref{sec:featureset}) and algorithms, both on single classifiers and ensemble learning (Sec~\ref{subsec:ensamble_learning}), that we used in our classification pipeline.

\subsection{Dealing with missing magnitudes: imputation}
\label{subsec:featureimput}
As already highlighted in Paper~\textsc{I}, the VEXAS coverage is not homogeneous across the different bands. This is particularly true for SM and in general for the infrared, or at least for the $Y$ and $H$ bands, given the strategy of the VHS survey, and for $W3$ and $W4$ from WISE. This means that, for many entries in the catalogue, one or more magnitudes in one or more bands can be missing, simply because this particular object has not been observed by the corresponding survey.  
Another reason is also that an object is detected in some bands but is too faint in others (e.g. $z\sim3$ objects lack flux in the $u$-band, \citealt{Steidel96}). Clearly, disentangling between these two cases would help in the classification of the sources with missing magnitudes. 

In general, simply removing the sources which have partial magnitude information would 
severely reduce the VEXAS dataset. This can be seen from Table~\ref{tab:num_obj_percentag} were we list, for each input VEXAS table the percentage of objects with measured magnitude in each band. Even without considering the $Y$ and $H$ bands from VISTA and $W3$ and $W4$ from WISE, which we excluded because of their shallower coverage, we would lose roughly 10-15\% of the sources for VEXAS-DESW and VEXAS-PSW and almost 80\% of the sources in VEXAS-SMW (or 50\% discarding the $u$-band). 
Rather than limiting to only sources with photometry measured in each band, a strategy of handling missing magnitudes has to be adopted within our pipeline if we aim at maximising the multi-band coverage.

\begin{table}[]
    \centering
    \begin{tabular}{c|c|c|c} \hline \hline
\textbf{Band} & \textbf{VEXAS-DESW} & \textbf{VEXAS-PSW} & \textbf{VEXAS-SMW}  \\ 
\hline
\hline
$u$ &-- &-- &19.0\% \\ \hline
$g$ &98.5\% &89.3\% &51.9\% \\ \hline
$r$ &99.6\% &97.3\% &65.0\% \\ \hline
$i$ &99.6\% &100.0\% &84.9\% \\ \hline
$z$ &99.8\% &98.0\% &94.1\% \\ \hline
$y$ &99.4\% &95.7\% &-- \\ \hline
$Y$ &6.4\% &51.4\% &29.1\% \\ \hline
$J$ &98.0\% &98.8\% &99.7\% \\ \hline
$H$ &31.6\% &24.1\% &16.7\% \\ \hline
$K_S$ &91.5\% &94.2\% &99.4\% \\ \hline
$W1$ &100.0\% &100.0\% &100.0\% \\ \hline
$W2$ &88.6\% &87.6\% &96.0\% \\ \hline
$W3$ &20.0\% &18.4\% &17.9\% \\ \hline
$W4$ &4.7\% &4.5\% &4.3\% \\ 
\hline
\hline
\end{tabular}
    \caption{Percentage of objects with a measured magnitude in each of the listed bands for the VEXAS input tables.}
    \label{tab:num_obj_percentag}
\end{table}

Furthermore, we note that a simple imputation with mean, median or constant values would allow us to preserve the total number of sources, 
but it would not guarantee the reliability of the classification results for sources with missing magnitudes. A better way to handle this is to use an intelligent imputer, which is able to find a relationship between missed and measured magnitudes for each source, which should reflect its true underlying properties. 

We follow this approach here, using an autoencoder (AE) neural network. In general, an AE is trained to reproduce the input features, and as a consequence the hidden layers end up selecting combinations of input features 
that carry most of the underlying physical information. For this reason, AEs are typically used for dimensionality reduction, compressing the information through a hidden layer with lower dimension than the input feature space. 
In our case, we are interested in reconstructing missing magnitudes, so the input to our AE is a vector of magnitudes of which some of them are masked out at random, and the output target is the full (un-masked) magnitude vector. We give further detail on our AE architecture in Appendix~\ref{app:AEarchitecture}. 
In order to test the reliability of the imputation procedure, we first performed a test on a sample of objects with known magnitudes to quantify the goodness of imputed magnitudes. Appendix~\ref{app:AEtraining} gives further details about it, including the trustworthiness of the output magnitudes.

To test the effect of feature imputation on the classification, instead, 
we ran our pipeline sometimes with and sometimes without imputation (see Sec.~\ref{subsec:ensamble_learning}) and compared the results obtained in both cases.  
However, we caution the reader that, in the case of VEXAS-SMW, the imputation, especially for quasars, might not work properly for all the objects. This is possibly due to the fact that the number of quasars with  magnitudes measured in all bands present in this table is very low ($<8000$)\footnote{We have obtained this number \textit{a posteriori} from the classified table presented in Sec.~\ref{sec:results}}.  
The imputer therefore does not have enough knowledge on the spectral energy distribution of this class of  objects and thus does not predict good values for the missing magnitudes by looking at the measured ones. More details are provided in Appendix~\ref{app:mag_imp_sm}.  

In addition, for all the classes, the percentage of objects with missing magnitudes in the optical bands is much larger for VEXAS-SMW, as visible from Table~\ref{tab:num_obj_percentag}, and this makes imputation harder. 
For this reason, in the case of VEXAS-SMW, we obtained and released two different classifications: the fiducial one with imputation, and an independent one, without imputation and based on the decision-tree based CatBoost algorithm \citep{cat1,cat2}, which is able to deal with missing magnitudes but is not as complete and generalizable as the the ensemble learning META\_MODEL described in Secion~\ref{subsec:ensamble_learning}. We  anticipate that the two algorithms result in  an object classification which is the same for $99.5\%$ of the sources in the VEXAS-SMW table (see Sec.~\ref{sec:results} and Fig.~\ref{fig:class_comp_sm}).




\begin{table*}[t]
    \begin{tabular}{c|c|c|l|l|c|c}
    \hline\hline
\textbf{Mod.} & \textbf{Method} & \textbf{Classes} & \textbf{Input bands} & \textbf{Features} & \textbf{Imputation} & \textbf{Aux.} \\
\# &  & \textbf &  &  &  & \textbf{sample} \\ 
\hline
\hline
\multicolumn{7}{c}{\textbf{COMPLETE MODELS (one for each algorithm)}} \\
\hline
1 & ANN & \texttt{STAR, QSO, GAL} & $J, K_S, g, r, i, z, y/u\;\;\tablefootmark{a}, W1, W2$ & sc.mags, colours, $P_{STAR}$ & \texttt{True} & \texttt{True} \\ \hline
2 & kNN & \texttt{STAR, QSO, GAL} & $J, K_S, g, r, i, z, y/u\;\;\tablefootmark{a}, W1, W2$ & sc.mags, colours, $P_{STAR}$ & \texttt{True} & \texttt{True} \\ \hline
3 & CB & \texttt{STAR, QSO, GAL} & $J, K_S, g, r, i, z, y/u\;\;\tablefootmark{a}, W1, W2$ & sc.mags, colours, $P_{STAR}$ & \texttt{True} & \texttt{True} \\
\hline\hline
\multicolumn{7}{c}{\textbf{DIFFERENT BAND-SETS}} \\
\hline
4 & ANN & \texttt{STAR, QSO, GAL} & $J, K_S, g, r, i, z, y/u\;\;\tablefootmark{a}$ & sc.mags, colours, $P_{STAR}$ & \texttt{True} & \texttt{True} \\ \hline
5 & kNN & \texttt{STAR, QSO, GAL} & $J, K_S, g, r, i, z, y/u\;\;\tablefootmark{a}$ & sc.mags, colours, $P_{STAR}$ & \texttt{True} & \texttt{True} \\ \hline
6 & CB & \texttt{STAR, QSO, GAL} & $J, K_S, g, r, i, z, y/u\;\;\tablefootmark{a}$ & sc.mags, colours, $P_{STAR}$ & \texttt{True} & \texttt{True} \\
\hline
7 & ANN & \texttt{STAR, QSO, GAL} & $g, r, i, z, y/u\;\;\tablefootmark{a}, W1, W2$ & sc.mags, colours, $P_{STAR}$ & \texttt{True} & \texttt{True} \\ \hline
8 & kNN & \texttt{STAR, QSO, GAL} & $g, r, i, z, y/u\;\;\tablefootmark{a}, W1, W2$ & sc.mags, colours, $P_{STAR}$ & \texttt{True} & \texttt{True} \\ \hline
9 & CB & \texttt{STAR, QSO, GAL} & $g, r, i, z, y/u\;\;\tablefootmark{a}, W1, W2$ & sc.mags, colours, $P_{STAR}$ & \texttt{True} & \texttt{True} \\
\hline
10 & ANN & \texttt{STAR, QSO, GAL} & $J, K_s, g, r, i, z, W1, W2$ & sc.mags, colours, $P_{STAR}$ & \texttt{True} & \texttt{True} \\
\hline
11 & KNN & \texttt{STAR, QSO, GAL} & $J, K_s, g, r, i, z, W1, W2$ & sc.mags, colours, $P_{STAR}$ & \texttt{True} & \texttt{True} \\
\hline
12 & CB & \texttt{STAR, QSO, GAL} & $J, K_s, g, r, i, z, W1, W2$ & sc.mags, colours, $P_{STAR}$ & \texttt{True} & \texttt{True} \\
\hline\hline
\multicolumn{7}{c}{\textbf{DIFFERENT CLASSIFICATION PROBLEMS}} \\
\hline
13 & ANN & \texttt{STAR, QSO+GAL} & $J, K_S, g, r, i, z, y/u\;\;\tablefootmark{a}, W1, W2$ & sc.mags, colours, $P_{STAR}$ & \texttt{True} & \texttt{True} \\ \hline
14 & kNN & \texttt{STAR, QSO+GAL} & $J, K_S, g, r, i, z, y/u\;\;\tablefootmark{a}, W1, W2$ & sc.mags, colours, $P_{STAR}$ & \texttt{True} & \texttt{True} \\ \hline
15 & CB & \texttt{STAR, QSO+GAL} & $J, K_S, g, r, i, z, y/u\;\;\tablefootmark{a}, W1, W2$ & sc.mags, colours, $P_{STAR}$ & \texttt{True} & \texttt{True} \\ \hline
16 & ANN & \texttt{STAR+GAL, QSO} & $J, K_S, g, r, i, z, y/u\;\;\tablefootmark{a}, W1, W2$ & sc.mags, colours, $P_{STAR}$ & \texttt{True} & \texttt{True} \\ \hline
17 & kNN & \texttt{STAR+GAL, QSO} & $J, K_S, g, r, i, z, y/u\;\;\tablefootmark{a}, W1, W2$ & sc.mags, colours, $P_{STAR}$ & \texttt{True} & \texttt{True} \\ \hline
18 & CB & \texttt{STAR+GAL, QSO} & $J, K_S, g, r, i, z, y/u\;\;\tablefootmark{a}, W1, W2$ & sc.mags, colours, $P_{STAR}$ & \texttt{True} & \texttt{True} \\ \hline
19 & ANN & \texttt{STAR+QSO,GAL} & $J, K_S, g, r, i, z, y/u\;\;\tablefootmark{a}, W1, W2$ & sc.mags, colours, $P_{STAR}$ & \texttt{True} & \texttt{True} \\ \hline
20 & kNN & \texttt{STAR+QSO,GAL} & $J, K_S, g, r, i, z, y/u\;\;\tablefootmark{a}, W1, W2$ & sc.mags, colours, $P_{STAR}$ & \texttt{True} & \texttt{True} \\ \hline
21 & CB & \texttt{STAR+QSO, GAL} & $J, K_S, g, r, i, z, y/u\;\;\tablefootmark{a}, W1, W2$ & sc.mags, colours, $P_{STAR}$ & \texttt{True} & \texttt{True} \\ \hline\hline
\multicolumn{7}{c}{\textbf{THE EFFECT OF IMPUTATION AND AUXILIARY SAMPLE }}\\
\hline
22 & ANN & \texttt{STAR, QSO, GAL} & $J, K_S, g, r, i, z, y/u\;\;\tablefootmark{a}, W1, W2$ & sc.mags, colours, $P_{STAR}$ & \texttt{True} & \texttt{False} \\
\hline
23 & kNN & \texttt{STAR, QSO, GAL} & $J, K_S, g, r, i, z, y/u\;\;\tablefootmark{a}, W1, W2$ & sc.mags, colours, $P_{STAR}$ & \texttt{True} & \texttt{False} \\
\hline
24 & CB & \texttt{STAR, QSO, GAL} & $J, K_S, g, r, i, z, y/u\;\;\tablefootmark{a}, W1, W2$ & sc.mags, colours, $P_{STAR}$ & \texttt{True} & \texttt{False} \\
\hline
25 & CB & \texttt{STAR, QSO, GAL} & $J, K_S, g, r, i, z, y/u\;\;\tablefootmark{a}, W1, W2$ & sc.mags, colours, $P_{STAR}$ & \texttt{False} & \texttt{True} \\ 
\hline
26 & CB & \texttt{STAR, QSO, GAL} & $J, K_S, g, r, i, z, y/u\;\;\tablefootmark{a}, W1, W2$ & sc.mags, colours, $P_{STAR}$ & \texttt{False} & \texttt{False} \\ 
\hline\hline
\multicolumn{7}{c}{\textbf{CLASSIFICATION WITHOUT STELLARITY}}\\
\hline
27 & ANN & \texttt{STAR, QSO, GAL} & $J, K_S, g, r, i, z, y/u\;\;\tablefootmark{a}, W1, W2$ & sc.mags, colours & \texttt{True} & \texttt{True} \\
28 & kNN & \texttt{STAR, QSO, GAL} & $J, K_S, g, r, i, z, y/u\;\;\tablefootmark{a}, W1, W2$ & sc.mags, colours & \texttt{True} & \texttt{True} \\
29 & CB & \texttt{STAR, QSO, GAL} & $J, K_S, g, r, i, z, y/u\;\;\tablefootmark{a}, W1, W2$ & sc.mags, colours & \texttt{True} & \texttt{True} \\
\hline \hline
\multicolumn{7}{c}{\textbf{CLASSIFICATION WITHOUT MAGNITUDES}}\\
\hline
30 & ANN & \texttt{STAR, QSO, GAL} & $J, K_S, g, r, i, z, y/u\;\;\tablefootmark{a}, W1, W2$ & colours, $P_{STAR}$  & \texttt{True} & \texttt{True} \\
31 & kNN & \texttt{STAR, QSO, GAL} & $J, K_S, g, r, i, z, y/u\;\;\tablefootmark{a}, W1, W2$ & colours, $P_{STAR}$  & \texttt{True} & \texttt{True} \\
32 & CB & \texttt{STAR, QSO, GAL} & $J, K_S, g, r, i, z, y/u\;\;\tablefootmark{a}, W1, W2$ & colours, $P_{STAR}$  & \texttt{True} & \texttt{True} \\

\hline\hline

    \end{tabular}
    \tablefoottext{a}{We use the $y$ band for PS and DES, and $u$ band for SM.}

    \caption{Description of the 32 individual classifiers that we combine into the ensemble learning, varying the algorithms, classification problem, input bands, feature sets, the use of imputation and the auxiliary training sample. 
    }
    \label{tab:classifiers}
\end{table*}

\subsection{Feature set}
\label{sec:featureset}
The basic principle of our feature set construction is that \texttt{STAR}, \texttt{QSO}, and \galaxy are different in their morpho-photometric properties. Thanks to the broad wavelength range covered by the VEXAS tables, we have at our disposal a large number of magnitudes ($u, g, r, i, z, y, J, K_s, W1, W2$)\footnote{We do not consider the $Y$ and $H$ bands from VISTA, nor the $W3$ and $W4$ from WISE, as their coverage is sparse and very limited, see Table~\ref{tab:num_obj_percentag}.}, 
and at least one morphological parameter highlighting the "stellarity" for each entry in the catalogue. 

In the following, we indicate the source index as $i$, and the magnitude indices as $j,k$ and we describe the input feature space that we used for the system classification, which is  based on: 

\textit{Colour indices}. Different types of sources have different spectral energy distributions (SEDs) and so they lie in different regions in colour-colour diagrams. 
Having a large set of magnitudes, we can form all the possible colours as pairwise differences between two magnitudes: $m_{i,j} -  m_{i,k}$. For the models where feature imputation is not used,  if one of the two considered magnitudes is not measured for a given source, also the colour obtained from that magnitude will be missing for that source (see below). 
  
\textit{Scaled magnitudes}. The physical characteristics of the sources can not be retrieved directly from the raw magnitudes. However, magnitude information could be important for some of the classifiers, which could create `their own` features space (e.g., artificial neural networks), or learn relationship between magnitudes and classes with distance-based approach (e.g., $k$ nearest neighbours). In both cases, the shape of the SEDs is expressed with a set of magnitudes. 
Rather than using 
standard magnitudes, we 
used 'scaled magnitudes' ($m'$), better suited to capture the relative "intensity" in each band for each source, independently on the intrinsic differences in brightness between the different objects. In particular, for each source, we identified the maximum magnitude across all bands and we re-scaled all the other bands to that value: $m'_{i,j} = m_{i,j}/max({m_{i})}$. 
This scaling helps increase the classification reliability on very faint sources.

\textit{Stellarity index}. As additional morphological information, we used the VISTA $P_{STAR}$ parameter which is an index of the `stellarity` of the sources, and has been proved to be a very important feature in objects classification (e.g., \citealt{Khramtsov19}).
This parameter has values between 0 and 1, where 1 correspond to a point-like source, and 0 -- to an extended source. It is more generally available than other indicators, e.g. \texttt{psf} \textit{vs} \texttt{model} magnitudes, as it is obtained directly at the level of image segmentation.

\subsection{Ensemble learning}
\label{subsec:ensamble_learning}
In the presence of noisy datasets (as ours are, Appendix~\ref{app:noisy_spec}), an ensemble of different classifiers performs better than a single trained and optimized classifier \citep{ensemble_book, Myoung2006, Lessmann2015}.
For this reason, here we used an ensemble of classifiers based on different algorithms.

In general, in ensemble learning, the final classification is obtained by joining the predictions of the individual classifiers. In particular, we adopted a stacking procedure, in which a meta-classifier uses the predictions of a number of individual classifiers as the input features, and learns to predict the classes using the input probabilities derived from single models. 

We defined 32 different models based on the following three algorithms:
\begin{enumerate}
    \item Artificial neural network (ANN). ANN learns in an iterative way the function, connecting input features with the target values, producing the weighted sum of neurons output, which are connected with each other by layers. Our ANN is made of seven layers; the first one accepts as input the raw features, and the following six transform them into $5\times k$ features, where $k$ is ranging from $k=5$ to $k=1$. The last layer consists of neurons, the number of which corresponds to the number of output classes. 
    To prevent overfitting and to speed-up the learning, we used the batch-normalization and dropout techniques; all layers are connected via a rectified linear unit activation function. The ANN minimizes the binary cross-entropy with the \textit{Adam} optimizer (\citealt{adam}; with a learning rate equal to $10^{-2}$) over 10 epochs, and the weights are saved for the epoch with the lowest loss function value on the validation dataset. 
    \item $k$-nearest neighbours ($k$NN). The $k$NN classifies a source by selecting the most prevalent class within the $k$ nearest neighbours. We selected here $k=15$ as number of neighbours to be used in the $k$NN algorithm and used Euclidean distance as the distance metric between sources in the feature space.  
    \item CatBoost \citep{cat1,cat2}. CatBoost is a Gradient Boosting \citep{gradboost} decision-tree ensemble algorithm. To train CatBoost, we used all defaults parameters, except for the number of iterations, that in our case is equal to 8\,000. We adopted 500 early-stopping steps, meaning that we stop the growth of trees as soon as the validation score does not increase after 500 steps. 
\end{enumerate}

Based on these three algorithms, we built a total of 32 different models which were then trained  
with different choices of input features and to solve slightly different classification problems. In particular, we changed the number of output classes, e.g. merging together stars and quasars (i.e. classifying point-like versus extended source), or merging together quasars and galaxies (i.e. extra-galactic \textit{vs} galactic sources). 
Moreover, in some cases we employed feature imputation (see Sec.~\ref{subsec:featureimput}) and/or the auxiliary training sample (see Sec.~\ref{sec:training}), while in some other cases we did not. 
The characteristics of these models are listed in Table~\ref{tab:classifiers}, where we also highlight the effect we aimed to study with each group of models.

In principle, one could then combine the models using a simple averaging. However, we preferred to use a slightly more sophisticated weighted averaging. 
We used Logistic Regression (LR) as meta-classifier, trained to predict \stars, \qso and \galaxy classes based on the probability scores derived from all the single models we set up. LR is closely related to a linear regression, but it is constructed using the Sigmoid function to limit the result in the range $[0; 1]$:
\begin{equation}
\hat{y}(\bf{C}, \bf{P}) = (1 + \exp^{-\bf{C}^T \cdot \bf{P}})^{-1} 
\end{equation}
where $\bf{P}$ is the vector of input probabilities, $\bf{C}$ is the corresponding coefficient, and $\bf{C}^T \cdot \bf{P} = \sum_i C_i P_i$ is a scalar product of both vectors, with $P_i$ being the probability to belong to a certain class, given by a single model, and $C_i$ the corresponding regression coefficient.

The regression model uses $L_1$-regularization: 
\begin{equation}
    L = \sum_j \big(y - \hat{y}(C, P)\big)^2 + \lambda \sum_i | C_i |
\end{equation}
where $L$ is a loss function, $\lambda$ is a regularization parameter (which we set to $\lambda=1$). In the above equation, summing over $j$, one proceeds through all training objects, while summing over $i$ one goes through all the coefficients. 
This type of regularization pushes the coefficients corresponding to useless input features to be zero, or very close to zero, hence helping us to select the most informative input features (in our case, probabilities of individual models).  
This is not the case for e.g. the 
$L_2$-regularization ($\lambda \sum_i C_i^2$), which would make the selection of important input features more difficult. 

Our META\_MODEL, trained to separate the classes from each other\footnote{We note that, while each individual classifier was trained on the training sample, the meta-classifier was trained on the predictions on the validation sample. This helps in preventing overfitting.}, solves a 3-class problem  
in a `One-\textit{vs}-Rest' scheme. This means that, in output, we have three LR models, one for each class, and thus three ``meta probabilities'' for each object to belong to each of three classes (one for \stars versus \qso + \galaxy, one for \qso versus \stars + \galaxy and one for \galaxy versus \stars + \qso). 
We note  that strictly speaking, it would have been more accurate to solve the multi-classes problem. 
However, for  computational simplicity, we restricted ourselves to the `One-\textit{vs}-Rest' scheme but we highlight that this is a very good approximation as the the three output probabilities always sum to 1 within better than 10$^{-5}$.

We also emphasize that the META\_MODEL is more general and comprehensive than a single model and allows us to better extrapolate the classification results to regimes that may not be covered by the current spectroscopic training set.  

\section{Testing the quality of the classification pipeline}
\label{sec:quality_results}
Before running the pipeline on the full VEXAS input tables, we performed a number of quality checks on the `unseen' test sample, to understand the strengths and limitations of our classification. An extensive description of these tests and their results both on the META\_MODEL and on the single ones, based on five different classification metrics and on regression coefficients, are given on the Github VEXAS repository.  
There we also provide a visualization of the relative importance of each classification feature for each single model, confirming that the VISTA $P_{STAR}$ parameter has a high classification power. However we stress that the fact that one feature is relatively important in a single model, it does not necessarily imply that the same feature is driving the ensemble classification.
The main conclusion we can draw analysing the performance on each single models on the three different VEXAS tables, in terms of classification quality metrics, 
is that joining different classification methods, based on different algorithms and input parameters significantly improved the classification results. In addition, we note that for some models, the ANN algorithm performs worse than the other two, especially in classifying stars and quasars. We have also tried to change the configuration of the ANN to obtain a better score but we were not successful. We nevertheless keep this algorithm in the final ensemble learning because this does not happen for all the models and all the input tables in the same way and it makes the derived uncertainties on the classification more conservative.

Here, for simplicity, we only provide a visualization of the distribution of predictions across the classes in the form of confusion matrices. This is one of the most commonly used classification quality metrics for ML based algorithm in astrophysics.

In general terms, a confusion matrix shows the fraction of sources belonging to a certain 'true' class and classified within a predicted class. For an $m$-class problem, the confusion matrix is defined as the $m\times m$ table, where each row represents the number of instances in a predicted class, and each column represents the number of the instances in an ground-truth  class\footnote{transposing the table will not affect the results.}. Each cell of the table represents the number of the sources from a class $i$, classified as class $j,$ and the diagonal shows how many objects (often given in percentage) have been correctly classified.

\begin{figure}[h]
    \centering
    \includegraphics[scale=0.21]{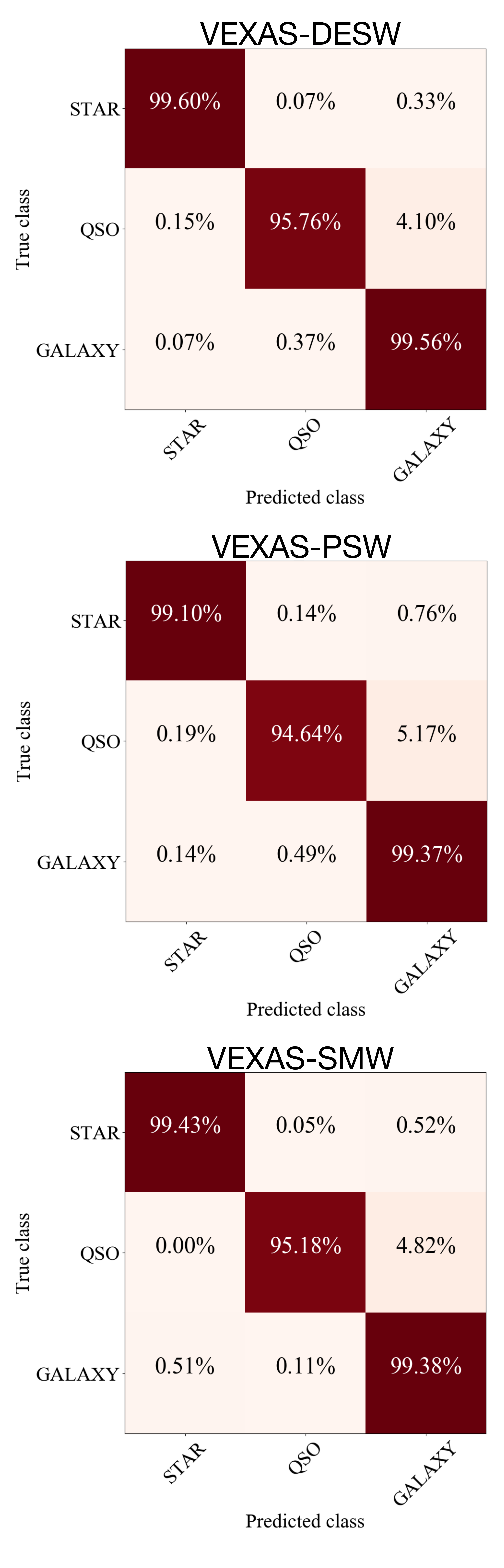}
    \caption{Confusion matrices obtained for the META\_MODEL on the test samples (20\% of the SDSS spectroscopic dataset) for the three tables.} 
    \label{fig:confmatr}
\end{figure}

The confusion matrices for each of the three different VEXAS test tables, computed from the META\_MODEL, are plotted in Figure~\ref{fig:confmatr}.

The scores obtained for the objects classified correctly (on the diagonal) are very high in all cases. We always obtained $>99\%$ of correct identifications on the \stars and \galaxy classes, and $>94\%$ on the \qso class (which is the least populated class). 
We also note that, for all tables, the largest fraction of mis-classified objects occured for \qso classified as \galaxy,  reflecting the fact that we can have both central-engine-dominated and host-dominated emission in these classes.

\begin{figure*}
    \centering
    \includegraphics[scale=0.55]{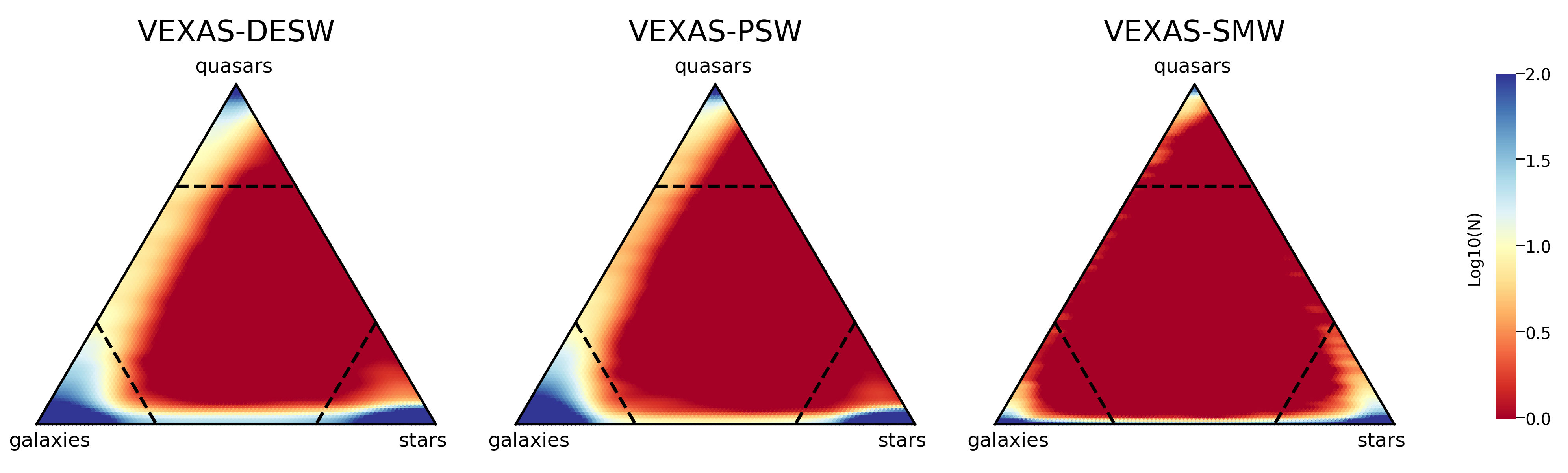}
    \caption{Density plot of the number of objects as a function of probability. The colour-bar indicates the $\log_{10}(N)$ per each probability cell of size $0.01\times0.01$. The black horizontal dashed lines indicate where the threshold $p_{\rm class}$ is $\ge 0.7$.}
    \label{fig:probability}
\end{figure*}

An interesting fact we can deduce from the confusion matrices is that the model trained on the VEXAS-DESW dataset (top panel) performs slightly better in splitting the sources in galactic and extragalactic (i.e. the contamination from stars in galaxies and quasars is the lowest: only $\approx0.4\%$ \stars are misclassified as \qso or \galaxy, versus $\ge 0.6\%$ for the other two tables). 

Unfortunately, the presence of the $u$-band in VEXAS-SMW does not seem to help in classifying quasars, however we stress again that only 19\% of the full table has $u-$band detections and the imputation procedure is sub-optimal in SMW. We will tackle this problem in forthcoming releases of VEXAS, extending the cross-match with the second data release of the  NOAO Source Catalog (NSC, \citealt{NSC_DR2_2020}), comprising 3.9 billion objects over $\sim35000$ square degrees with $ugrziY$ precise photometry. 
Finally, we also observe that the fraction of \stars classified as \qso in VEXAS-SMW (bottom panel) is equal to zero and the fraction of \qso classified as \stars is also minimal (0.05\%). However, we believe that this is mainly a limitation of the AE imputation on quasars, as we will show in the next Section and in Appendix~\ref{app:mag_imput_tec}. 

\begin{figure*}
    \centering
    \includegraphics[scale=0.23]{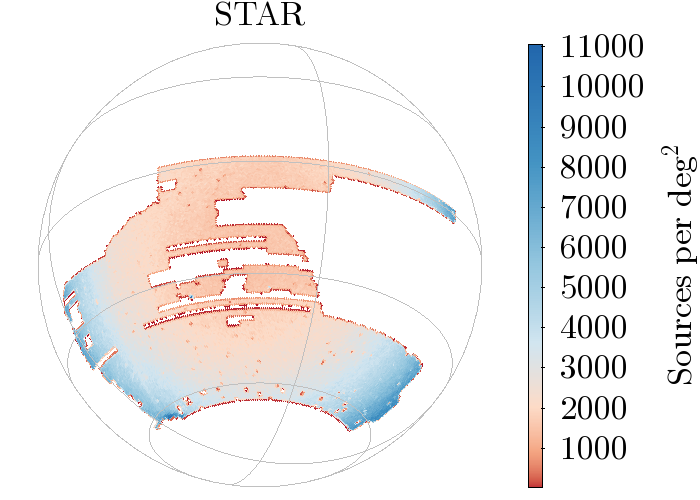}
    \includegraphics[scale=0.23]{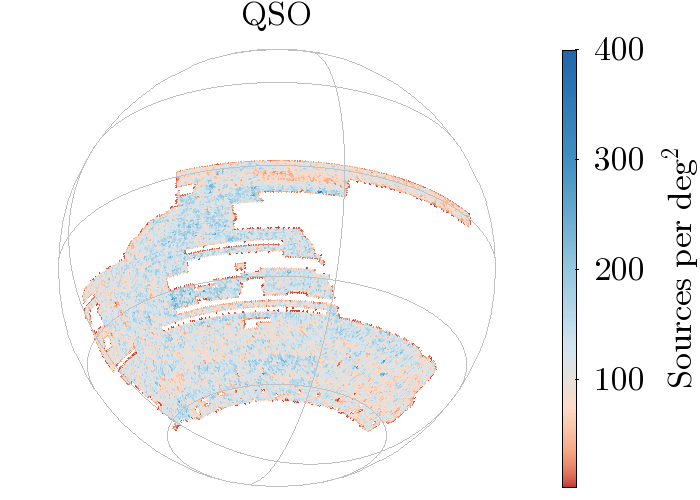}
    \includegraphics[scale=0.23]{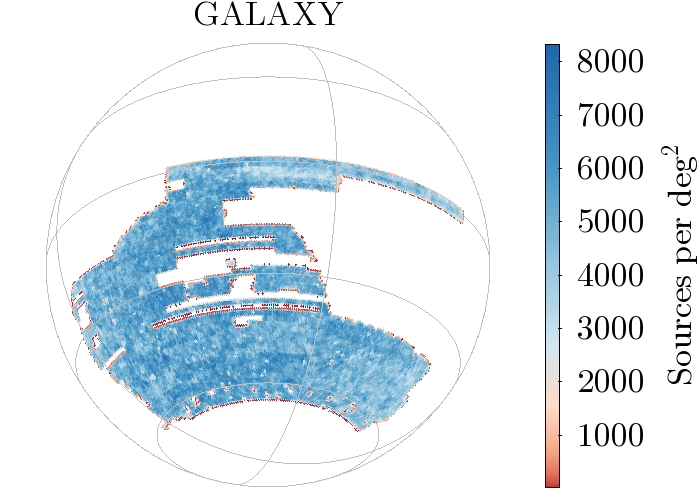}
    \includegraphics[scale=0.23]{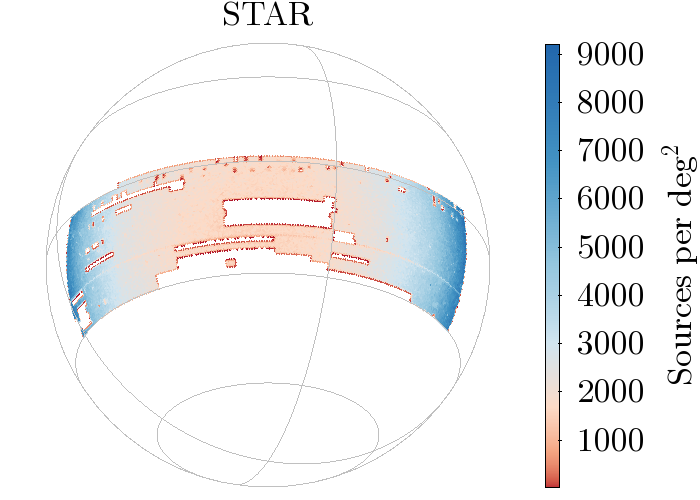}
    \includegraphics[scale=0.23]{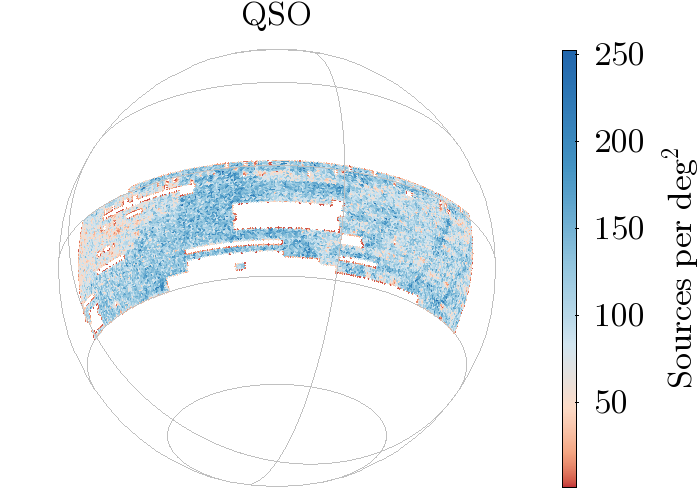}
    \includegraphics[scale=0.23]{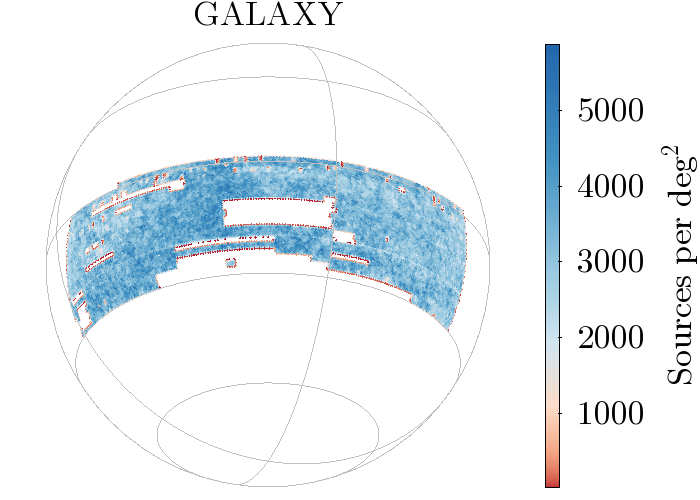}
    \includegraphics[scale=0.23]{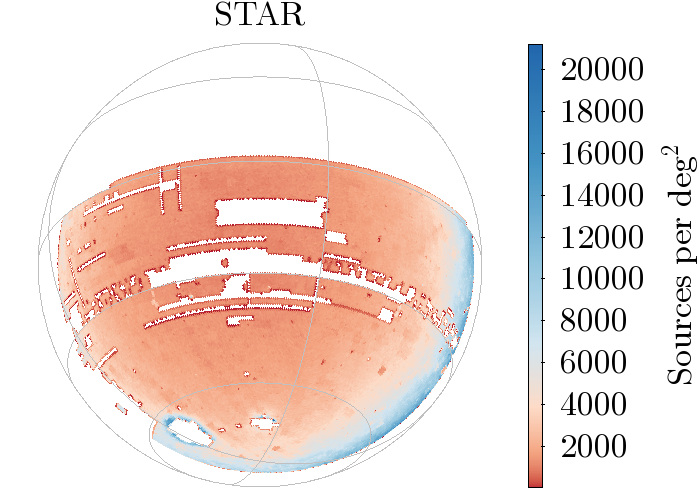}
    \includegraphics[scale=0.23]{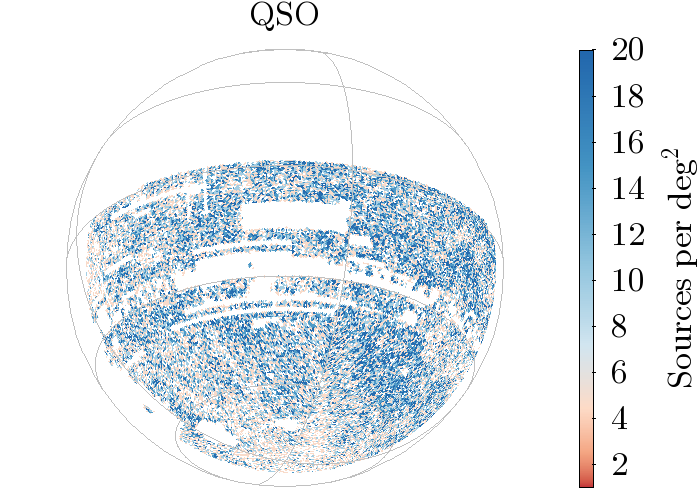}
    \includegraphics[scale=0.23]{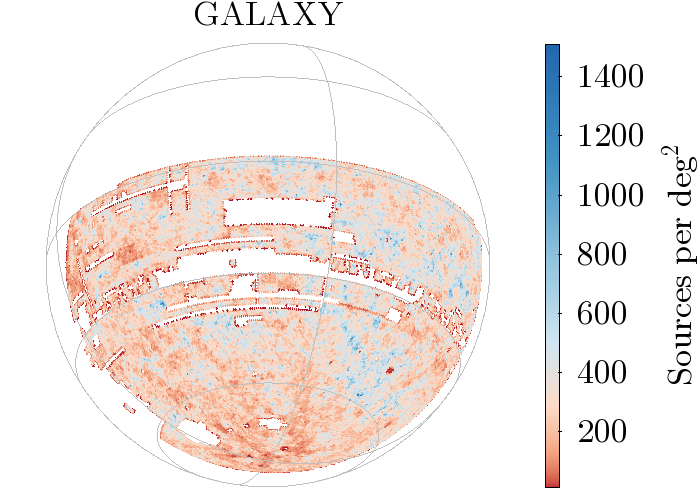}
    \caption{Spatial density of each class of objects in the three VEXAS tables. 
    The top raw shows results obtained for the VEXAS-DESW, the middle row refers to VEXAS-PSW and the bottom row shows the results obtained for VEXAS-SMW. Toward the borders the contribution from MW stars increase the density of \stars. For \qso and \galaxy the non uniform spatial distribution is mainly due to different depths reached by the surveys in that region. }
    \label{fig:sky_density}
\end{figure*}

\section{Classification results and validation}
\label{sec:results}
Our META\_MODEL returned, for each object in the VEXAS input catalogues, three
numbers, representing the probability of belonging to each of the three classes of objects: $p_{\texttt{STAR}}$, $p_{\texttt{GALAXY}}$, $p_{\texttt{QSO}}$, obtained solving a `One-\textit{vs}-Rest' problem. 

In general, a source belongs to a given class when the corresponding probability for that class is the highest. However, we considered as trustable classifications those for which the probability is $p_{\rm class}\ge 0.5$. 
With this simple assumption, starting from the input $\approx33$ million objects in VEXAS-DESW, $\approx22$ in VEXAS-PSW and $\approx32$ in VEXAS-SMW, we classify 1.6\% of the sources as quasars, 37.4\% as stars and 61.0\% as galaxies for DESW, 1.9\% quasars, 48.4\% stars and 49.7\% galaxies for PSW. For VEXAS-SMW instead, 91.4\% of the sources are classified as stars, 8.35\% as galaxies and only a 0.25\% are classified as quasars.  
The percentages relative to the classes are thus comparable for VEXAS-DESW and VEXAS-PSW, while very different for VEXAS-SM which is roughly two magnitude shallower than the other two surveys. 

Depending on the scientific purpose one might have, it would be more appropriate to use a even more severe threshold, with the purpose of maximise as much as possible the purity of of the obtained sample, penalizing however the total number of selected objects. On the other hand, with the aim of assembling the largest possible sample of a given class of objects, one might use slightly lower probability thresholds, paying the price of a larger number of contaminants. This is why we list the number of objects classified in each class as a function of the probability threshold, for each VEXAS catalogue, in Table~\ref{tab:class_numb_prob} and we release the probabilities values. 

\begin{table}[]
    \centering
    \begin{tabular}{c|c|c|c}
\multicolumn{4}{c}{} \tabularnewline
\multicolumn{4}{c}{\textbf{VEXAS-DESW}}\tabularnewline
\hline\hline
\textbf{p} & \textbf{\stars} & \textbf{\qso} & \textbf{\galaxy}\\ 
\hline
0.5 & 13\,225\,467 & 552\,847 & 21\,593\,339 \\\hline
0.6 & 13\,207\,001 & 524\,388 & 21\,543\,204 \\\hline
0.7 & 13\,185\,692 & 496\,514 & 21\,479\,584 \\\hline
0.8 & 13\,156\,283 & 465\,135 & 21\,381\,884 \\\hline
0.9 & 13\,105\,319 & 422\,868 & 21\,145\,300 \\
\hline \hline
\multicolumn{4}{c}{} \tabularnewline
\multicolumn{4}{c}{\textbf{VEXAS-PSW}}\tabularnewline
\hline\hline
\textbf{p} & \textbf{\stars} & \textbf{\qso} & \textbf{\galaxy}\\ 
\hline
0.5 & 10\,586\,167 & 408\,598 & 10\,886\,179 \\\hline
0.6 & 10\,558\,641 & 380\,388 & 10\,832\,329 \\\hline
0.7 & 10\,526\,795 & 351\,648 & 10\,769\,276 \\\hline
0.8 & 10\,485\,364 & 318\,899 & 10\,680\,432 \\\hline
0.9 & 10\,410\,731 & 271\,353 & 10\,505\,175 \\
\hline \hline
\multicolumn{4}{c}{} \tabularnewline
\multicolumn{4}{c}{\textbf{VEXAS-SMW}}\tabularnewline
\hline\hline
\textbf{p} & \textbf{\stars} & \textbf{\qso} & \textbf{\galaxy}\\ 
\hline
0.5 & 29\,242\,620 & 80\,943 & 2\,673\,869 \\\hline
0.6 & 29\,214\,063 & 76\,406 & 2\,645\,062 \\\hline
0.7 & 29\,181\,114 & 72\,004 & 2\,612\,243 \\\hline
0.8 & 29\,135\,737 & 67\,144 & 2\,569\,647 \\\hline
0.9 & 29\,050\,417 & 59\,434 & 2\,494\,954 \\\hline
\hline
    \end{tabular}
    \caption{Number of objects classified in each class as a function of the probability threshold for the DESW, PSW and SMW final tables.}
    \label{tab:class_numb_prob}
\end{table}

A visualization of the class distribution of the objects in the output catalogues is plotted in Figure~\ref{fig:probability}, in the form of a triangle density plot. The colours indicate the $\log_{10}$ number of objects contained in a probability cell of size $0.01\times0.01$. Each corner represents the maximum probability to belong to a given class ($p_{\rm class}$ = 1) and the dashed horizontal black lines show the threshold level of $p_{\rm class}$ = 0.7, which we defined as threshold for the 'high-confidence objects' as it corresponds to objects which have a probability with a confidence of more or less one $\sigma$ (0.67). This is, in our opinion, the natural compromise between purity and completeness and is used throughout the remaining of the paper for all the plots and tables. 

We show instead the density of each class of object on sky in Figure~\ref{fig:sky_density} and report the mean values obtained for high-confidence objects in Table~\ref{tab:density_p07}.

Once again, we note that for VEXAS-DESW and VEXAS-PSW the numbers are similar, especially for stars  ($\approx 2850$-$3050\mathrm{/deg^{2}}$) and quasars ($\approx 110$-$115\mathrm{/deg^{2}}$). For galaxies, the density in DESW is higher with $\approx 4700\mathrm{/deg^{2}}$ (versus $\approx 3100\mathrm{/deg^{2}}$ in PSW). 
The situation is instead very different for VEXAS-SMW, where the density of stellar objects is comparable, but that of extragalactic sources is more than ten times smaller. As already previously said, we attribute this difference to the shallower depth of the Sky Mapper Survey. 

Since quasars and galaxies are more or less uniformly distributed on the sphere, the fact that the spatial distribution of these objects is not always uniform, is due to spatial changes in the surveys depth caused by  extinction, different exposure times, or different number of bands observed in that region. For stars, instead, toward the borders we approach the MW disk, and this, of course, increases of a factor of $\sim10$ the density of galactic objects. 

We remind the reader that our metrics for the classification accuracy are averaged over the sky. This means that there may be more cross-contamination towards the low  galactic-latitude edges of the VEXAS footprint where the MW stellar density is higher. 
In future, and with more demanding iterations, one may also fold-in some spatial information to re-balance the scores accounting for how many object for each class are expected at a given position. In this DR2, we preferred to simply give the mean output scores, because any further iteration depends on accurate coverage maps and on a proper model for the MW stellar density. 


\begin{table}[]
    \centering
    \begin{tabular}{c|c|c}\hline\hline
\textbf{Table} &  \textbf{Class} & \textbf{Density (\# /deg$^2$)} \\
\hline
\hline
\multirow{3}{*}{VEXAS-DESW} & \stars & 2889.3 \\
&\qso &111.0 \\
&\galaxy &4697.1 \\\hline
\multirow{3}{*}{VEXAS-PSW} &\stars &3037.8 \\
 &\qso &103.1 \\
 &\galaxy &3106.5 \\\hline
\multirow{3}{*}{VEXAS-SMW}  &\stars &3324.8 \\
 &\qso &10.7 \\
 &\galaxy &300.0 \\
\hline
\hline
\end{tabular}
    \caption{Density of high-confidence ($p_{\rm class}\ge 0.7$) \texttt{STAR}, \qso and \galaxy in the three output tables.} 
    \label{tab:density_p07}
\end{table}

\begin{figure*}
    \centering
    \includegraphics[scale=0.7]{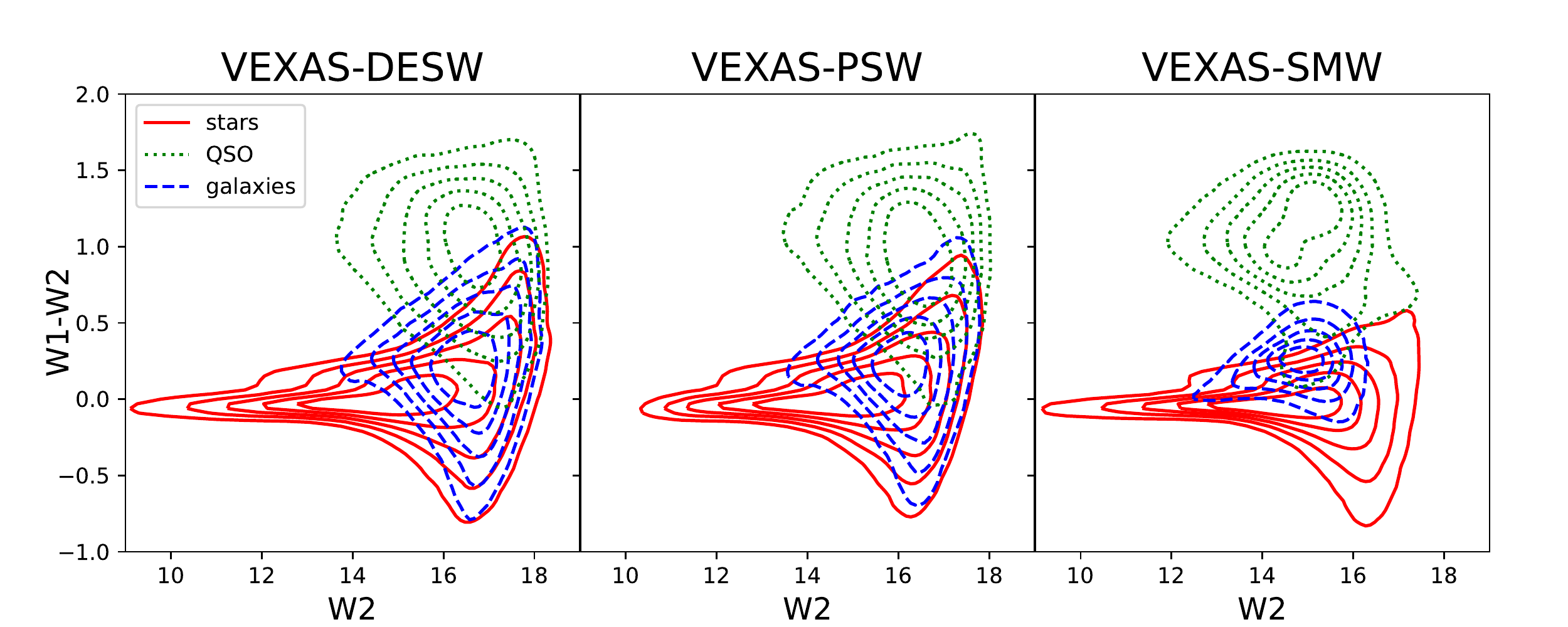}
    \includegraphics[scale=0.7]{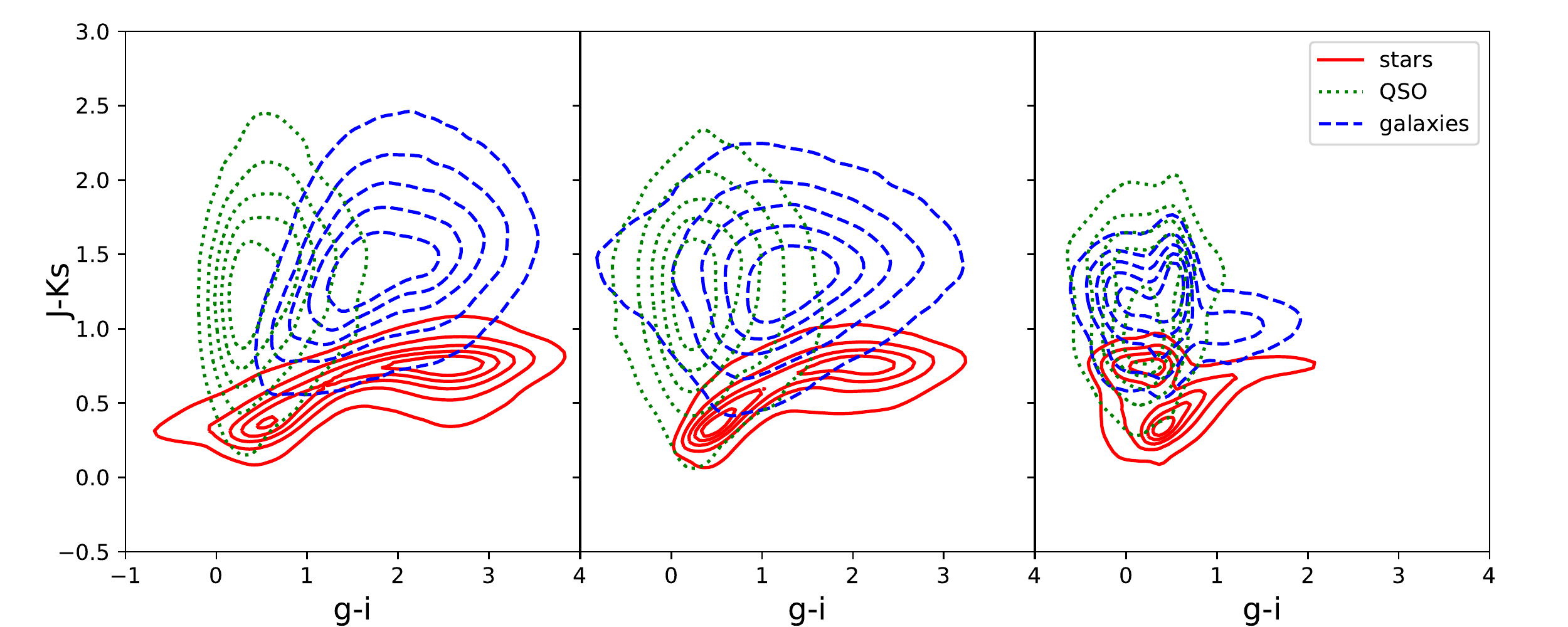}
    \caption{Selected colour-colour and magnitude-colour diagrams of VEXAS high-confidence ($p_{\rm class}=0.7$) \stars (red contours), \qso (green contours) and \galaxy (blue contours), over the three optical footprints (VEXAS-DESW, left column, VEXAS-PSW, middle column, and VEXAS-SMW  right column), split according to the predicted class. The bottom right panel shows the issue with the imputation for VEXAS-SMW, which is described in detail in Sec.~\ref{sec:results} and in Appendix~\ref{app:mag_imp_sm}.} 
    \label{fig:mag-col_plots}
\end{figure*}

To further assess the reliability of the classification results, in Figure~\ref{fig:mag-col_plots} we plot two among the most common infrared and optical magnitude-colour and colour-colour diagrams used in the literature to separate different class of objects \citep[e.g.][]{stern2005, stern2012, Assef13,Chehade18}. The contours show object density levels, with the different colours indicating the three families of objects, as specified in the caption. The stellar locus (with low $J-K_s$) and the galaxy and quasar regions (with higher $W2$ and $W1-W2$) are clearly visible. These plots also show that the three macro-classes cannot be separated with very high accuracy through simple colour cuts, emphasizing the need for full ML-based classification approaches. 

\begin{figure}
    \centering
    \includegraphics[scale=0.42]{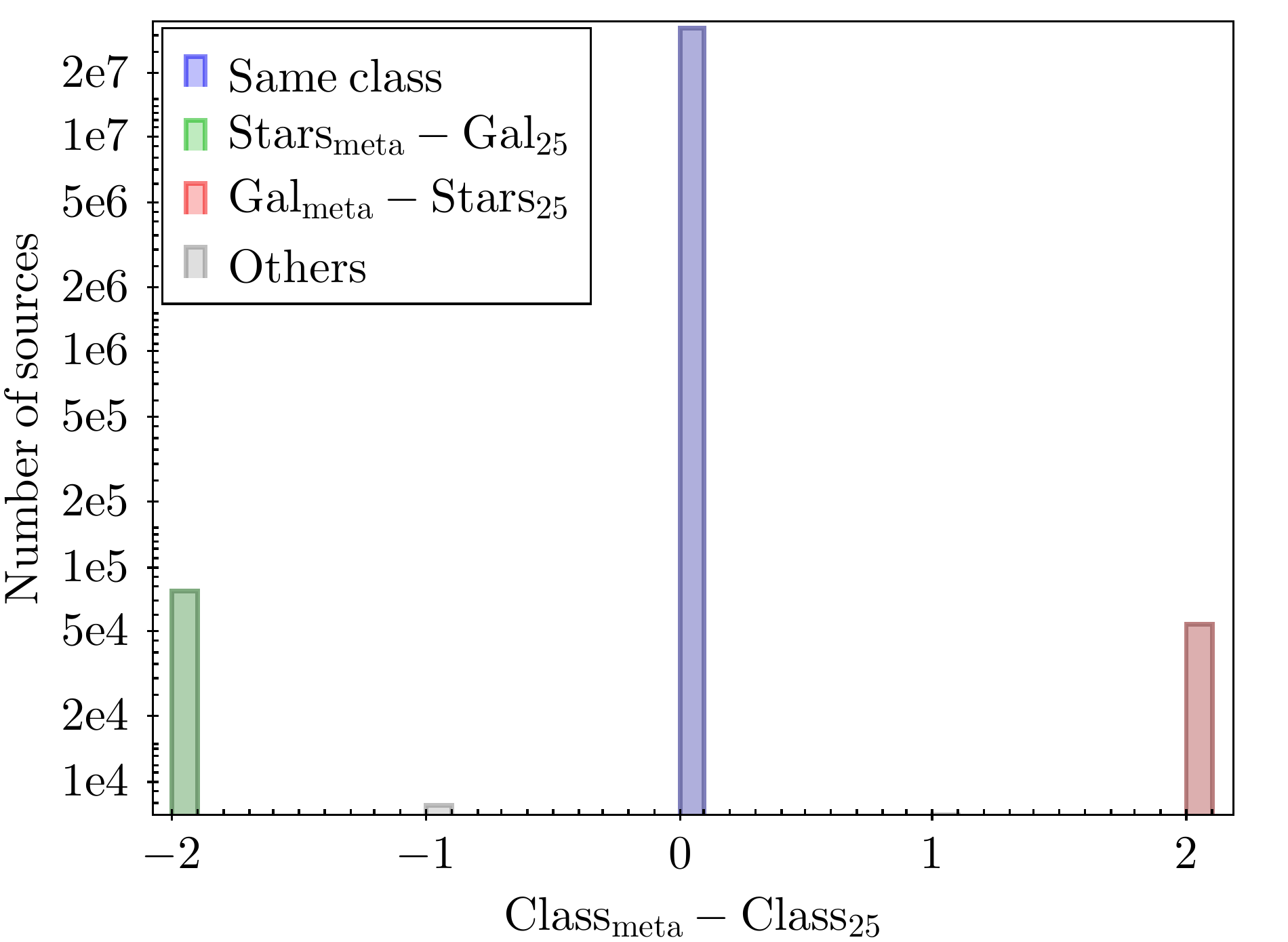}
    \caption{Comparison between the classification obtained with our fiducial META\_MODEL and with a single CatBoost Model ($25$). 99.5\% of the sources is classified in the same class.}
    \label{fig:class_comp_sm}
\end{figure}

For VEXAS-SMW, the three classes of objects do not show the same separation as in the other two tables (bottom-right panel, $J-Ks$ versus $g-i$). We show in Appendix~\ref{app:mag_imp_sm} that this is caused by the imputation, which suffers from the under-representation of quasars in the input sample. However, Figure~\ref{fig:class_comp_sm} 
demonstrates that the \textit{classification} is not affected by this problem.

\begin{figure*}
    \centering
    \includegraphics[scale=0.37]{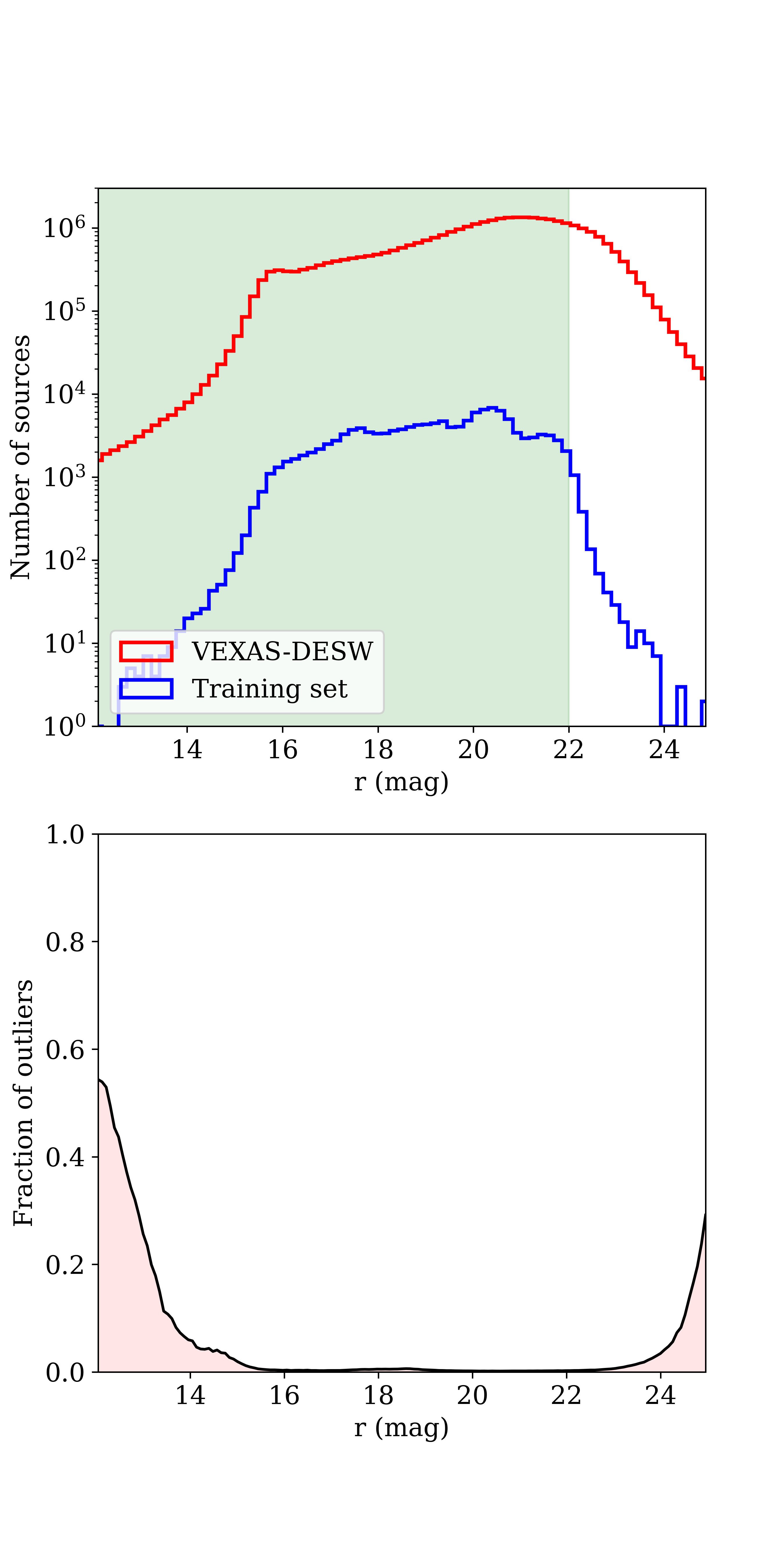}
    \includegraphics[scale=0.37]{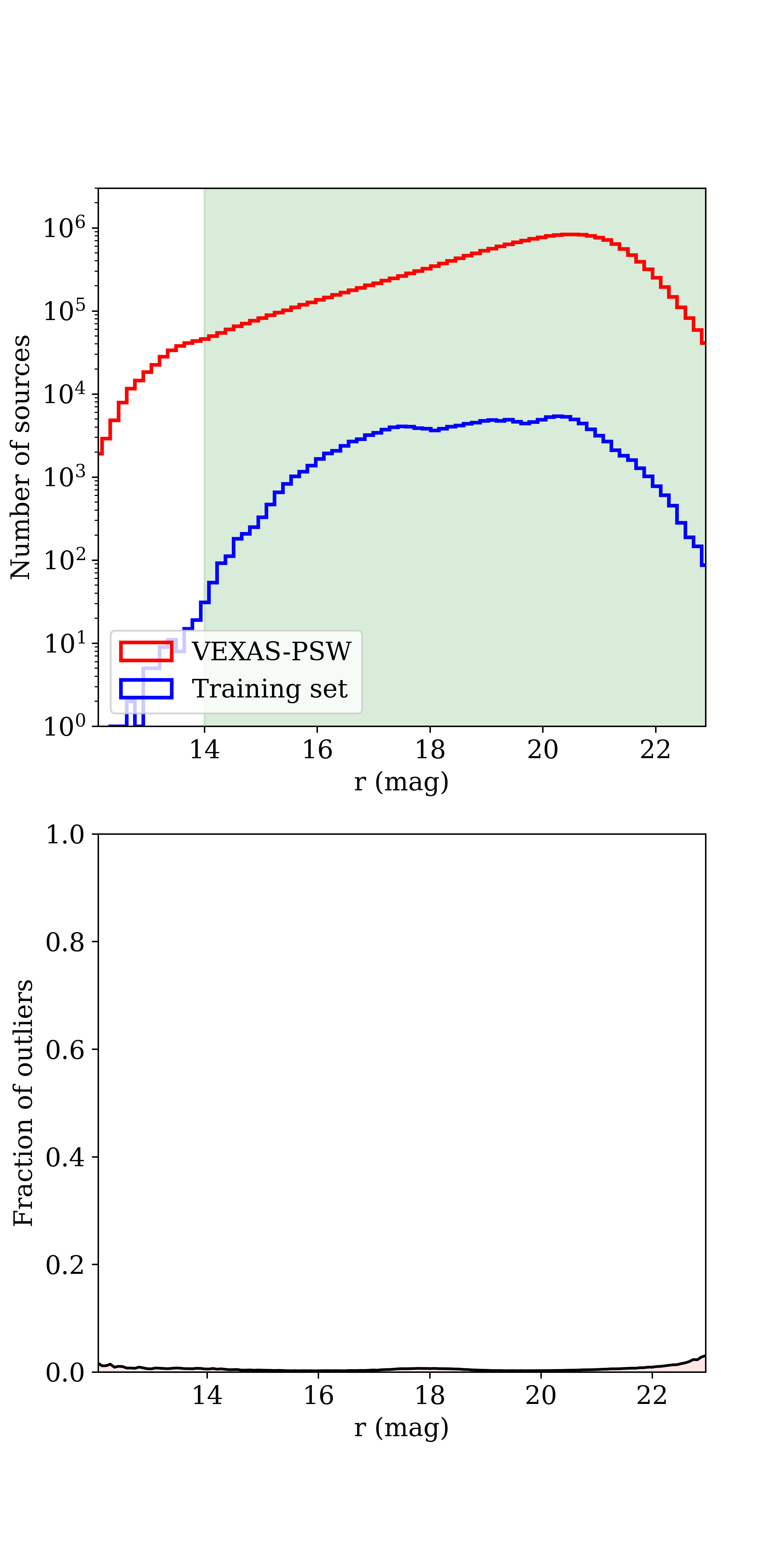}
    \includegraphics[scale=0.37]{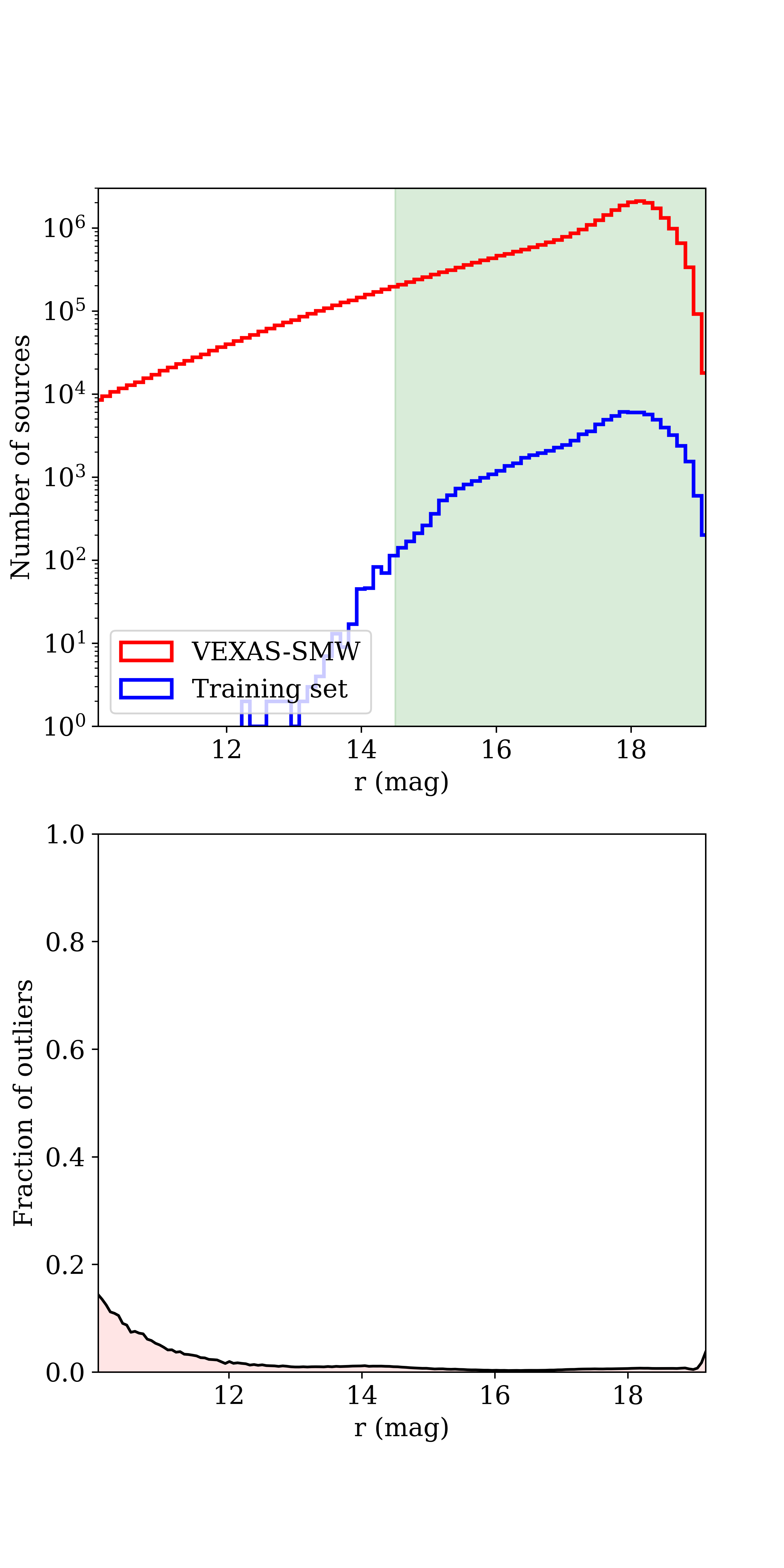}
    \caption{\textit{Top:} Histogram of the $r$ magnitudes for the input catalogues (blue) and the training sample (red) for each of the VEXAS tables. \textit{Bottom}: Fraction of outliers (see text for more details) as a function of the $r$ magnitudes for each of the VEXAS tables.}
    \label{fig:completeness_training}
\end{figure*}

The Figure shows the difference between the class of each object predicted by the META\_MODEL and that predicted by a single CatBoost model (\#25, without imputation). 
For 99.5\% of the objects, the difference is null, 
and the three $p_{\rm class}$ obtained from the two models are almost identical. The classification changes only for $\sim14.6\times10^4$ source ($\approx0.5\%$ of the total), mainly 
from \stars to \galaxy or viceversa (Class$_{\rm meta}-$Class$_{25}=\pm 2$). 
The classification for \qso is almost completely unaffected, which demonstrates that the META\_MODEL is not affected by the sub-optimal imputation for this class.  

For the sake of completeness, in this DR2, we release both sets of three probabilities (with and without imputation) in the classified VEXAS-SMW table. Moreover, in all three released VEXAS tables we added a flag column to identify the imputed magnitudes.


\subsection{Safe ranges, outliers and saturation}
\label{sec:saferange}
One of the most common issue in machine learning based classifications is that the 'depth' of the data is larger than that of the training set. The most common solution is to cut out the input tables, limiting the inference to bright objects only hence avoiding any extrapolation to unseen regions in the space of features (e.g. \citealt{Khramtsov19, Khramtsov20}).

However, this approach is not desirable in our case, since it would violate the main purpose of the VEXAS Project: collect as much information as possible in the multi-wavelength sky and thus classify the largest possible number of sources. Therefore, here we do not restrict the classification to the brightest sources only, but nevertheless caution the readers that at the very bright/faint end the classification might be less secure.

We plot the completeness of the training samples in the $r$-band for each VEXAS input table in the upper panel of  Figure~\ref{fig:completeness_training}. While for the VEXAS-PSW table a secure classification is ensured at almost all magnitudes, since the distribution of training and input are very similar,  for the VEXAS-SMW (VEXAS-DESW) table the training sample is incomplete at the bright (faint) end. 
To better quantify this, we define as 'safe ranges' the $r$-band region where the ratio between the training set and the corresponding input table is larger than 0.1\%. 
The safe ranges, which are highlighted with shaded green regions in the plots, are  $r<22^m$ for VEXAS-DESW, $14^m<r<23.2^m$ for VEXAS-PSW and $14.5^m<r<19.5^m$ for VEXAS-SMW.
For magnitudes outside these intervals, we caution the readers that the classification might be slightly bias due to an under-sampled training set.

A possibility to improve the coverage of a under-sampled training set has been provided in the literature based on a re-weighting procedure of the training sample (e.g. \citealt{Sanchez14, Bonnett16}).  Unfortunately a similar algorithm will only marginally help in our case since there are objects outside the safe range region with one or more colors which are completely "off-models" (i.e. they have absolute value larger than the maximum covered by the training set). In a sense, this region is not under-sampled by the training, but not sampled at all. 
We therefore define "outlier" a source for which at least one colour has a value smaller (larger) than the minimum (maximum) value allowed by the training set. 
The number of outliers as a function of the $r$-band magnitude is plotted in the bottom panels of Figure~\ref{fig:completeness_training}. 
For these objects the classification is likely to be biased. 
We note, however, that the number of outliers is much smaller than the number of objects and it represents only $\sim 0.5\%$ of the whole dataset for every VEXAS table.

Finally, for the VEXAS-DESW table, where the input is much deeper than the training sample, we report in Appendix~\ref{app:DESW_faint} a test which demonstrates that despite a under-sampled training set our classification reaches fair performances.

In conclusion, in the classified VEXAS tables released in this DR2, we insert a column containing a "warning flag" which takes value of 0 if the source is within the safe ranges, 1 if it is outside them but it is not an outlier, according to the distribution of its colors, and 2 if it is an outlier.

\begin{figure}
    \centering
    \includegraphics[scale=0.28]{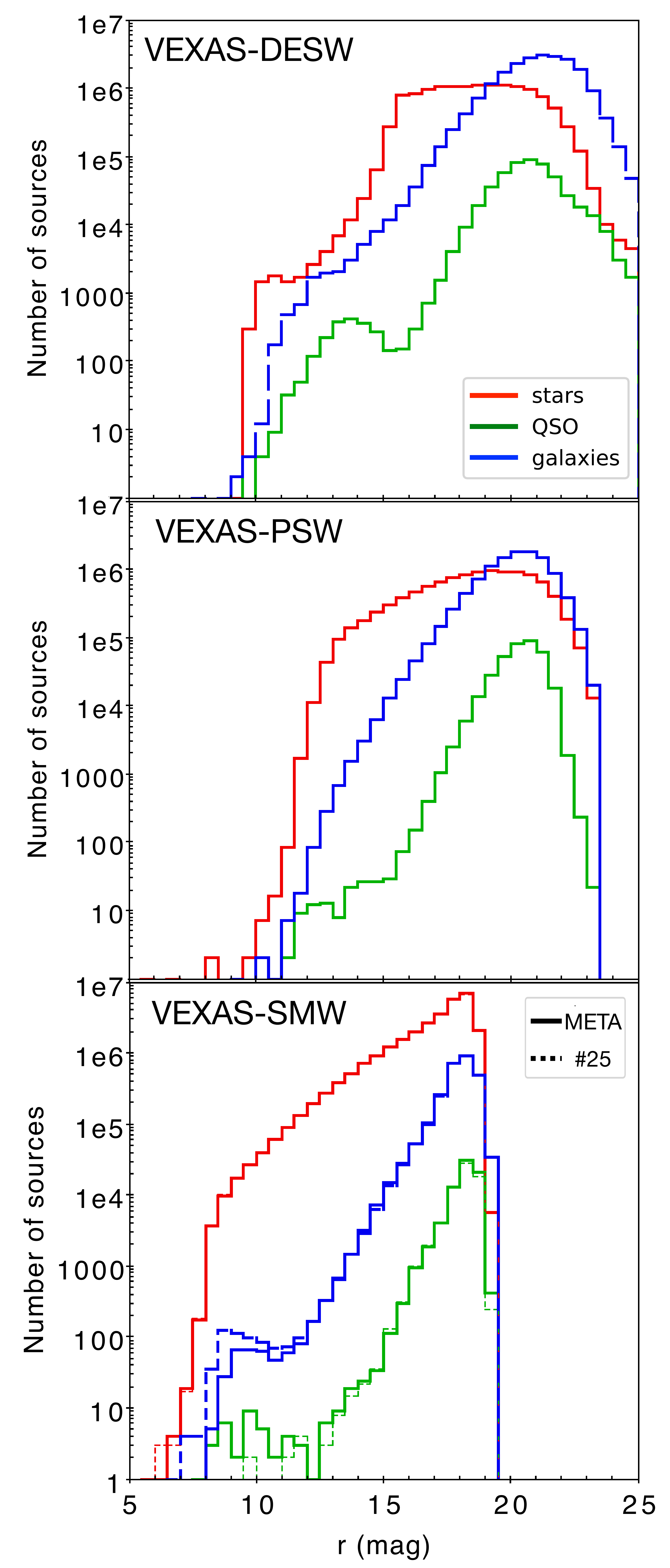}
    \caption{The $r$-magnitude distribution the sources classified in each class for each of the three classified tables. For the VEXAS-SMW, we show both the classification obtained from the META\_MODEL (solid lines) and from the single CatBoost model ($\#25$, dashed) which does not use imputation. }  
    \label{fig:distrib_rmag_sources}
\end{figure}

At the bright end, we note that in addition to a not-well sampled training set, there is the bigger problem of saturation, as we will demonstrate below.
Figure~\ref{fig:distrib_rmag_sources} shows the distribution of the $r$-band magnitude in the three VEXAS tables, for objects classified as high-confidence \stars (red), \galaxy  (green), or \qso (blue). A further confirmation that the sub-optimal imputation for the VEXAS-SMW does not bias the classification results, comes from the fact that the distribution of sources in $r-$band is almost identical using the $p_{\rm class}=0.7$ obtained from the META\_MODEL or these obtained from model $\#25$.  

All plots show similar behaviours of the luminosity functions: at the faint end they all have the same shape except for the depth cut-off, which is dominated by the WISE depth in DESW and PSW and by the optical in SMW; at the bright end, there are secondary bumps and cutoffs which are an artefact of saturation, which occurs at $r\approx15$ in DESW, $r\approx14$ in PSW and $r\approx11$ in SMW. The saturated objects amount to $\approx1\%$ in DESW, and an overall correction is
\begin{eqnarray}
    mag_{\mathrm{corr}} & = & 8.0+ 1.25(mag_{\mathrm{DES}}-12.0)\\  \nonumber & \text{if} &\ \  mag_{\mathrm{SM}}<mag_{\mathrm{DES}}+1\ \mathrm{and}\ mag_{\mathrm{SM}}<16.5
\end{eqnarray}
for $mag=(g,r,i),$ and 
\begin{eqnarray}
    z_{\mathrm{corr}} & = & 8.0+ 1.1(z_{\mathrm{DES}}-11.5)\\
    \nonumber & \text{if} &\ \  z_{\mathrm{SM}}<z_{\mathrm{DES}}+1\ \mathrm{and}\ z_{\mathrm{SM}}<16.5
\end{eqnarray}

Comparing SMW and PSW, there is an overall offset $gri_{\rm PS1}-gri_{\rm SM}=0.2$ and $z_{\rm PS1}-z_{\rm SM}=0.7$ on all magnitudes.

The issue of saturation on the performance of the classification at the bright end can also be seen in the astrometric properties of the objects, e.g. the proper motions of bright objects classified as quasars, which are discussed in the next Section. 

\subsection{Astrometric validation with \textit{Gaia}}
\label{sec:validation}

\begin{figure*}
    \centering
    \includegraphics[scale=0.42]{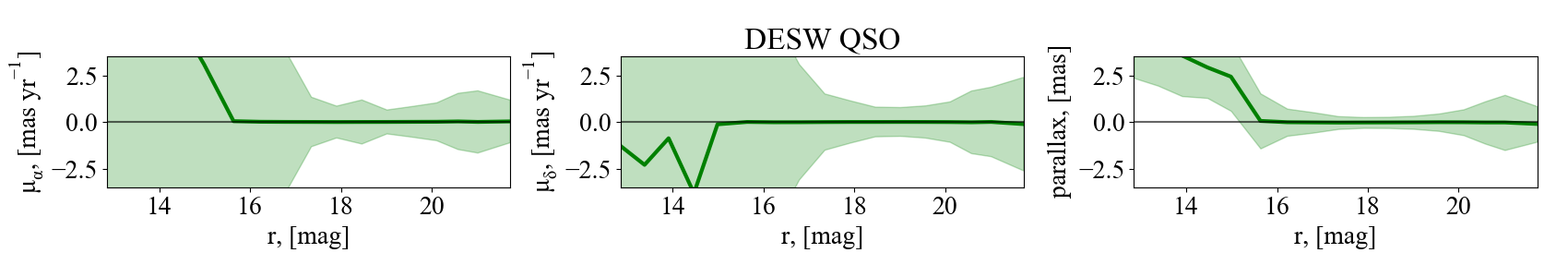}
    \includegraphics[scale=0.42]{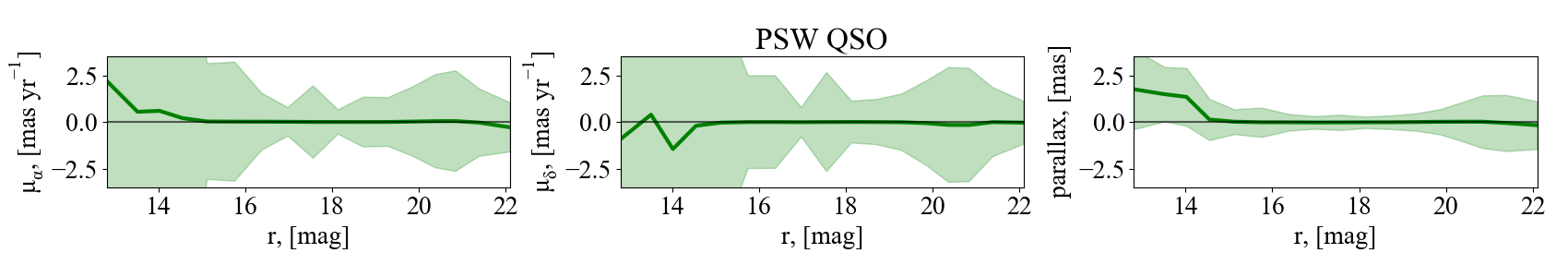}
    \includegraphics[scale=0.42]{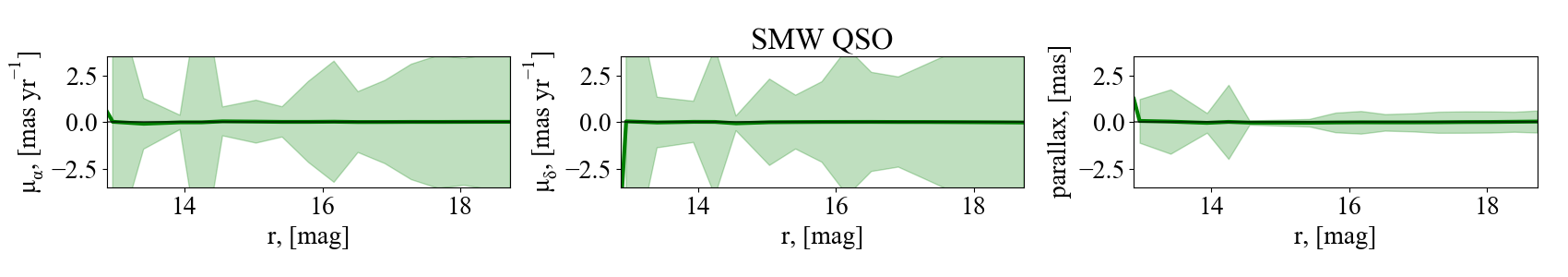}
    \caption{Dependencies of astrometric parameters on $r$-band magnitudes for objects classified as quasars ($p_{\qso}>0.7$) from each survey. From left to right we plot the two proper motion components ($\mu_{\alpha*}$, $\mu_{\delta}$), and the parallax. Solid lines and filled areas represent the median values and standard deviations of the  parameters in each magnitude bin.} 
    \label{fig:astrometry}
\end{figure*}
 
The latest data release of ESA-\textit{Gaia}, Early Data Release 3 \citep[EDR3,][]{gaiaedr3} 
provides five astrometric parameters (positions $\alpha, \delta$, proper motions $\mu_{\alpha}, \mu_{\delta}$, and parallaxes $\varpi$) for $\approx1.468$ billion sources, covering the whole celestial sphere up to $G\lesssim21^m$\footnote{This limit corresponds to $r\approx 21^m$ for quasars at $z\leq3$, \citep{proft2015}}. 
This makes the \textit{Gaia}-EDR3 an excellent means to test the purity of our catalogue, especially for quasars and stars and, indirectly, also for galaxies (see below).

We cross-matched each VEXAS output table with the \textit{Gaia} EDR3 catalogue using a $1''.5$ matching radius, and considered sources with defined astrometric parameters.  Table~\ref{tab:astrometry} lists the number of sources with a cross-match in \textit{Gaia} EDR3, split according to our classification pipeline (using the threshold $p_{class} \ge 0.7$). In all cases, more than $90\%$ of the matched objects were classified as \texttt{STAR}. 




To assess the purity of the \galaxy sample, we used the simple argument that, by construction, the \textit{Gaia} catalogue should contain very few galaxies \citep{Robin2012}, so very few of the objects with high $p_{\texttt{GALAXY}}$ should have a match in \textit{Gaia} EDR3. This is, of course, only a rough approximation since there might be a number of galaxies that \textit{Gaia} still detects, but without accurate proper motions and parallaxes, as the \textit{Gaia} astrometric solution fits only for point-like sources \citep{gaia_lindegren}. Indeed, resolved objects can be detected in \textit{Gaia} because their $G-$band magnitudes are $>0.1$mag higher than the synthetic magnitude $G_{\rm RB} = G_{\rm RP}-2.5\log_{10}[1+10^{0.4(G_{\rm RP}-G_{\rm BP})}]$ from the blue and red pass-bands, whenever these are available \citep{AS19}. 
From the cross-match with \textit{Gaia}, we found $\approx$1\,440\,000, $\approx$1\,100\,000 and $\approx$1\,910\,000 objects classified by our algorithm as \galaxy respectively in VEXAS-DESW, VEXAS-PSW and VEXAS-SMW.
Among these, only $\lesssim20\%$ sources have measured proper motions/parallaxes, and the majority are motionless, suggesting that they could indeed be extended sources at $z>0$ (galaxies). The remaining objects can instead be stars misclassified by our algorithm, or very compact galaxies.

To assess the purity of the \qso class, we analyzed the proper motions and parallaxes of the objects  classified as high confidence quasars as function of their $r$-band magnitude. As quasars are very distant sources, they have proper motions of only few micro-arcseconds, due to different cosmological effects \citep[][]{quasar_pm}, within the current accuracy of the ESA-\textit{Gaia} relativistic solution \citep[][]{gaia_lindegren}. 
Hence, we checked the proper motions and parallaxes of all the objects classified as quasars and with a match in \textit{Gaia}, to test the assumption that they are indeed zero-proper motion and zero-parallax sources, within the systematic errors. In Figure~\ref{fig:astrometry} we plot the two proper motion components ($\mu_{\alpha*}$, $\mu_{\delta}$) and the parallax as function of the $r$-band magnitude for high confidence quasars ($p_{\rm qso}\ge 0.7$) in each of the three cross-matches between VEXAS-DR2 and \textit{Gaia}-EDR3. Clearly, for magnitudes fainter than the saturation limits ($\sim15$ for VEXAS-DESW, top row; $\sim14$ for VEXAS-PSW, middle row; and $\sim11$ for VEXAS-SMW, top row; see also Fig.~\ref{fig:distrib_rmag_sources}), \qso are perfectly consistent to be motionless. 
This is not true for \stars and \galaxy, as we show in Table~\ref{tab:astrometry} were we report the Root-Mean-Square Error (RMSE) on the parallax and on each component of the proper motion for all the objects with a match in \textit{Gaia}, split in the three classes. 

\begin{table*} 
\begin{tabular}{c|c|r|r|r|c|c|c} 
\hline\hline
{\bf Table} &{\bf Class} & {\bf Total} & $\mathbf{N_{GOOD}}$
& $\mathbf{N_{GOOD}}$ &$\mu_{\alpha*}$ &$\mu_{\delta}$ &$\varpi$\\
& & {\bf matches} & & ($r>15^m$) & (mas yr$^{-1}$) & (mas yr$^{-1}$) & (mas)\\
 \hline\hline
\multirow{3}{*}{ VEXAS-DESW } & \stars & 12\,397\,335 & 11\,815\,841 & 10\,740\,398 & 5.566$\pm$11.464 & -3.722$\pm$11.891 & 1.012$\pm$1.119 \\ \cline{2-8}
& \qso & 281\,681 & 237\,191 & 235\,311 & 0.041$\pm$1.117 & -0.051$\pm$1.208 & -0.020$\pm$0.751 \\ \cline{2-8}
& \galaxy & 1\,445\,987 & 169\,198 & 158\,700 & 1.367$\pm$5.782 & -1.354$\pm$5.907 & 0.067$\pm$1.343 \\ \hline
\multirow{3}{*}{ VEXAS-PSW } & \stars & 10\,108\,741 & 9\,580\,815 & 8\,885\,919 & 3.810$\pm$11.345 & -6.014$\pm$10.819 & 0.986$\pm$1.142 \\ \cline{2-8}
& \qso & 205\,789 & 172\,019 & 171\,940 & 0.155$\pm$2.035 & -0.544$\pm$2.464 & 0.007$\pm$0.842 \\ \cline{2-8}
& \galaxy & 1\,098\,769 & 171\,344 & 170\,264 & 1.635$\pm$6.326 & -3.428$\pm$7.059 & 0.304$\pm$1.308 \\  \hline
\multirow{3}{*}{ VEXAS-SMW } & \stars & 29\,551\,196 & 29\,182\,681 & 26\,150\,059 & 2.750$\pm$10.299 & -4.633$\pm$10.748 & 0.872$\pm$0.995 \\ \cline{2-8}
& \qso & 68\,857 & 65\,442 & 65\,364 & 0.322$\pm$3.544 & -0.633$\pm$3.774 & 0.074$\pm$0.561 \\ \cline{2-8}
& \galaxy & 1\,911\,503 & 408\,894 & 407\,107 & 1.766$\pm$8.58 & -3.146$\pm$9.498 & 0.681$\pm$1.380 \\  \hline \hline
\end{tabular}
\caption{Cross-match with \textit{Gaia} EDR3 for the three VEXAS classified tables and for each class of objects.  From left to right we list:the total number of identified sources, the number of sources with measured astrometric parameters (N$_{\rm GOOD}$), the same number but only for sources that are not saturated ($r>15^{m}$), the two proper motion components and the parallax, with the corresponding standard deviations. }\label{tab:astrometry}
\end{table*}


\subsection{Internal validation}
\label{sec:int_validation}
In this Section, we present a qualitative validation of the classification results obtained comparing the probabilities derived from the different tables on common objects. 
We cross-matched the three output tables using a radius of $1.5\arcsec$.  There are 7\,817\,243 objects in common between VEXAS-DESW and VEXAS-PSW, 7\,743\,844 objects in common between VEXAS-PSW and VEXAS-SMW and, finally, 9\,912\,142 objects in common between VEXAS-SMW and VEXAS-DESW.

\begin{figure*}
    \centering
    \includegraphics[scale=0.25]{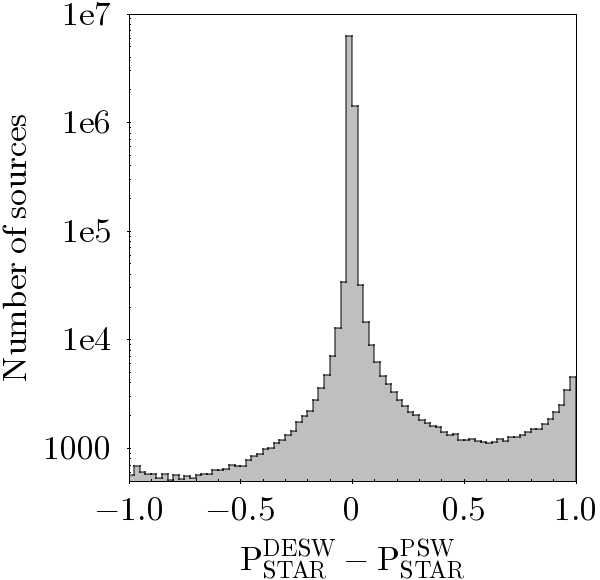}
    \includegraphics[scale=0.25]{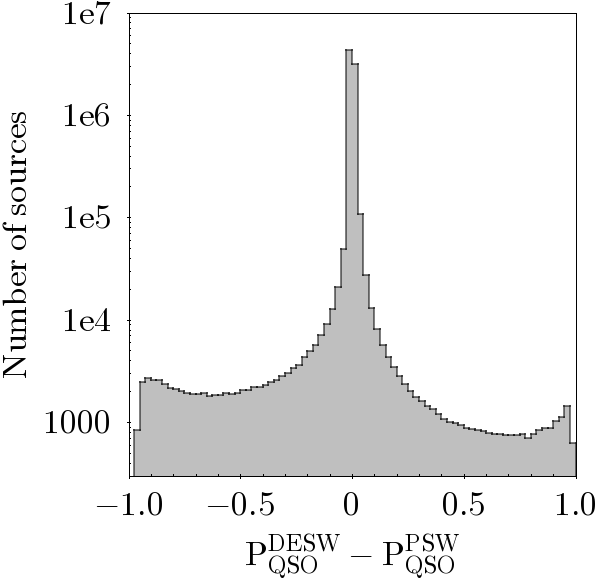}
    \includegraphics[scale=0.25]{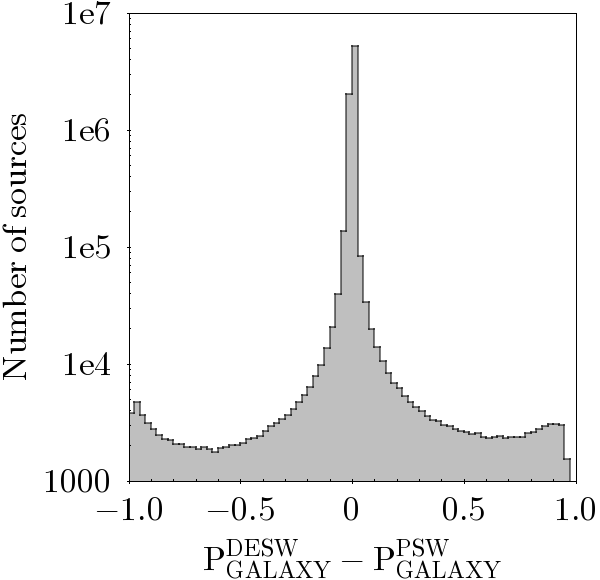}
    
    \includegraphics[scale=0.25]{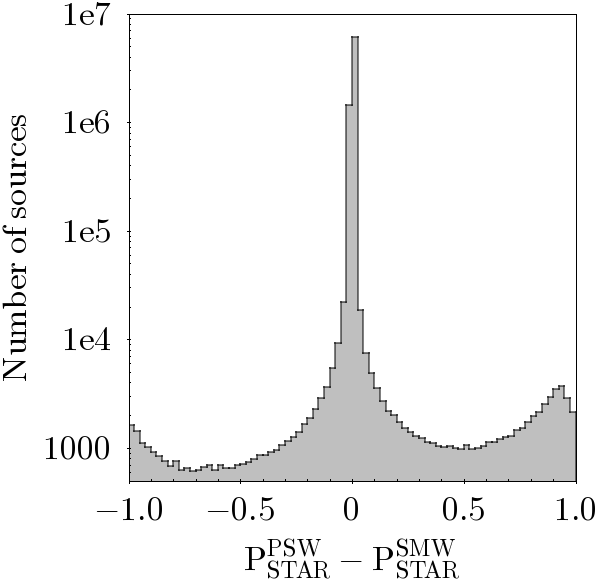}
    \includegraphics[scale=0.25]{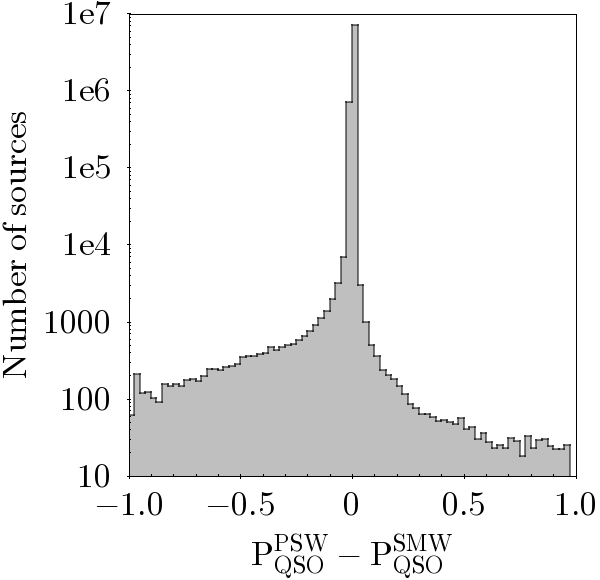}
    \includegraphics[scale=0.25]{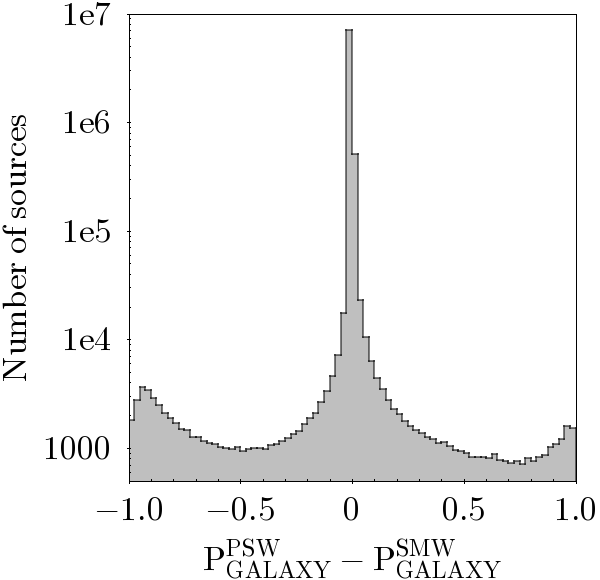}

    \includegraphics[scale=0.25]{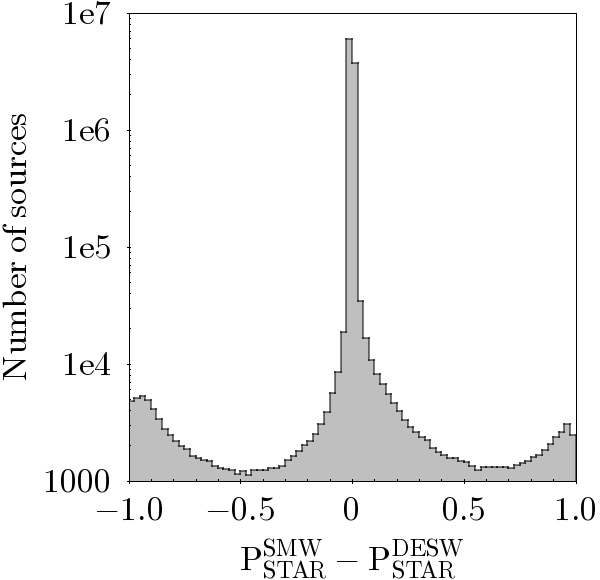}
    \includegraphics[scale=0.25]{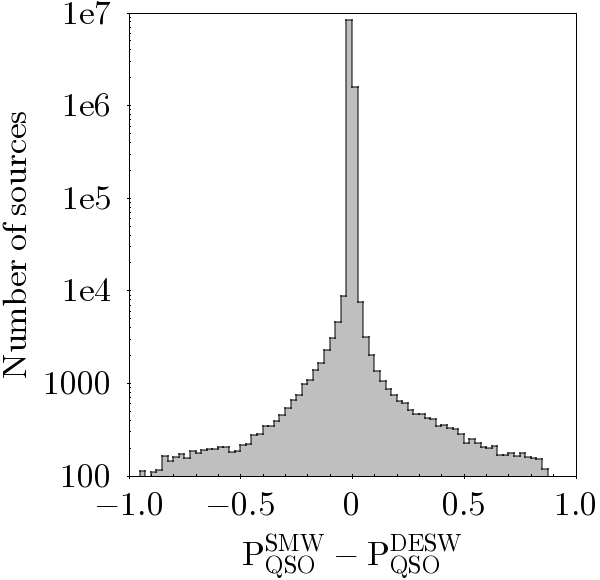}
    \includegraphics[scale=0.25]{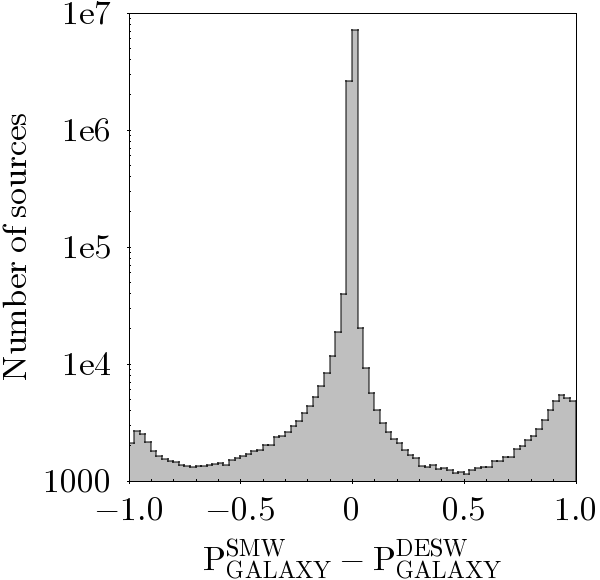}

    \caption{Distribution of the probability differences on common objects in pairs of tables:  top row, VEXAS-DESW $\times$ VEXAS-PSW, middle row, VEXAS-PSW $\times$ VEXAS-SMW, bottom row, VEXAS-SMW $\times$ VEXAS-DESW. Each column shows one class of objects (left \stars, middle \qso, right \galaxy).}
    \label{fig:histo_prob}
\end{figure*}

In Figure~\ref{fig:histo_prob} we provide a visualization of this internal validation. For each class of objects (different columns) and for each cross-match between pair of surveys (different lines), we plot
the histograms of the difference between the probabilities computed by either. 

The agreement among classification results is excellent, as the histograms always peak around 0. 
More quantitatively, in all cases and for all classes, $<3\%$ of the objects in common have probabilities that disagree by more than 0.1 and only between 0.5 and 1.2\% of the sources have probability that differ by more than 0.5. Thus, for more than 95\% of the common sources, the independent classifications obtained from the three tables are in perfect agreement. This simple test also provides further evidence that the sub-optimal imputation for the VEXAS-SMW table does not play a role during the classification step for the great majority of the sources (99.5\%).  

\section{Conclusions and outlook}
\label{sec:conclusions}
The increase in depth, footprint, and sharpness of wide-field imaging surveys posits as many opportunities as challenges. 
Multiple endeavours, both current and upcoming, aim at collecting large spectroscopic samples to map the Milky Way through its stellar content, and the Large-Scale Structure of the Universe through spatial correlations of galaxies or quasars \citep{Kollmeier17, deJong2019}.  An outstanding issue in these fiber-fed spectroscopic surveys is the target pre-selection, which was already acknowledged in the preparation of earlier quasar samples \citep{2QZ, Dawson2013}, and is even more pressing in the Southern Hemisphere, where $u-$band coverage is much shallower. More generally, the use of simple colour cuts to select objects of interest can significantly affect the overall completeness of spectroscopic follow-up samples, with direct consequences on the science. All of this is complicated by the \textit{patchiness} and partial overlap of different surveys. Here, we have deployed a collection of different machine-learning techniques to circumvent these issues, and provide membership probabilities in macro-classes.

We have trained a total of 32 different classifiers, with different techniques (ANN, kNN, trees) and input features, using magnitudes from optical surveys (DES, PS1, SkyMapper), VISTA ($J,$ $K_{s}$), and WISE magnitudes ($W1,$ $W2$). To deal with missing entries, we have 
used a machine-learning based feature imputation. The final classification is an aggregation of all the different classifiers. Each of them has undergone extensive vetting, to identify the physical properties that drive its performance, which in turn is quantified in multiple ways to find robust thresholds in the classification scores. While a clear separation can be made between most stars and extragalactic objects, the separation between galaxies and quasars is less abrupt, reflecting the range of central-engine-dominated and host-dominated emission in these objects. As simple colour-magnitude diagnostics show, our classification generalises simple colour-magnitude cuts that have been proposed in the literature \citep{stern2005, stern2012, Assef13,Chehade18},  
but also deals with the overlap of different classes, especially towards the faint end in WISE magnitudes and towards higher-redshift quasars.

The three VEXAS optical+IR classified tables, with object IDs, coordinates, optical and infrared magnitudes, including the imputed ones (which are flagged), and the probability to belong to each class, are publicly released and available for the scientific community as part of the VEXAS Data Collection (DR2) via the ESO Phase 3\footnote{The active link to VEXAS Collection  through the \href{https://archive.eso.org/scienceportal/home?data_collection=VEXAS}{Science Portal} and the  \href{https://www.eso.org/qi/}{Catalog Interface} are given in this footnote in the online version. While we are processing the Phase 3 documentation and format, we release the table via a temporary repository (\href{https://drive.google.com/drive/folders/18IjYlkKrvEB2AEcj6UZiHTOUKbijULpf?usp=sharing}{here}).}. 
The machine-learning classification pipeline and the code are available on the VEXAS Github repository\footnote{\url{https://github.com/VEXAS-team/VEXAS-DR2}}. 

The density of extragalactic objects varies with the survey depth, quite expectedly. Considering only high-confidence objects, i.e. those for which the probability to belong to the class is $p_{\rm class}\ge 0.7$, our classifiers yielded  $111$ $\mathrm{qso/deg^{2}}$ 
for the VEXAS-DESW footprint ($\approx4900$ $\mathrm{deg}^{2}$), 
$103$ $\mathrm{qso/deg^{2}}$ for the VEXAS-PSW footprint ($\approx3800$ $\mathrm{deg}^{2}$). These numbers roughly meet the requirements of the 4MOST selection for Baryon Acoustic Oscillations measurements. The density drops to $\approx 10$ $\mathrm{qso/deg}^{2}$ for the VEXAS-SMW footprint ($\approx9300$ $\mathrm{deg}^{2}$). 

All in all, the combined survey depth is the limiting factor. The VEXAS-SMW footprint is limited by the shallow optical coverage of SkyMapper, which also has less uniform coverage than DES and PanSTARRS. A solution would be to consider only NIR and midIR magnitudes in the classification, which in turn requires more uniform coverage in the NIR and smaller uncertainties in the WISE magnitudes. This may be obtained thanks to unWISE, a re-processing of WISE and NeoWISE imaging \citep{Lang2014, Schlafly19}, or with forced photometry (on SkyMapper and unWISE cutouts) based on VISTA detections. 

We caution that these samples are only the very first step for cosmological measurements, which also require spectroscopic redshifts and well characterised coverage maps. The spectroscopic follow-up is already the aim of Southern surveys (4MOST, \citealt{deJong12}; SDSS-\textsc{V}, \citealt{SDSS-V}). Since this paper is focused on the classification techniques and their application to VEXAS, the coverage maps are outside the scope of this work and will be part of a subsequent release.
\vspace{2.4in}


 \begin{acknowledgements} \\
The authors wish to thank the anonymous referee for a very constructive and useful report, which improved the quality of the final manuscript. \\
CS is supported by a Hintze Fellowship at the Oxford Centre for Astrophysical Surveys, which is funded through generous support from the Hintze Family Charitable Foundation.  
AA is supported by a grant from VILLUM FONDEN (project number 16599). This project is funded by the Danish council for independent research under the project `Fundamentals of Dark Matter Structures', DFF - 6108-00470.\\
This research has made use of the services of the ESO Science Archive Facility and of the cross-match service provided by CDS, Strasbourg. The authors are thankful to Laura Mascetti and the ESO Archive Science Group Team, led by Magda Arnaboldi, for the precious help in making the VEXAS tables Phase-3 compliant and releasing them through the Phase-3 Science Archive.\\
\medskip
Funding for the Sloan Digital Sky Survey IV has been provided by the Alfred P. Sloan Foundation, the U.S. Department of Energy Office of Science, and the Participating Institutions. SDSS-IV acknowledges support and resources from the Center for High-Performance Computing at the University of Utah. 
SDSS-IV is managed by the Astrophysical Research Consortium for the Participating Institutions of the SDSS Collaboration including the Brazilian Participation Group, the Carnegie Institution for Science, Carnegie Mellon University, the Chilean Participation Group, the French Participation Group, Harvard-Smithsonian Center for Astrophysics, Instituto de Astrof\'isica de Canarias, The Johns Hopkins University, Kavli Institute for the Physics and Mathematics of the Universe (IPMU) / University of Tokyo, the Korean Participation Group, Lawrence Berkeley National Laboratory, Leibniz Institut f\"ur Astrophysik Potsdam (AIP), Max-Planck-Institut f\"ur Astronomie (MPIA Heidelberg), Max-Planck-Institut f\"ur Astrophysik (MPA Garching), Max-Planck-Institut f\"ur Extraterrestrische Physik (MPE), National Astronomical Observatories of China, New Mexico State University, New York University, University of Notre Dame, Observat\'ario Nacional / MCTI, The Ohio State University, Pennsylvania State University, Shanghai Astronomical Observatory, United Kingdom Participation Group, Universidad Nacional Aut\'onoma de M\'exico, University of Arizona, University of Colorado Boulder, University of Oxford, University of Portsmouth, University of Utah, University of Virginia, University of Washington, University of Wisconsin, Vanderbilt University, and Yale University.\\
Wigglez acknowledges financial support from The Australian Research Council (grants DP0772084, LX0881951 and DP1093738 directly for the WiggleZ project, and grant LE0668442 for programming support), Swinburne University of Technology, The University of Queensland, the Anglo-Australian Observatory, and The Gregg Thompson Dark Energy Travel Fund at UQ.\\
GAMA is a joint European-Australasian project based around a spectroscopic campaign using the Anglo-Australian Telescope. The GAMA input catalogue is based on data taken from the Sloan Digital Sky Survey and the UKIRT Infrared Deep Sky Survey. Complementary imaging of the GAMA regions is being obtained by a number of independent survey programmes including GALEX MIS, VST KiDS, VISTA VIKING, WISE, Herschel-ATLAS, GMRT and ASKAP providing UV to radio coverage. GAMA is funded by the STFC (UK), the ARC (Australia), the AAO, and the participating institutions. 
\\
Funding for the DEEP2 Galaxy Redshift Survey has been provided by NSF grants AST-95-09298, AST-0071048, AST-0507428, and AST-0507483 as well as NASA LTSA grant NNG04GC89G.\\
This paper uses data from the VIMOS Public Extragalactic Redshift Survey (VIPERS). VIPERS has been performed using the ESO Very Large Telescope, under the "Large Programme" 182.A-0886. The participating institutions and funding agencies are listed at \url{http://vipers.inaf.it}.\\
This research uses data from the VIMOS VLT Deep Survey, obtained from the VVDS database operated by Cesam, Laboratoire d'Astrophysique de Marseille, France.\\
This work has made use of data from the European Space Agency (ESA) mission
{\it Gaia} (\url{https://www.cosmos.esa.int/gaia}), processed by the {\it Gaia}
Data Processing and Analysis Consortium (DPAC,
\url{https://www.cosmos.esa.int/web/gaia/dpac/consortium}). Funding for the DPAC
has been provided by national institutions, in particular the institutions
participating in the {\it Gaia} Multilateral Agreement.\\
Based on observations made with ESO Telescopes at the La Silla or Paranal Observatories under programme ID(s) 179.A-2005(A), 179.A-2005(B), 179.A-2005(C), 179.A-2005(D), 179.A-2005(E), 179.A-2005(F), 179.A-2005(G), 179.A-2005(H), 179.A-2005(I), 179.A-2005(J), 179.A-2005(K), 179.A-2005(L), 179.A-2005(M), 179.A-2005(N), 179.A-2005(O) 
\end{acknowledgements}


\begin{appendix}

\section{Testing the classification at the faint end of VEXAS-DESW}
\label{app:DESW_faint}
 As described in details in Sec.~\ref{sec:saferange}, the $r$-band magnitude ranges covered by the input tables are larger than those covered by the corresponding tables. This is particularly true for the VEXAS-DESW table which reaches magnitudes as faint as $r\sim25^m$ (see Fig.~\ref{fig:completeness_training}).  
Therefore, in this Appendix, we carry out a test on this table, to assess the performance of our ML classification at such faint magnitudes.

Unfortunately the number of faint objects with footprint within VEXAS-DESW and with a secure spectroscopical classification is very limited and mainly made of galaxies.  
We considered the following spectroscopic surveys:
\begin{itemize}
  \item DEEP2 redshift survey Data Release 4 \citep{deep2};
  \item VIMOS Public Extragalactic Redshift Survey (VIPERS) Data Release~2 \citep{vipers};
  \item VIMOS VLT Deep Survey (VVDS) final data release: Deep and Wide subsamples \citep{vvds};
  \item SDSS test subsample (see Section~\ref{sec:splittingdata})
\end{itemize}
and selected all objects with magnitudes fainter than $22^{m}$, i.e. all objects outside the safe ranges defined in Section~\ref{sec:saferange}.

We then applied some criteria to filter out unreliable spectroscopic measurements. In particular, for the DEEP2 we required that the spectroscopic class is not empty; for VIPERS we required that integer part of $zflg$ flag is $>2$\footnote{See \href{http://vipers.inaf.it/data/pdr2/catalogs/PDR2_SPECTRO_TABLES.html}{the survey documentation} provided as link in the online version of this manuscript.}. For VVDS we accepted as galaxies sources with $ZFLAGS$ equals to $3,4$ or $9$, as quasars sources with $ZFLAGS$ equals to $13,14$ or $19$, and as stars sources with redshift $z<0.001$.

Finally, we matched the coordinates of all sources with secure spectroscopic classification for each survey with the VEXAS-DESW classified table, using a matching radius of $1.5\arcsec$.  We obtained in this way a set of 7\,279 \galaxy, 115 \qso and 85 \stars with $r>22^m$.

\begin{figure}
    \centering
    \includegraphics[scale=0.32]{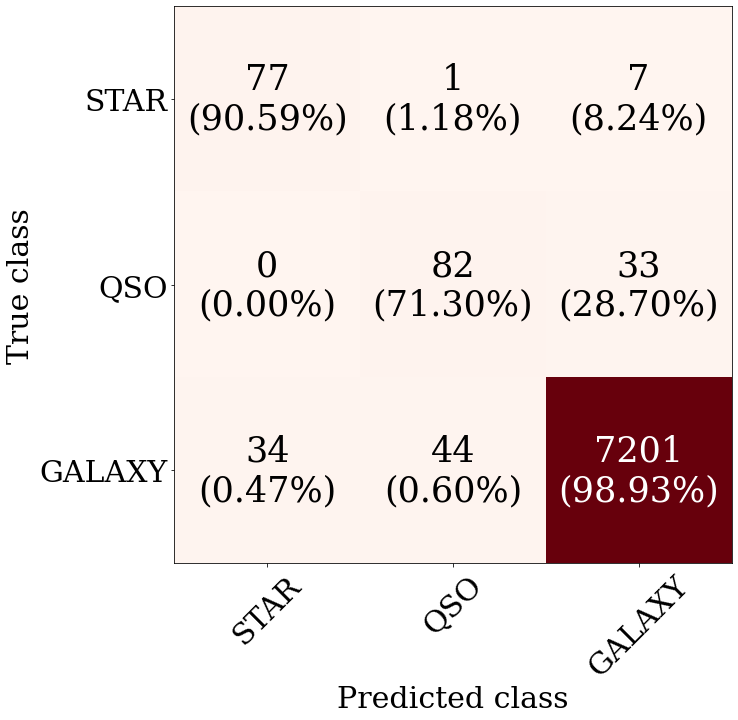}

     \caption{Confusion matrix obtained for the META\_MODEL on the deep external spectroscopic samples within  VEXAS-DESW.}
    \label{fig:deep_test}
\end{figure}

Figure~\ref{fig:deep_test} shows the comparison between the ML predicted and spectroscopically measured classes, in the form of a confusion matrix. Almost all galaxies were classified correctly from our pipeline ($\sim99\%$ of the total number of objects classified as such). For stars and quasars the performance of our ensemble learning is less optimal but still acceptable, with $\sim10\%$ of \stars and $\sim30\%$ of \qso misclassified (as galaxies mostly). We note that the lower performance obtained in the \qso class might due to the fact that the spectroscopic surveys we used here are targeting galaxies. This means that 
probably the few quasars have bright host galaxies, which contaminate their colors, making the classification harder.

In conclusion we can state that, with respect to the classification obtained within the safe regions (see Fig.~\ref{fig:confmatr}, top panel), the performance of our ML classification is worse, as expected, but still highly reliable, at least for the \galaxy class.



\section{Noisy spectroscopic datasets}
\label{app:noisy_spec}

In the main text of this manuscript we argued that the spectroscopic classification sample that was used to train the ensemble learning was not perfect, despite the manual cleaning we applied.  Here we show an experiment to confirm this statement.  
For this experiment, we only used the VEXAS-PSW table, as an example, but this result could be easily replicated to the other cases, assuming the uniformity of the label noise in spectroscopic surveys across the sky. 

\begin{figure}
    \centering
    \includegraphics[scale=0.13]{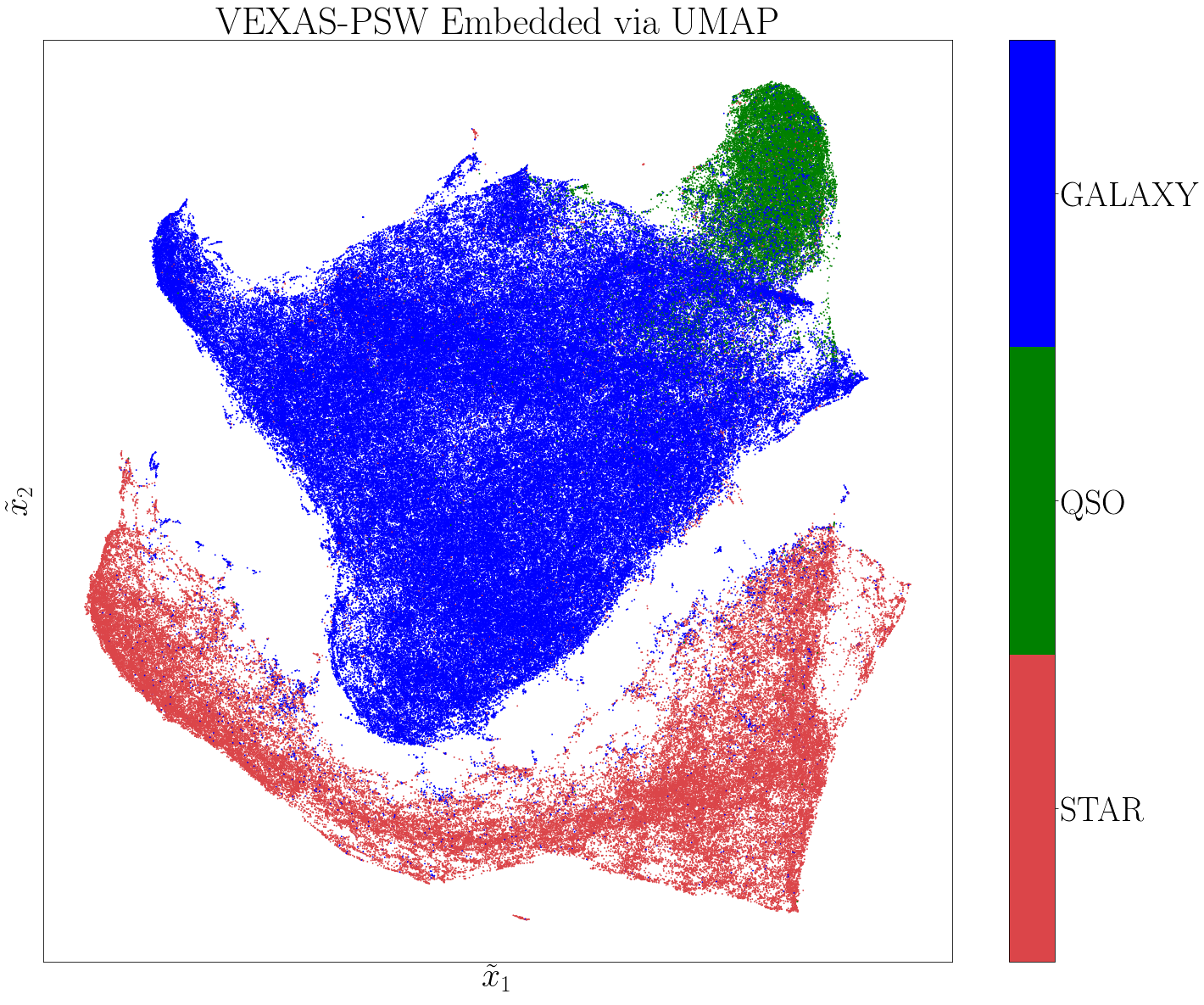}
    \includegraphics[scale=0.18]{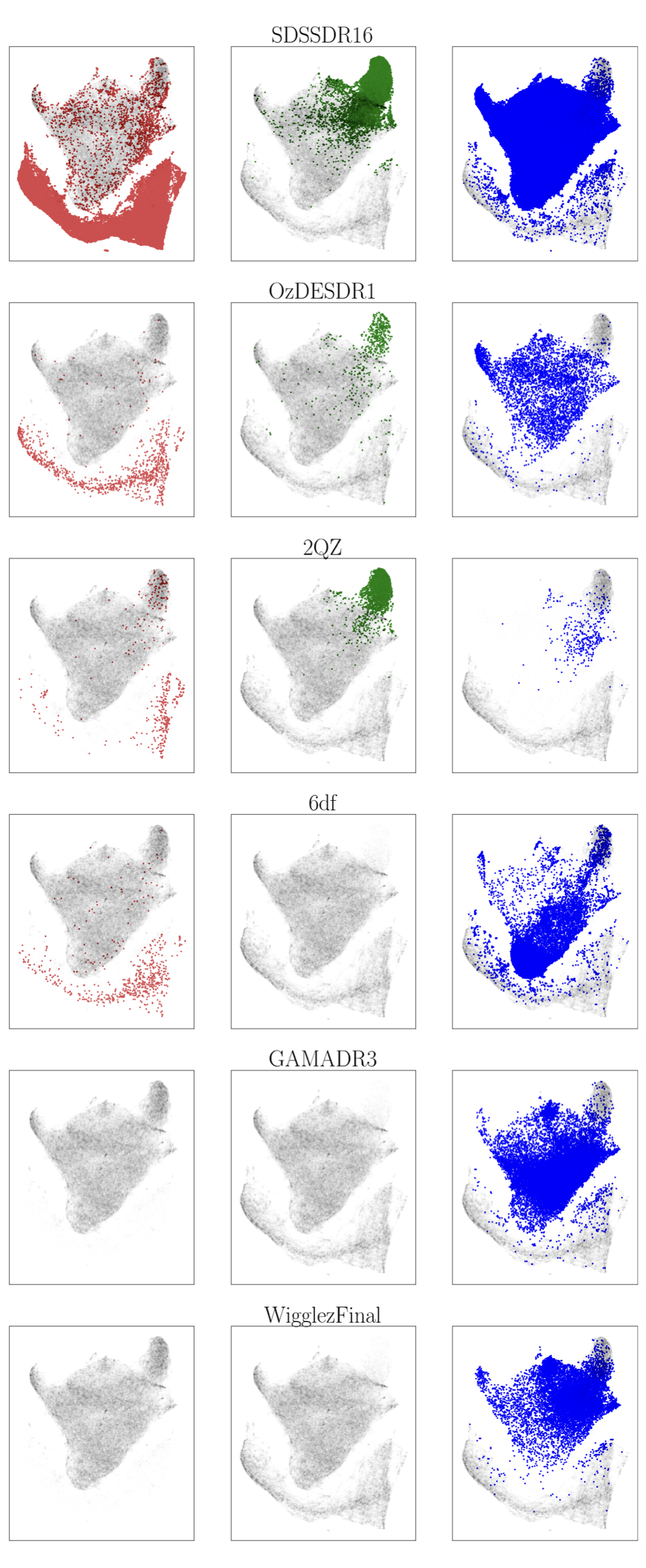}
    \caption{Two-dimensional projection of the VEXAS-PSW sources with a spectroscopic match created with UMAP. Each point represents a source within the two-dimensional reduced space, and it is colour coded according to its spectroscopic classification. In the bottom, lower panel, each line of three-plots is relative to a different spectroscopic survey, as indicated by the titles. Each panel within a line highlight a different class of objects.    Grey points represent instead the sources belonging to the other two classes.}
    \label{fig:umap}
\end{figure}

We used a dimensionality reduction algorithm called Uniform Manifold Approximation and Projection \citep[UMAP, ][]{umap}. 
The utility of such type of algorithms has already been shown in many publications, e.g., \citet{Nakoneczny2019, Clarke2019}. 
We applied the UMAP algorithm to the imputed optical+IR magnitudes from the VEXAS-PSW table, reducing the input parameter space of 9 magnitudes ($grizy$ from PS, $K_S$ and $J$ from VISTA, $W1$ and $W2$ from WISE) to only two values. 
This makes possible to produce 2D figures where the objects belonging to different class lie on different regions, as shown in Figure~\ref{fig:umap} for VEXAS-PSW (upper, bigger panel) and 
for the six spectroscopic tables used in Section~\ref{sec:training}. We colour-coded the three classes consistently with the main body of the paper: \stars are shown in red, \qso in green and \galaxy in blue.  
In the upper panel all spectroscopic matches are plotted and, visually, each family is clustered in a different region in the UMAP reduced parameter space. However, when we splits the two dimensional projection in three different panels, one for each of the different classes, as we do in the smaller panels, survey by survey, we can clearly see that there are objects belonging to a given class but laying in a region corresponding to another one. This is particularly true for stars (red points), and partially for galaxies (blue points). 

\section{Magnitude imputation}
\label{app:mag_imput_tec}

In the main body of the paper, we have imputed magnitudes in $grizy(u)JK_SW1W2$ using an AE artificial neural network. 
The need for an intelligent feature imputation is an essential step to prevent reducing dramatically the size of the VEXAS DR2 classified tables, which will prevent us to fulfill the final VEXAS purpose of building a homogeneous multi-band photometric catalogue with a sky coverage as large as possible. 
However, we have also showed that a possible issue with the imputation of the VEXAS-SMW table, for which the percentage of missing magnitudes in the optical is much larger than that in the other two tables, might exist.  
In this Appendix we first describe in more details the AE architecture, then provide results on the magnitude imputation carried out on a training sample for which the 'true' magnitudes were known a priori. Finally, we focus on the VEXAS-SMW case, showing evidence supporting the fact that the object classification based on the ensemble learning is not affected by the non-optimal imputation.

\subsection{AE architecture}
\label{app:AEarchitecture}
With respect to the traditional AE architecture, our imputer presents  a few modifications. 
First, we added a skip-connection\footnote{The idea of skip-connection is introduced in U-Net network for the segmentation of images.},
which concatenates the output AE layer (with size equal to the number of input magnitudes) and the input magnitudes. The skip-connection simply creates a layer of size $2\times$ that of the number of magnitudes. We also added one neuron to this layer, which contains the stellarity parameter $P_{STAR}$. This layer is then connected with the final layer, where each output represents an imputed magnitude. 
The activation function of each layer is a scaled exponential unit, \citep[SELU, ][]{selu2017}. Finally, we also added the additional information on the stellarity of the objects ($P_{STAR}$) at the input layer, to further help the AE to learn useful  representations and obtain, as result, a more accurate magnitude imputation. The AE architecture is summarized in Table~\ref{tab:ae_architecture}.

\begin{table}[]
\centering
    \begin{tabular}{c|c|l}
    \hline\hline
    Description & Size & Activation \\\hline
    Input & 10 & -- \\\hline
    \multirow{2}{*}{ Encoder } & 7 & SELU \\
    & 5 & SELU \\\hline
    \multirow{2}{*}{ Decoder } & 7 & SELU \\
    & 9 & SELU \\\hline
    Skip-connection & 19 & -- \\\hline
    Output & 9 & Linear \\\hline\hline
    \end{tabular}
\caption{AE imputer architecture construction. Each row represents a layer with its size (number of neurons), and activation function. The Input layer refers to the input set of magnitudes with missed values ($grizy(u)JK_SW1W2$) and additional features ($P_{STAR}$ in our case), and the Output layer to the to the imputed magnitudes. The Skip-connection concatenates the Input and output of decoder and has size equal to $2\times$ the number of magnitudes plus an additional input value ($P_{STAR}$). } 
    \label{tab:ae_architecture}
\end{table}

The input magnitude entries that are masked at random are replaced with zeros. For computational ease, we divide all magnitudes by 25, so that they all belong to the $0<(mag/25)<1$ interval. 
We trained the AE with a \texttt{logcosh} loss function and the \textsc{NAdam} optimizer\footnote{\url{http://cs229.stanford.edu/proj2015/054_report.pdf}}.


\subsection{Training and testing the performance of the AE imputation}
\label{app:AEtraining}
To train the AE imputer, we created three tables, one for each survey, comprising of only objects with the entire set of measured magnitudes ($g,r,i,z,y,J,K_S,W1,W2$ for the DESW and PSW, and $u,g,r,i,z,J,K_S,W1,W2$ for the SMW).
Then we masked out these magnitudes for a number of objects proportional to the complementary fraction given in Table~\ref{tab:num_obj_percentag}, with the exception of the $W1$-band, for which we assumed a rate of measurements equal to 99.9\% and thus masked out this magnitude for only 0.1\% of the tables. This was done to keep the training of the AE as general as possible. 
We retrieved in this way  28\,537\,891 (81\% of the full table) 15\,592\,387 (71\%) 5\,923\,761 (19\%) sources for DESW, PSW, and SMW respectively. 
As already noted above, the portion of sources with the entire set of measured magnitudes is much smaller for the VEXAS-SMW, compared to the other two tables. This is probably causing a sub-optimal performance for the imputation in SMW, as we will discuss in the next Section.

The sample was then split into training and validation samples in a proportion of 85\% -- 15\%.
At this point, the AE was trained over 150 epochs, with batch size equal to 256 sources. We scheduled the learning rate value to converge deeper into the loss minimum: if our loss on the validation sample did not change during seven epochs, we decreased the value of the learning rate of a factor of 0.1. 

\begin{figure*}
    \centering
    \includegraphics[scale=0.17]{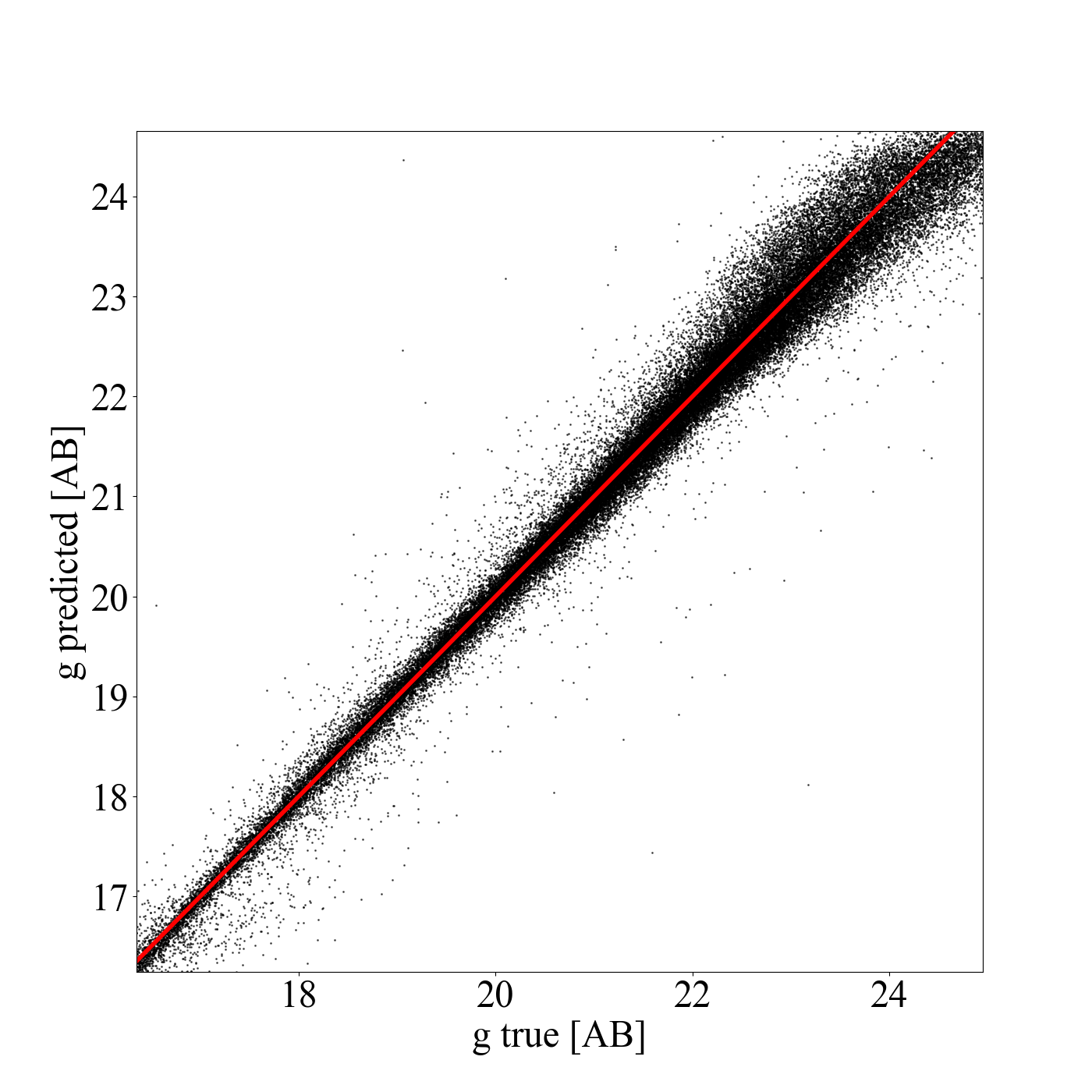}
    \includegraphics[scale=0.17]{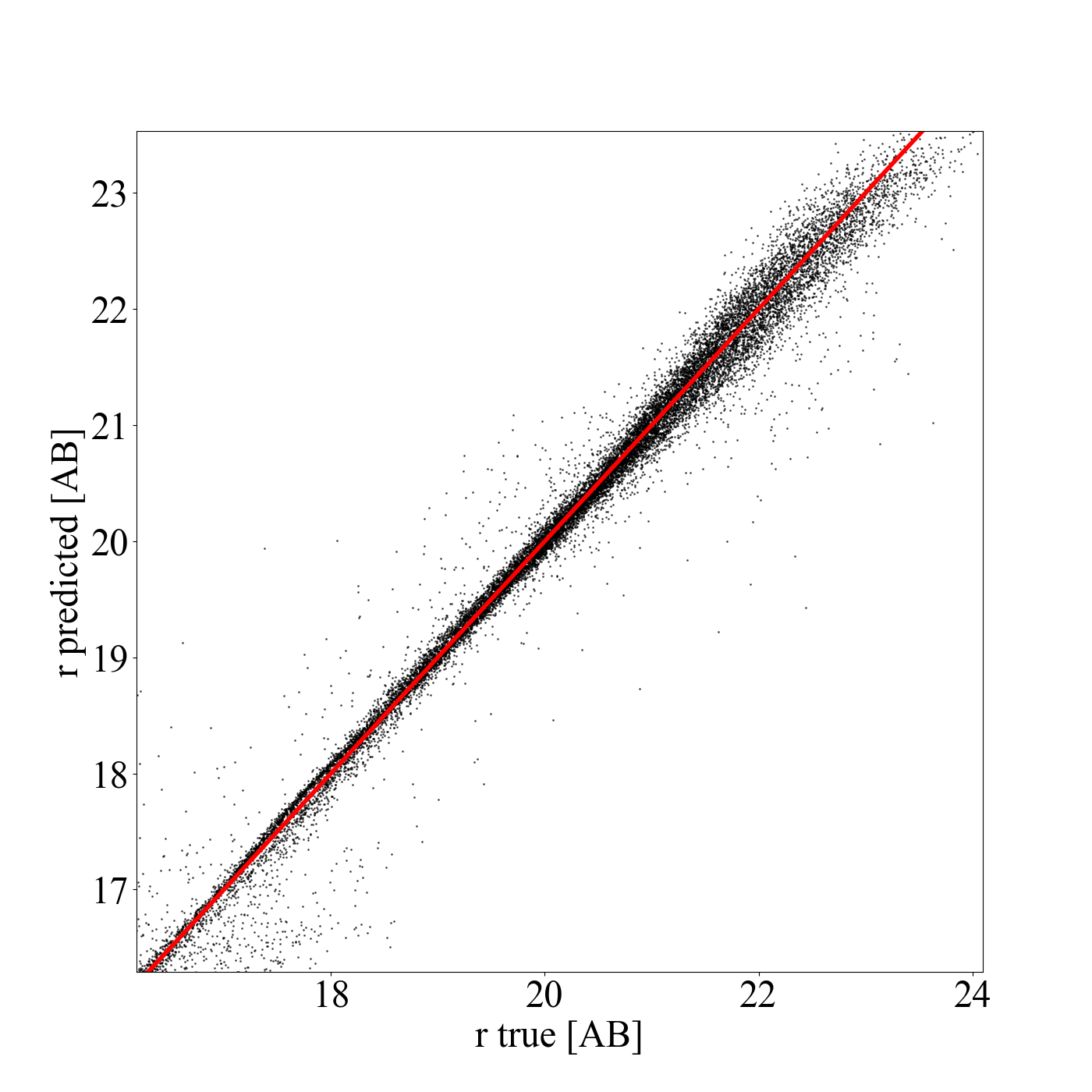}
    \includegraphics[scale=0.17]{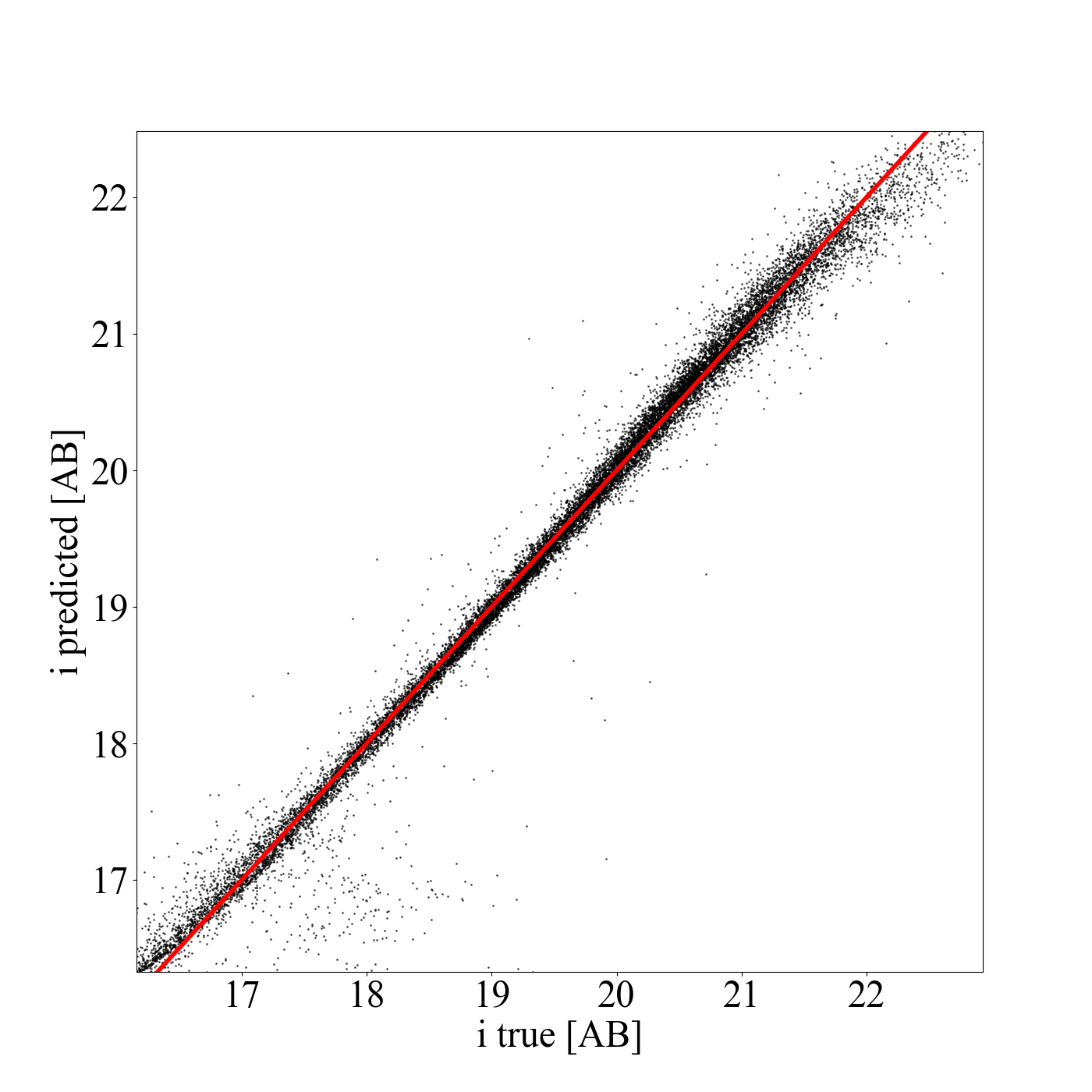}
    \includegraphics[scale=0.17]{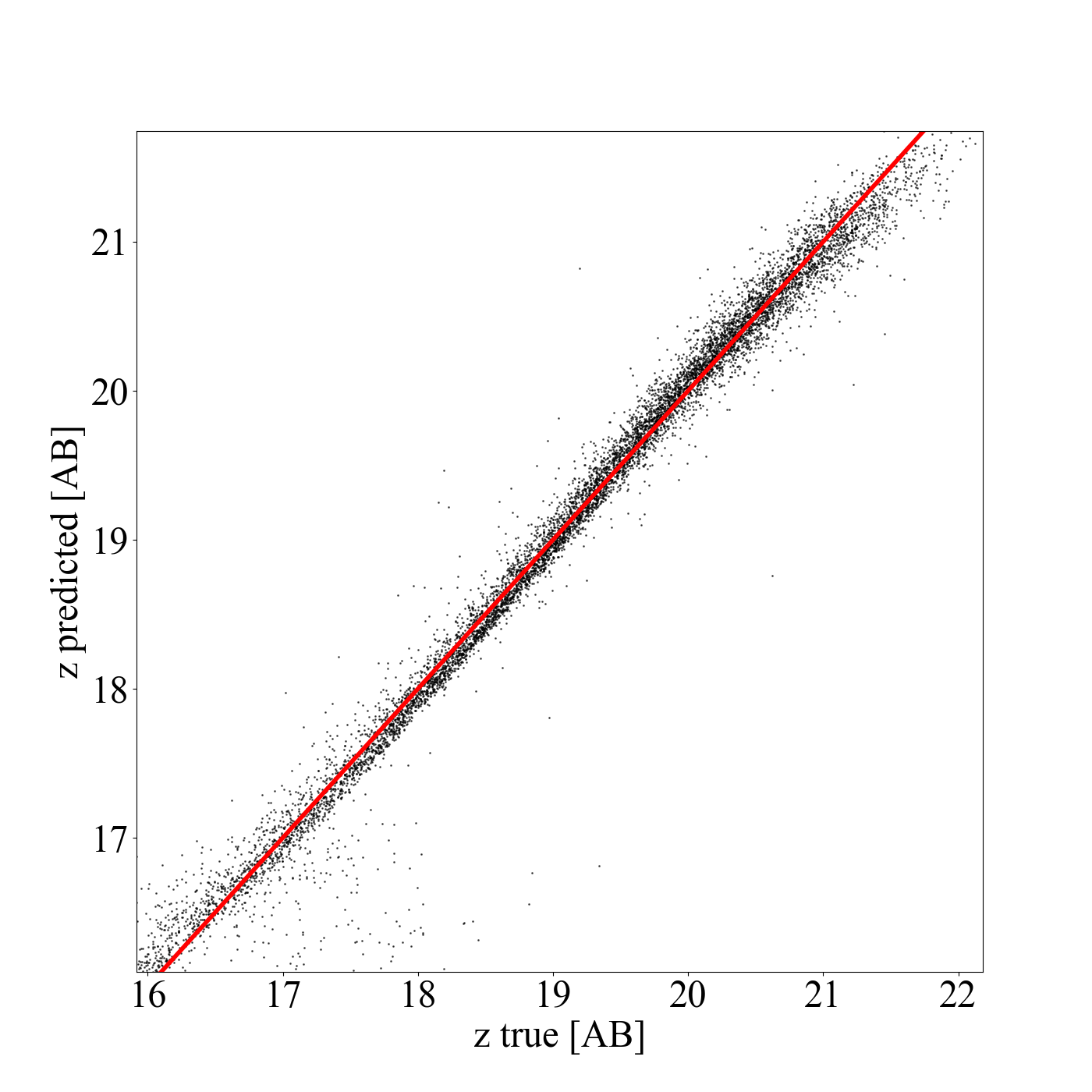}
    \includegraphics[scale=0.17]{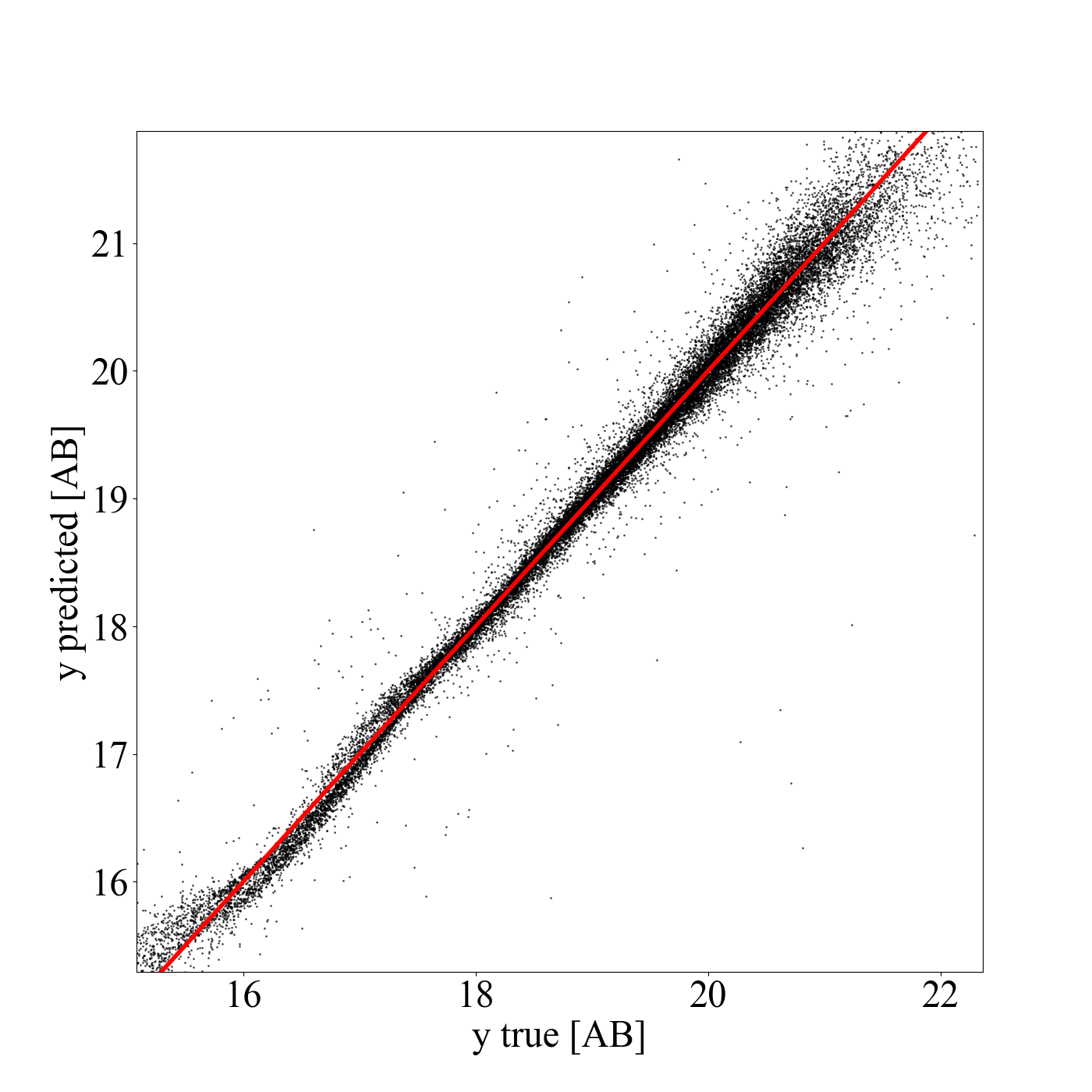}
    \includegraphics[scale=0.17]{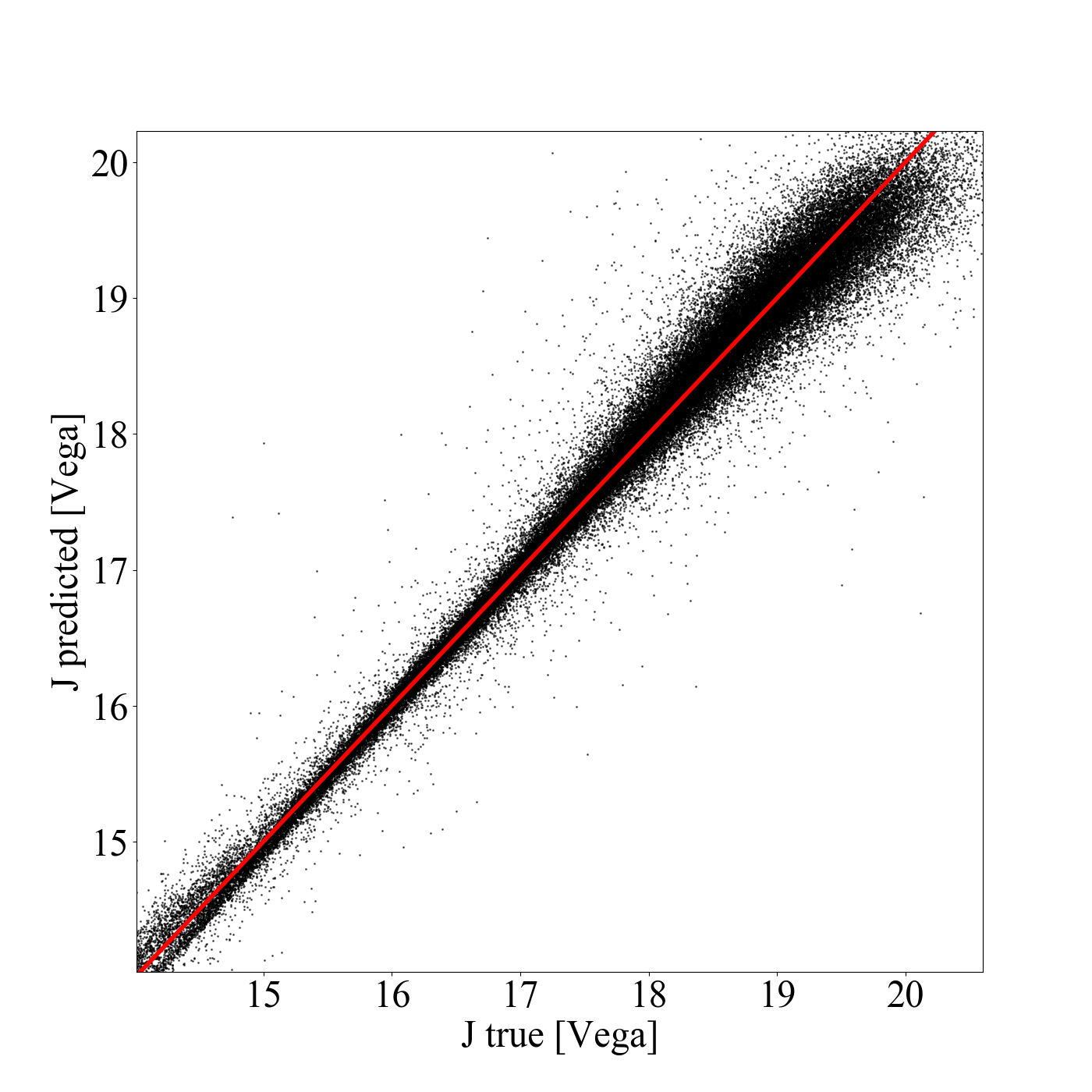}
    \includegraphics[scale=0.17]{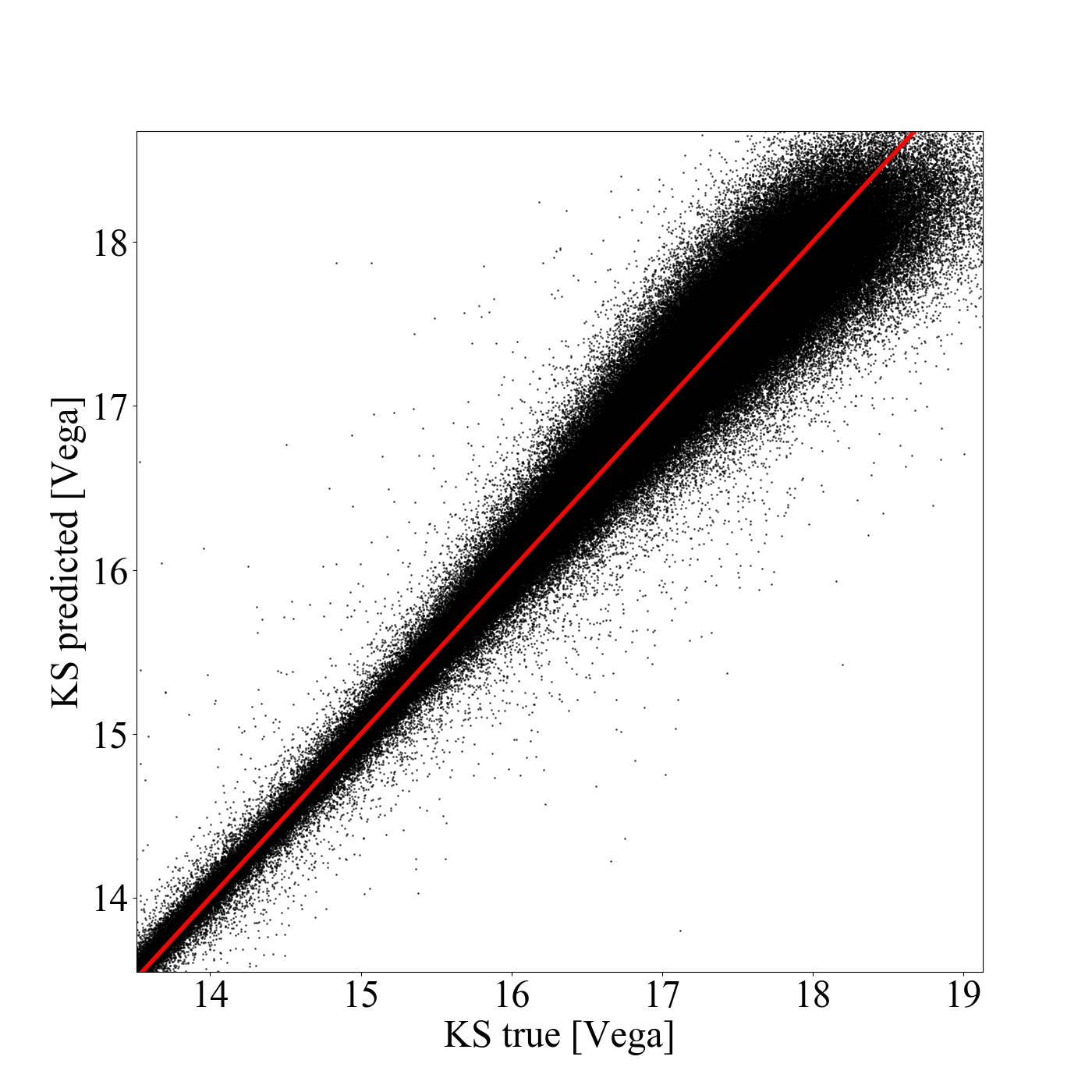}
    \includegraphics[scale=0.17]{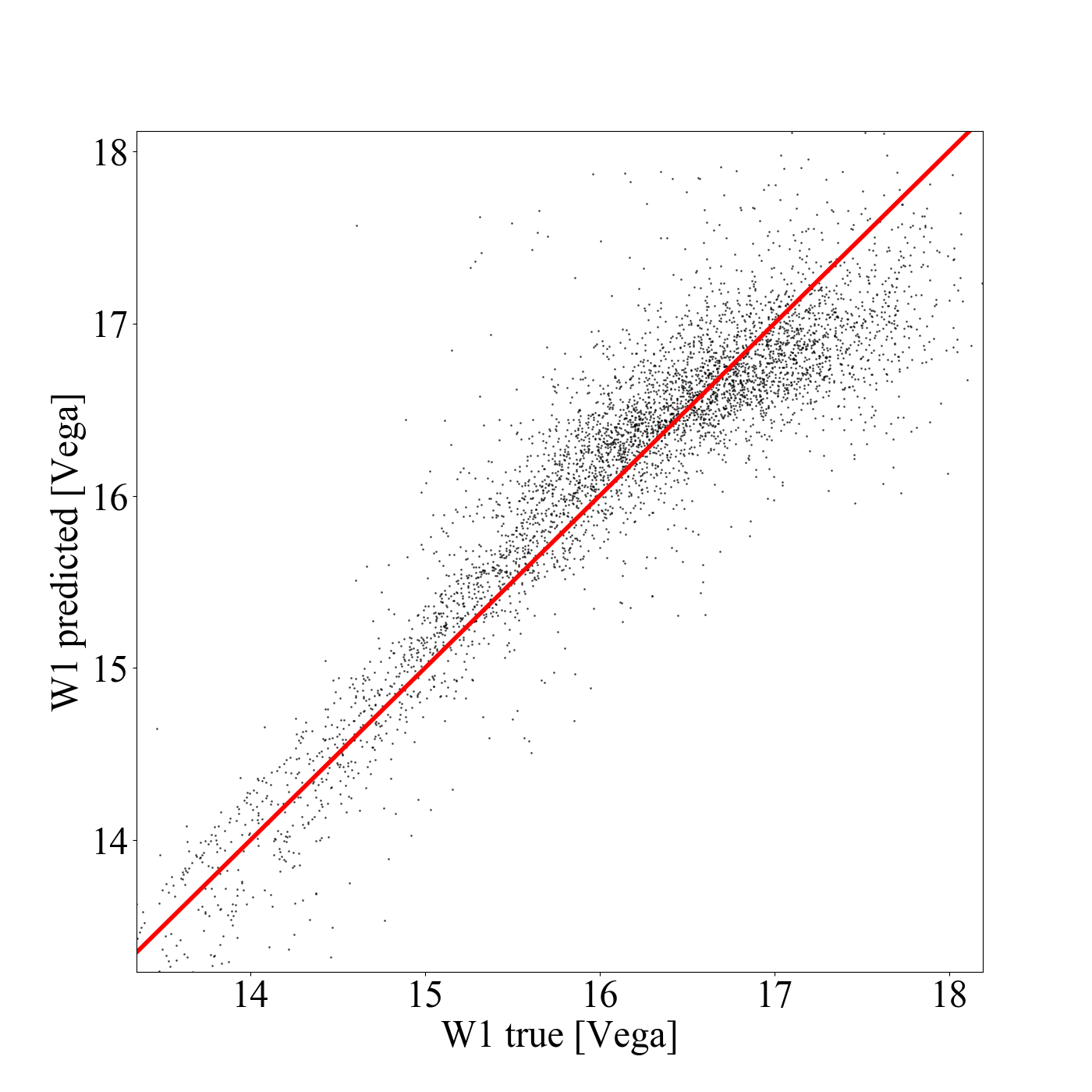}
    \includegraphics[scale=0.17]{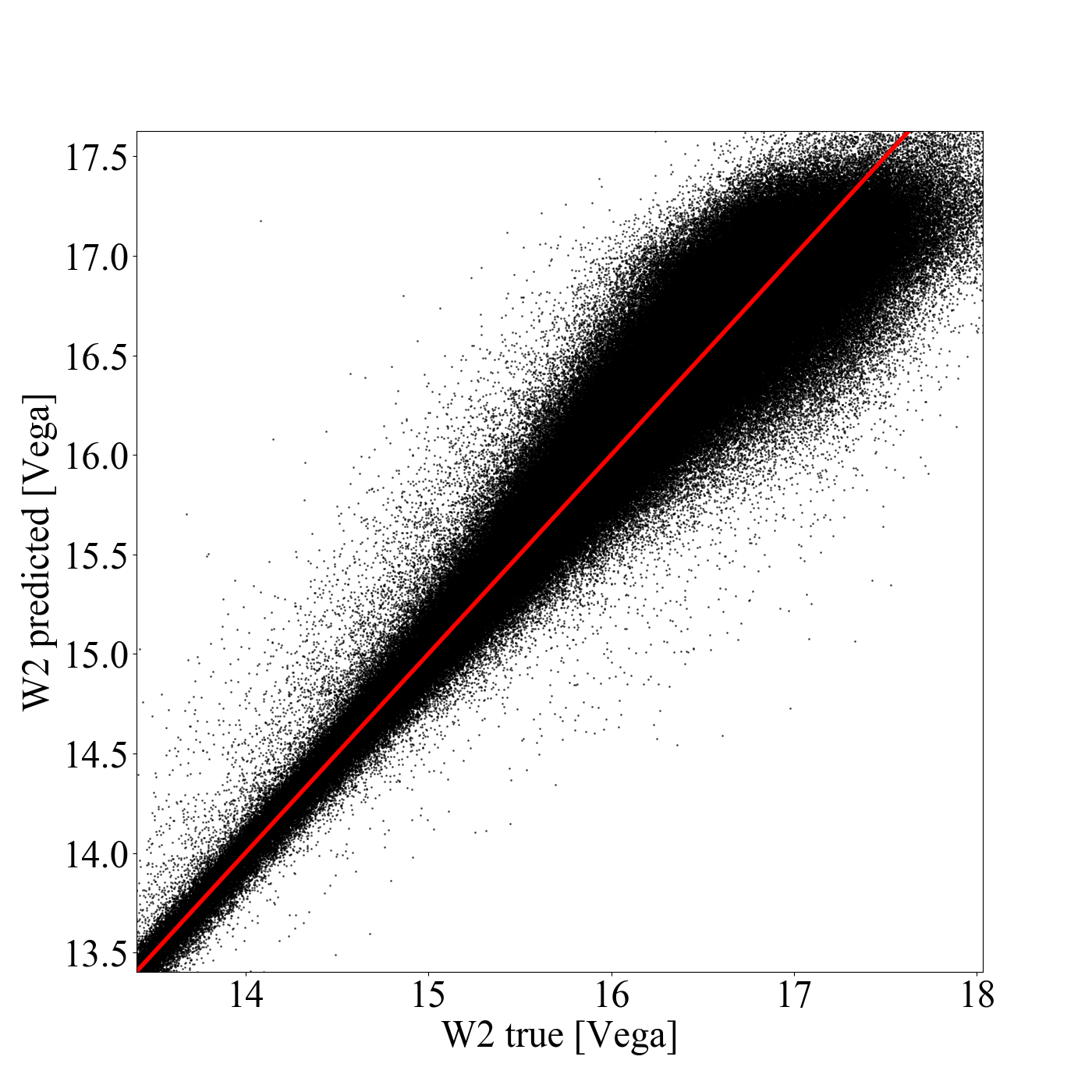}
    \caption{Results of the magnitude imputation on the training sample for the VEXAS-DESW table. The red line shows the one-to-one correlation.}
    \label{fig:imput_DESW}
\end{figure*}

\begin{figure*}
    \centering
    \includegraphics[scale=0.17]{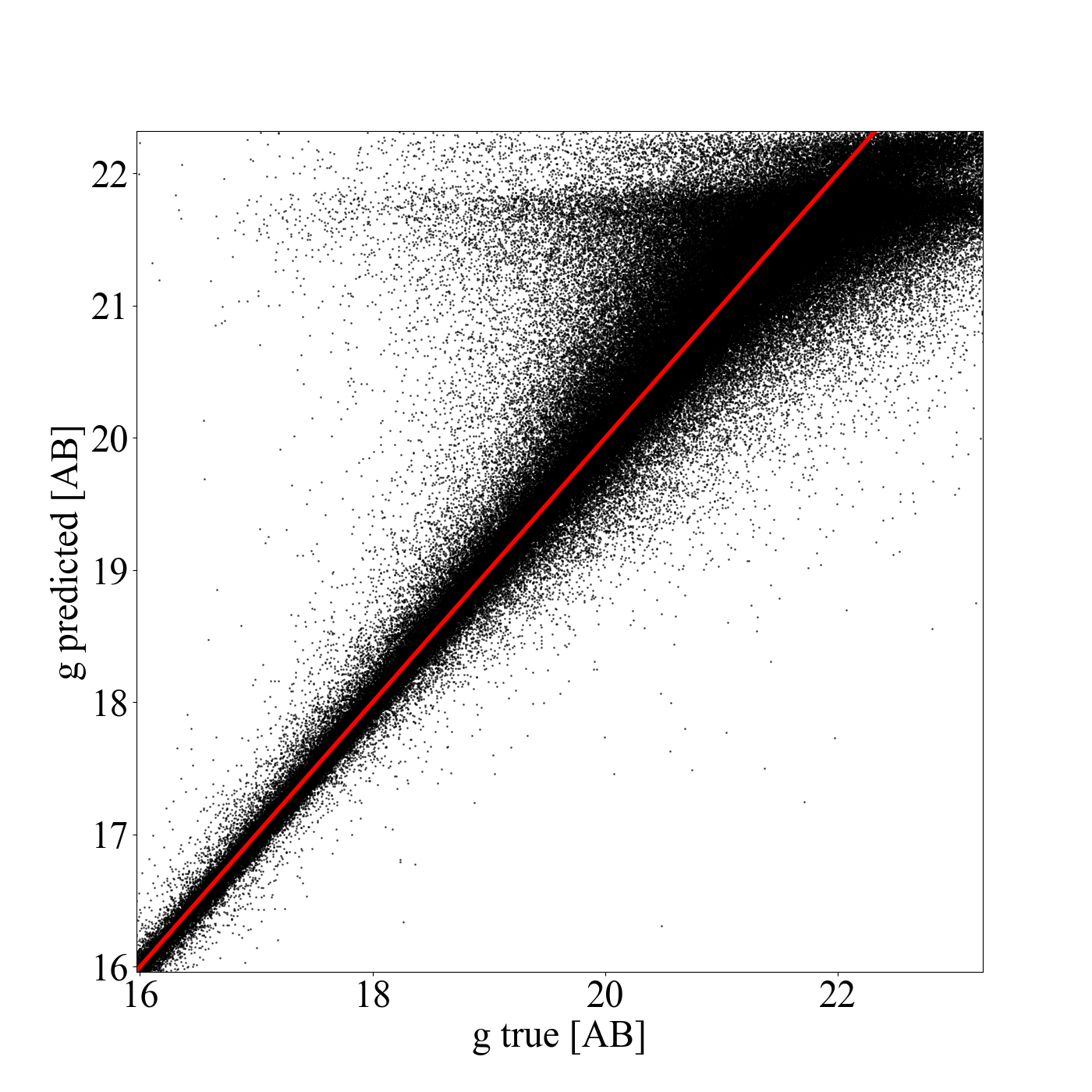}
    \includegraphics[scale=0.17]{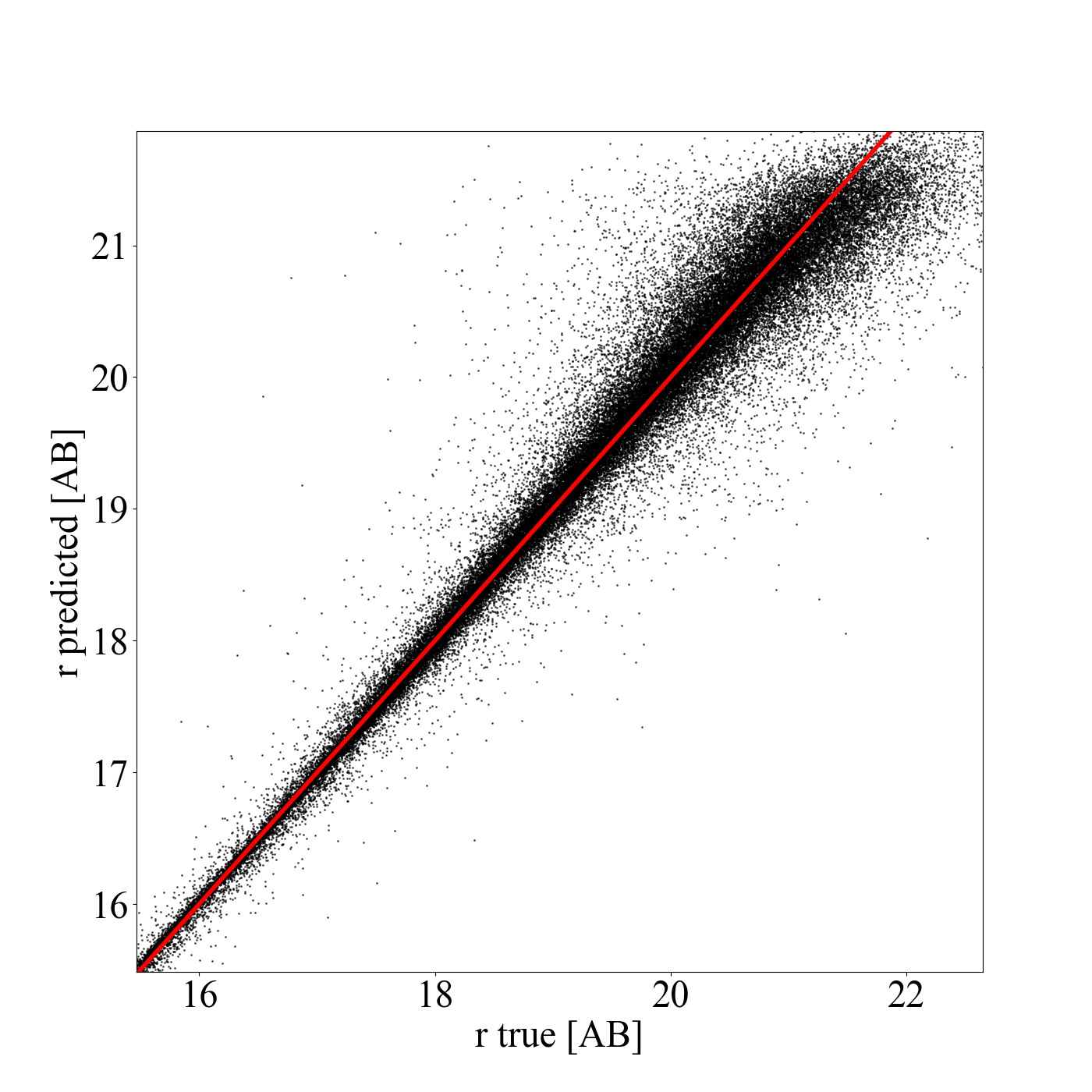}
    \includegraphics[scale=0.17]{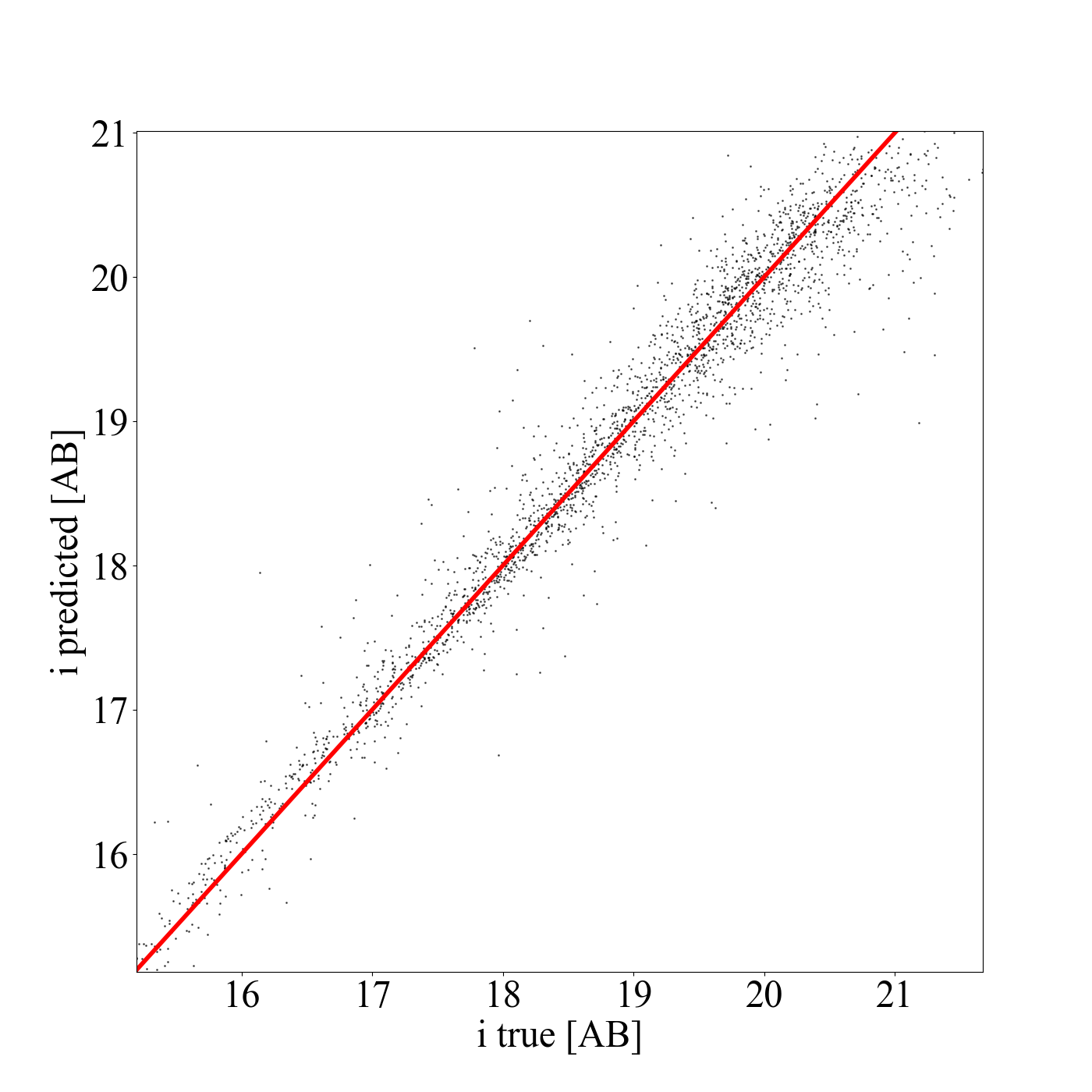}
    \includegraphics[scale=0.17]{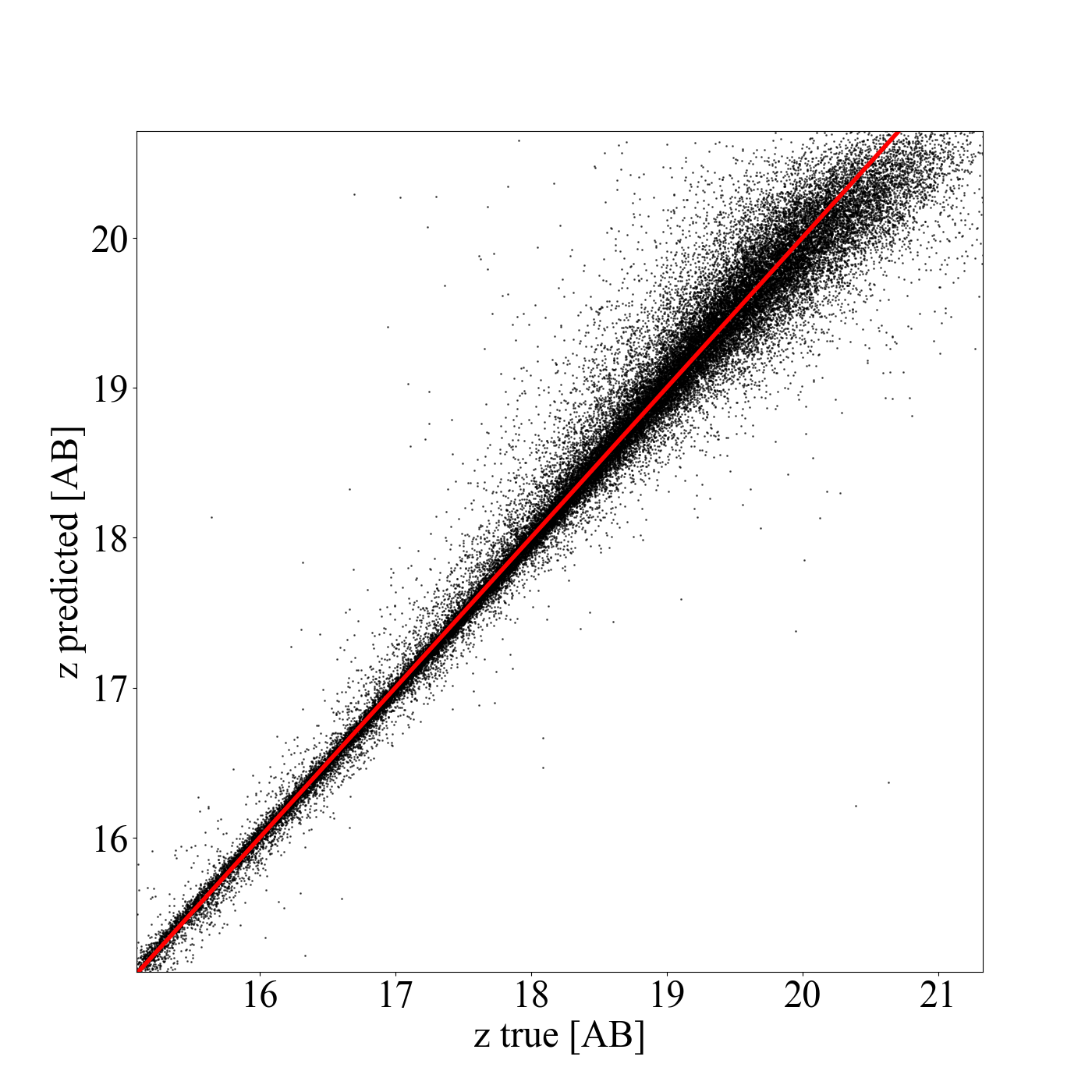}
    \includegraphics[scale=0.17]{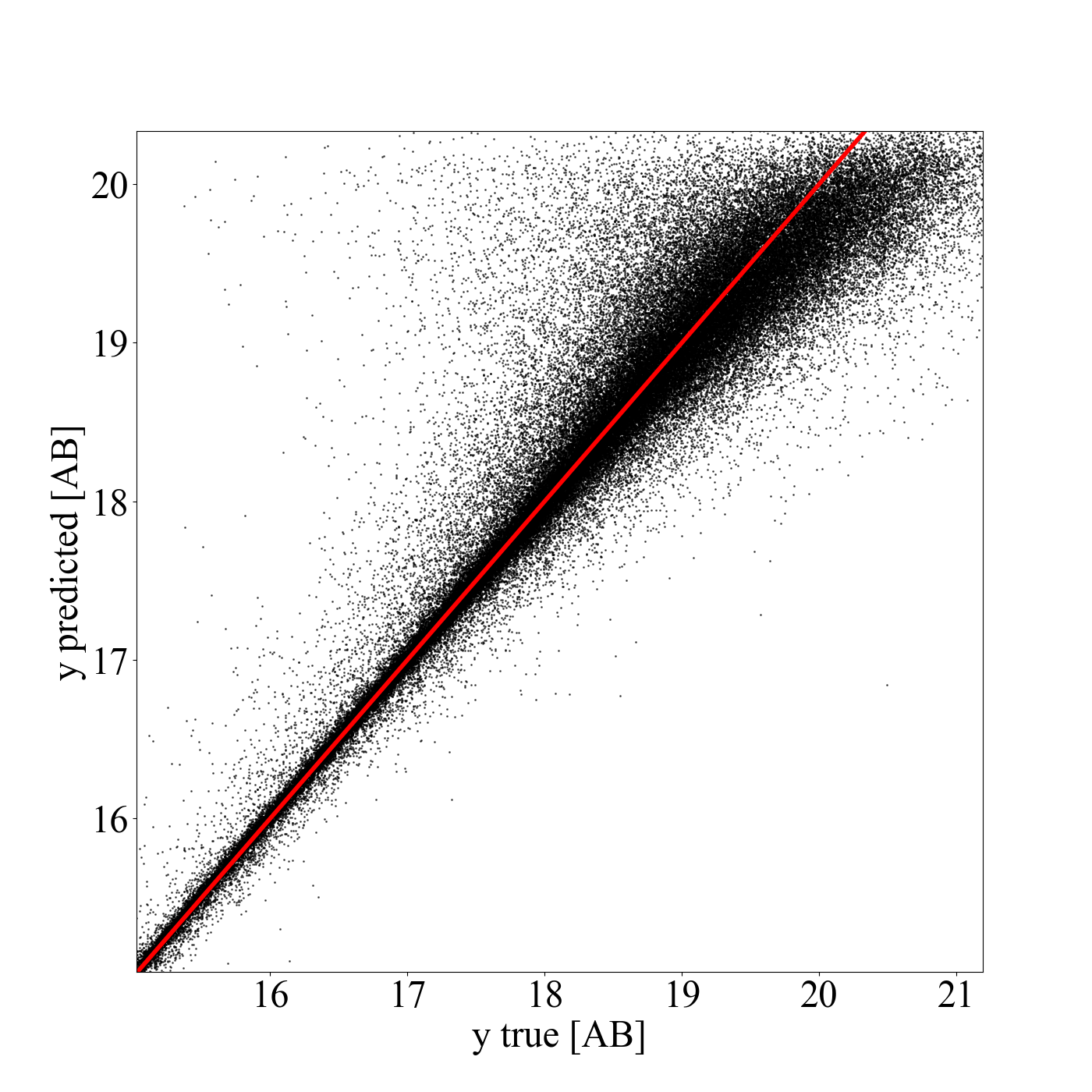}
    \includegraphics[scale=0.17]{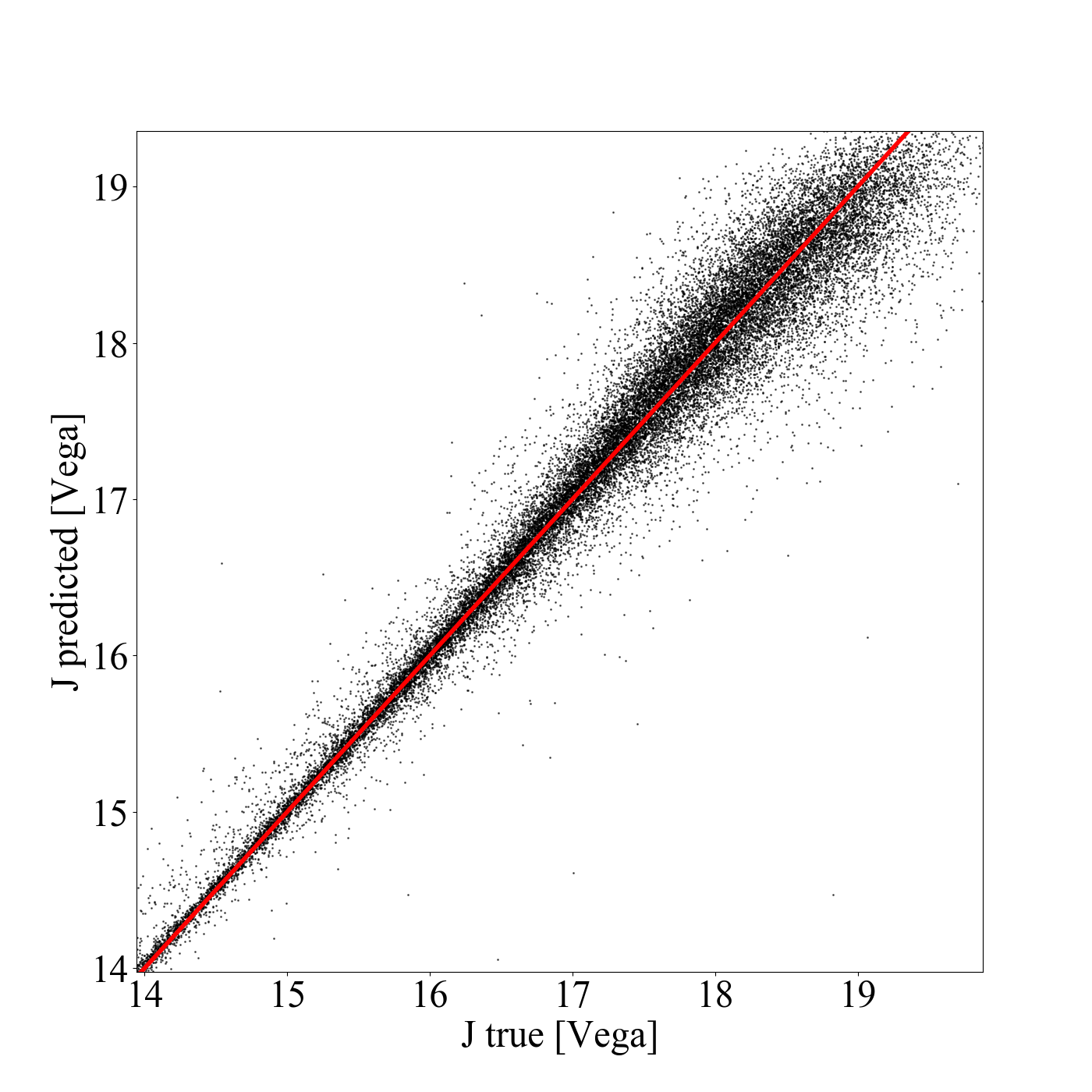}
    \includegraphics[scale=0.17]{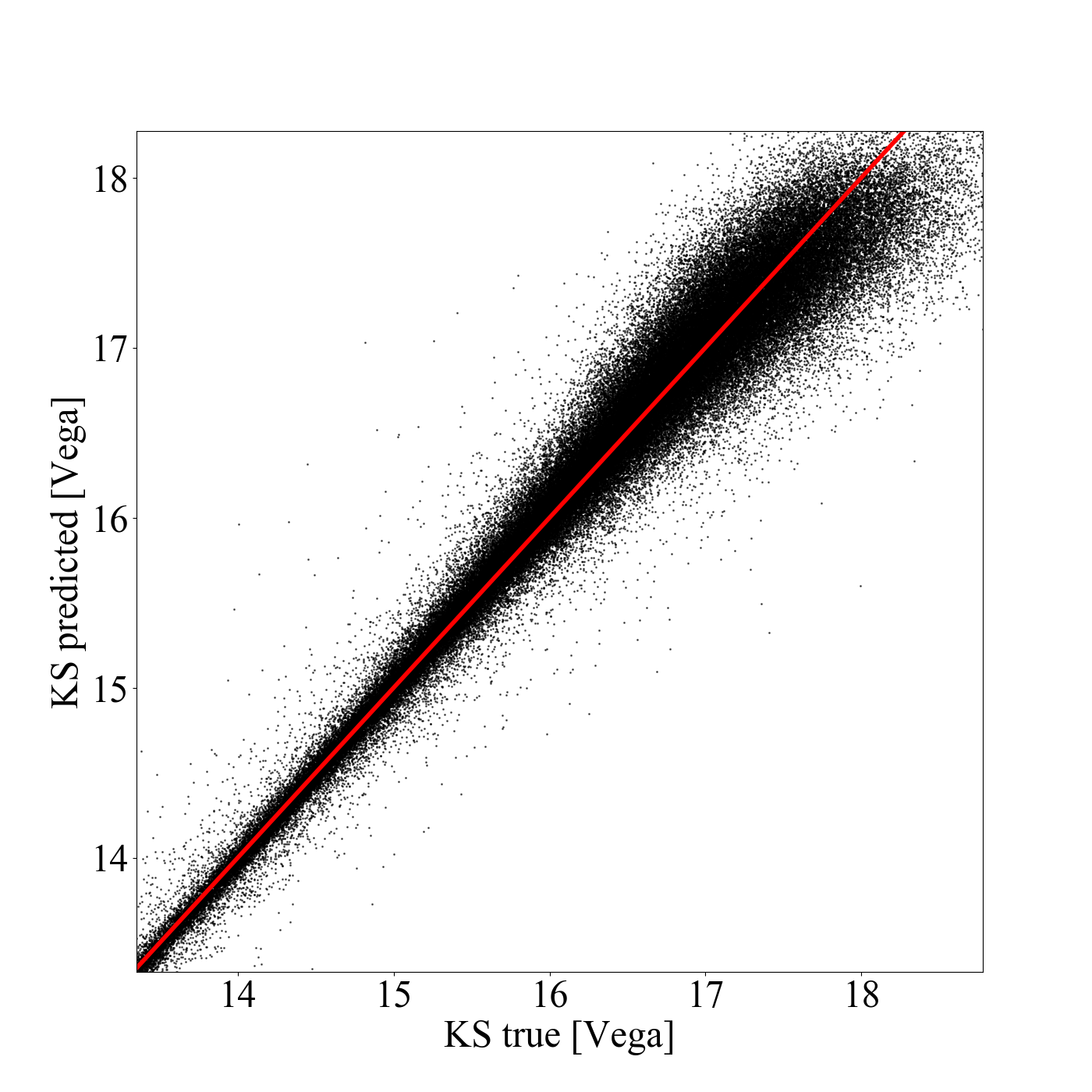}
    \includegraphics[scale=0.17]{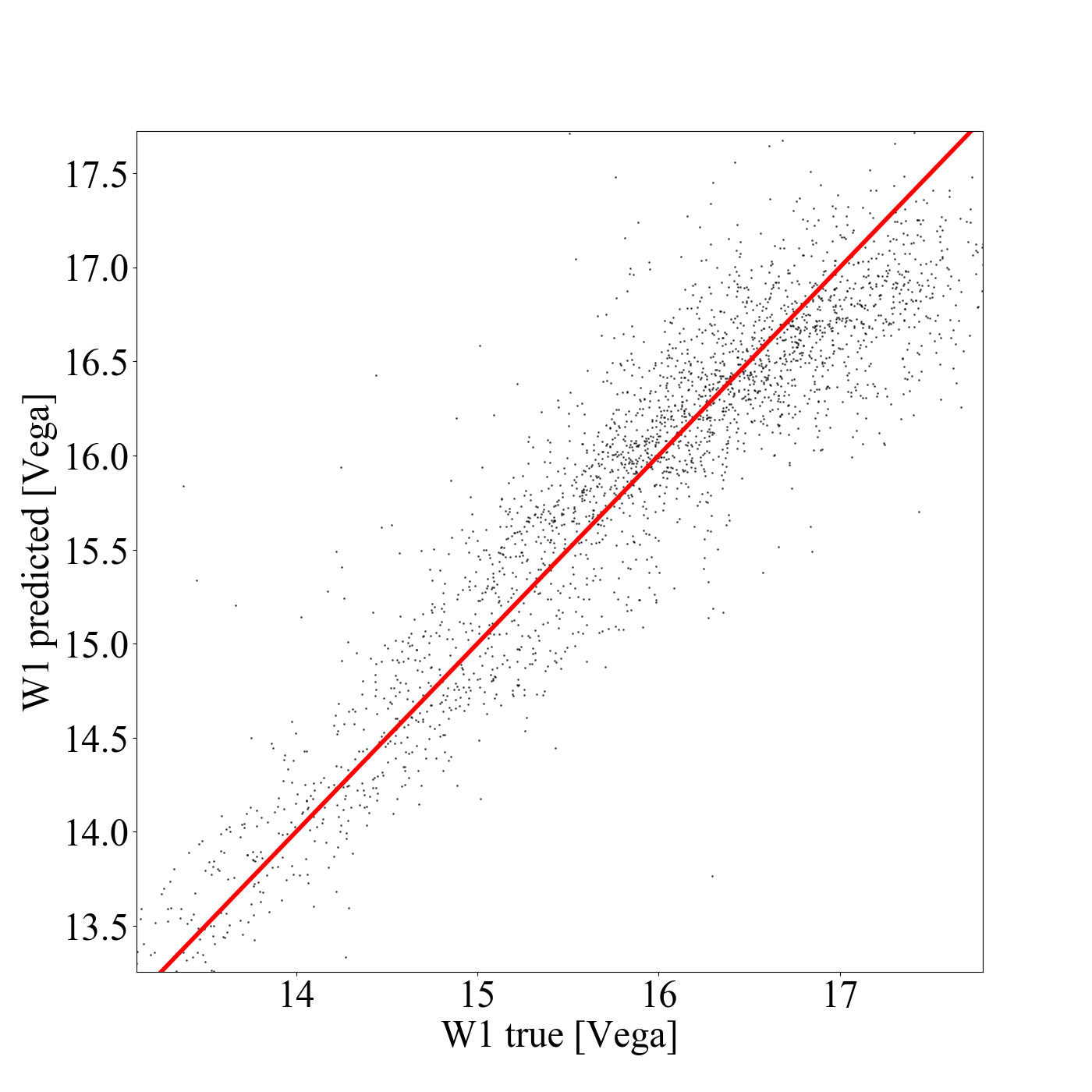}
    \includegraphics[scale=0.17]{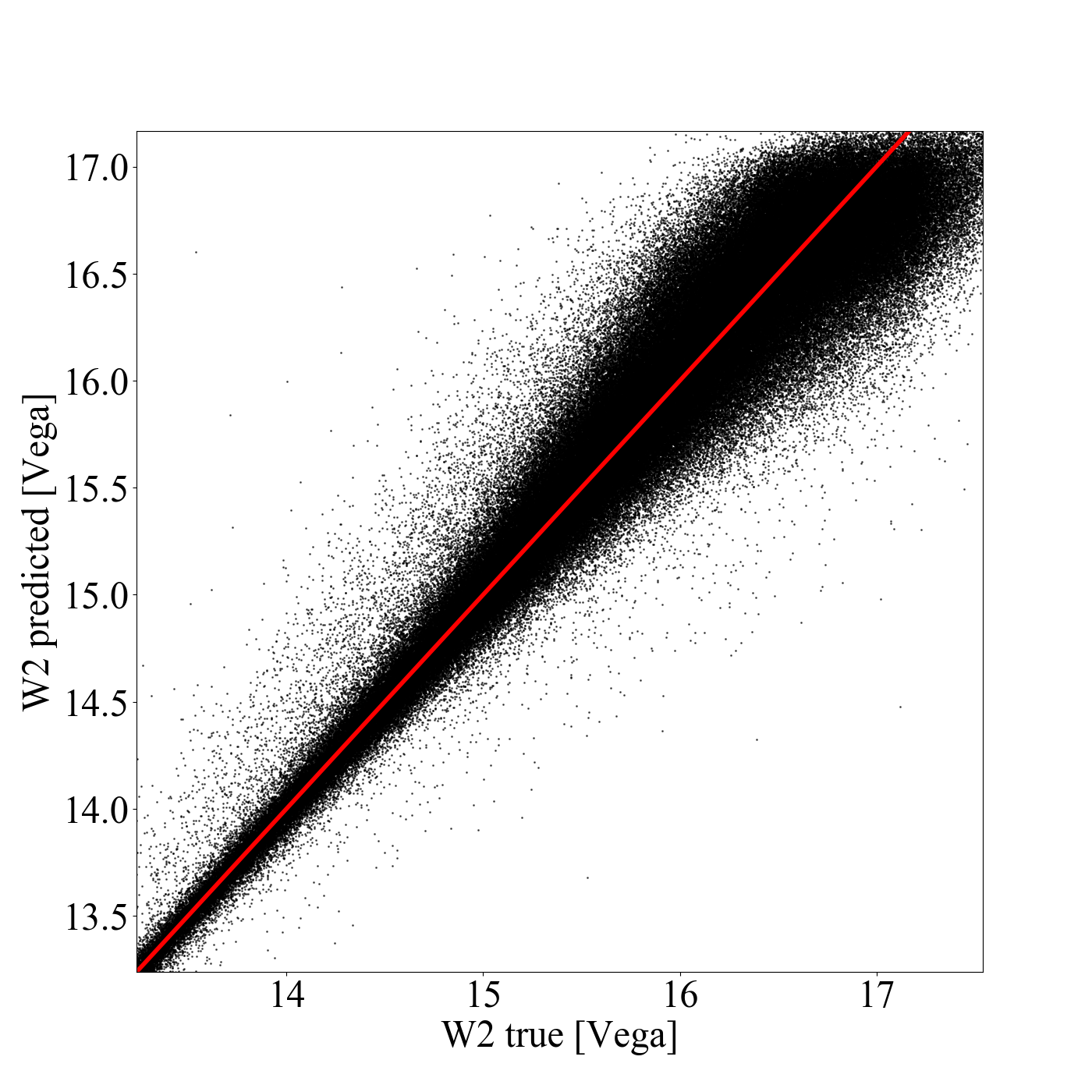}
    \caption{ Results of the magnitude imputation on the training sample for the VEXAS-PSW table. The red line shows the one-to-one correlation. }
    \label{fig:imput_PSW}
\end{figure*}

\begin{figure*}
    \centering
    \includegraphics[scale=0.17]{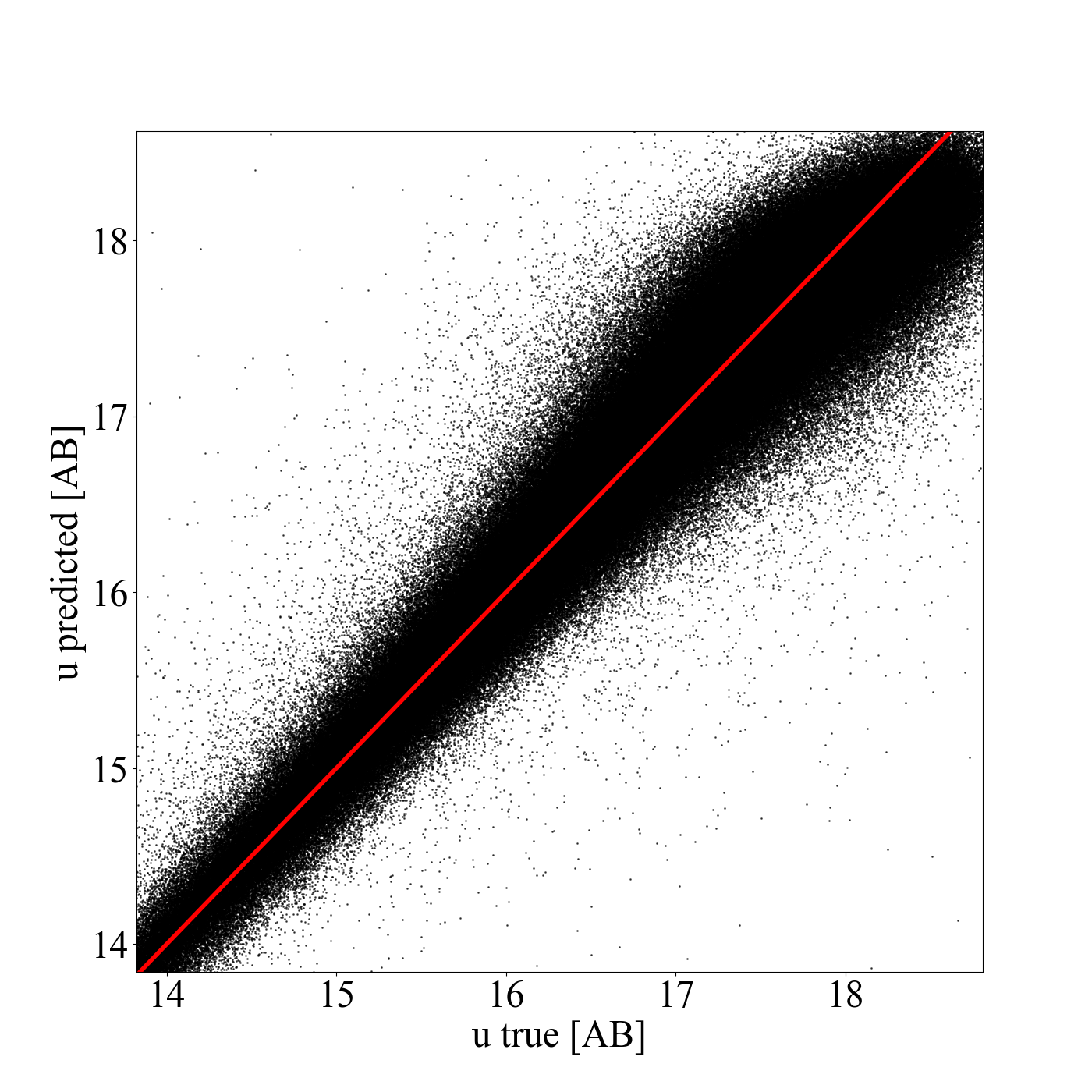}
    \includegraphics[scale=0.17]{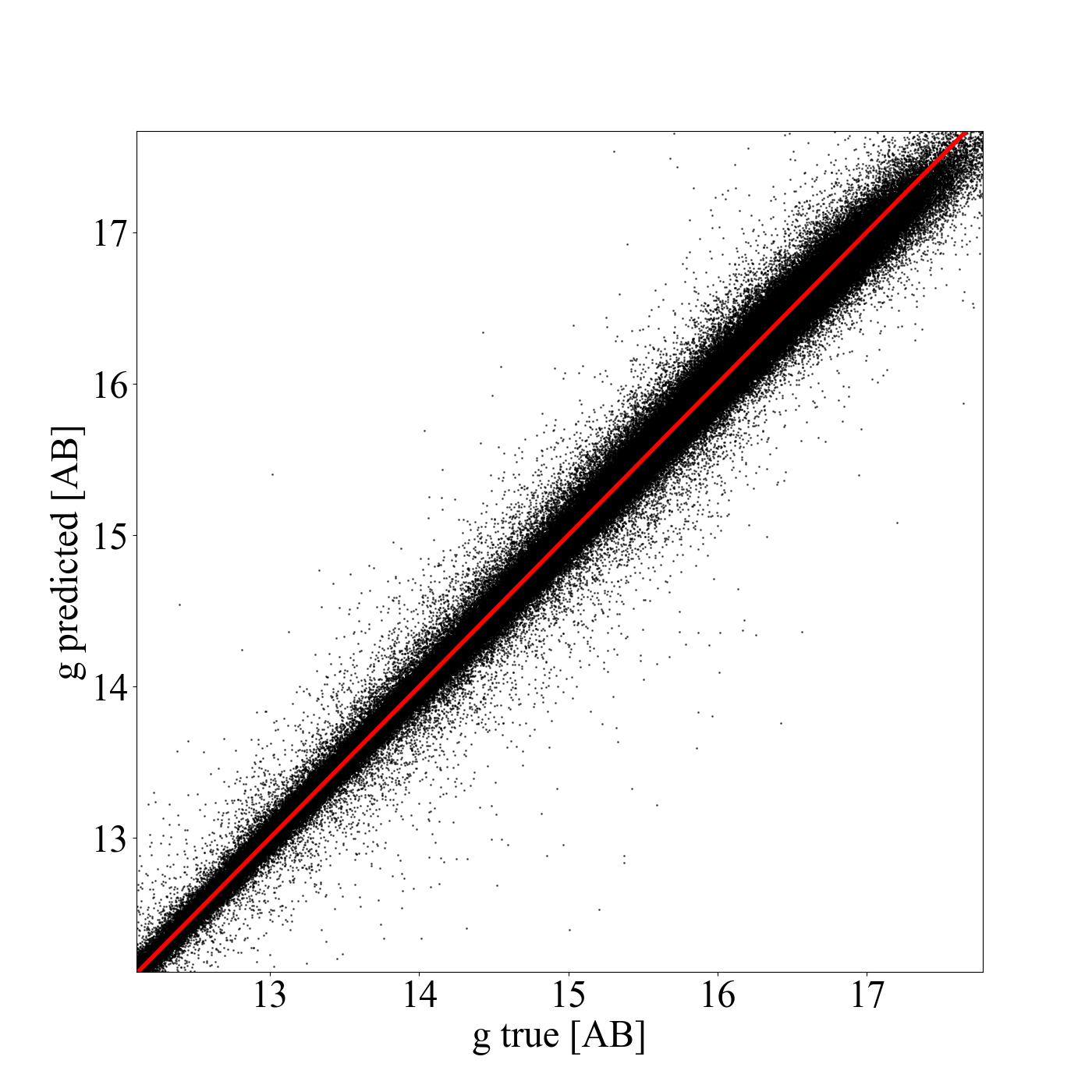}
    \includegraphics[scale=0.17]{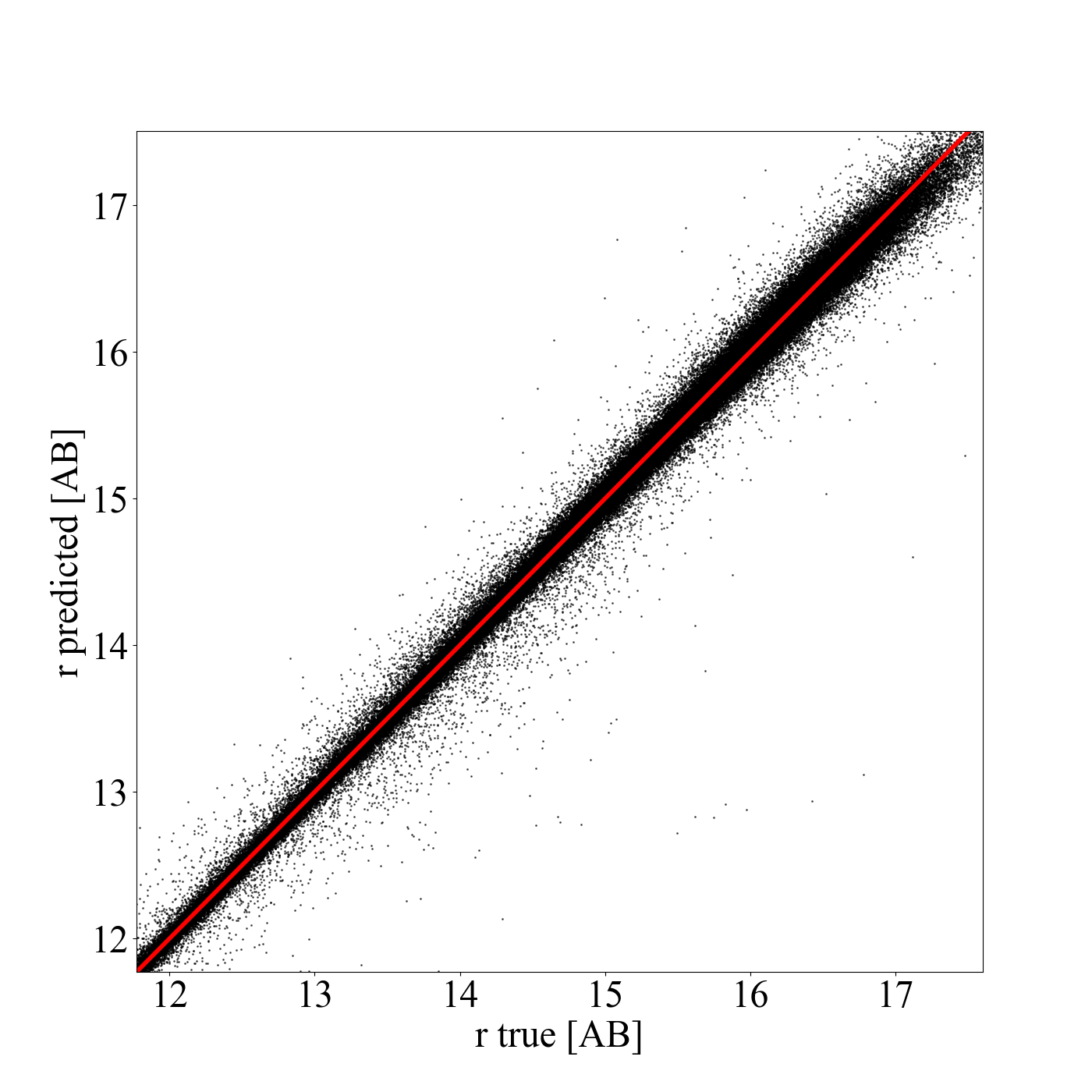}
    \includegraphics[scale=0.17]{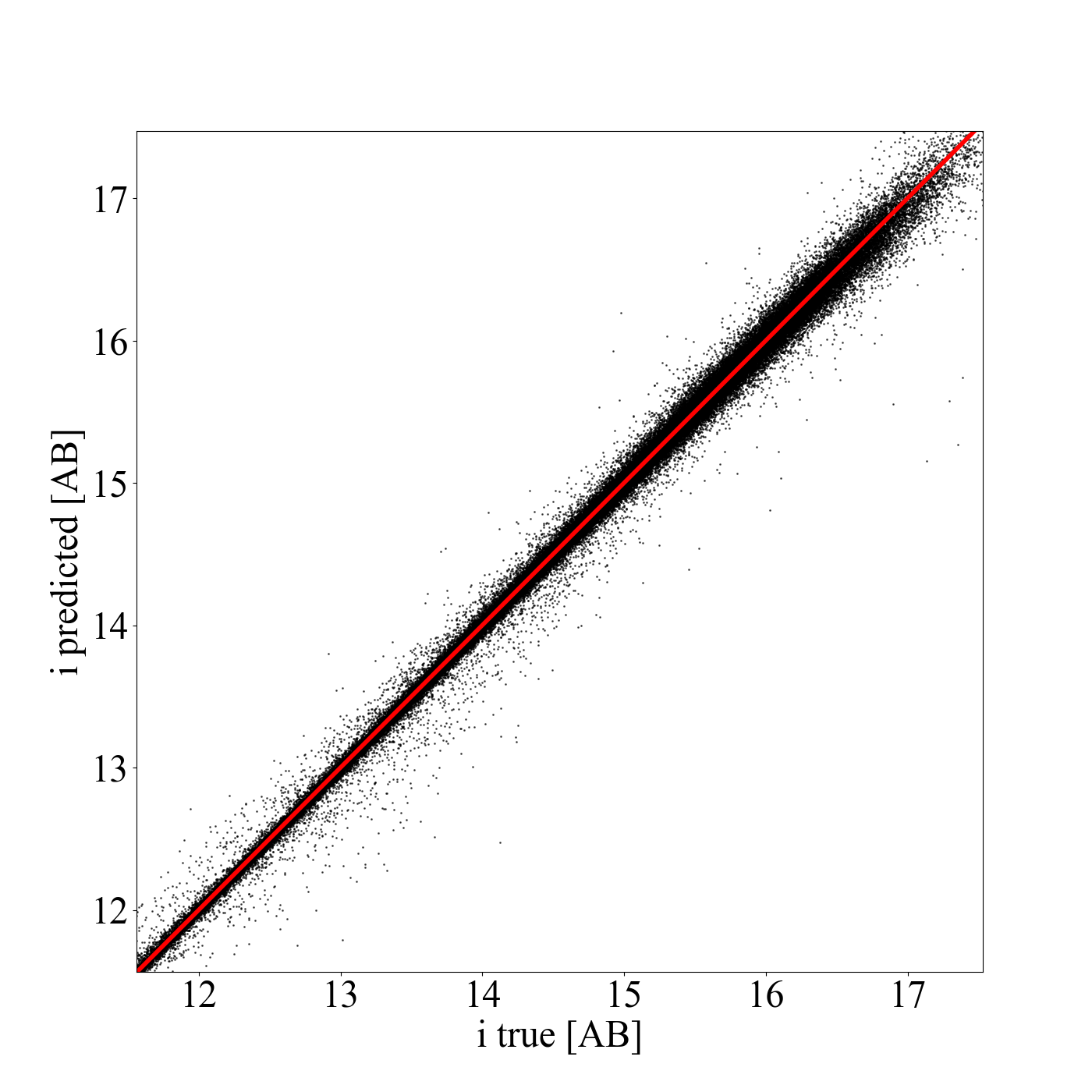}
    \includegraphics[scale=0.17]{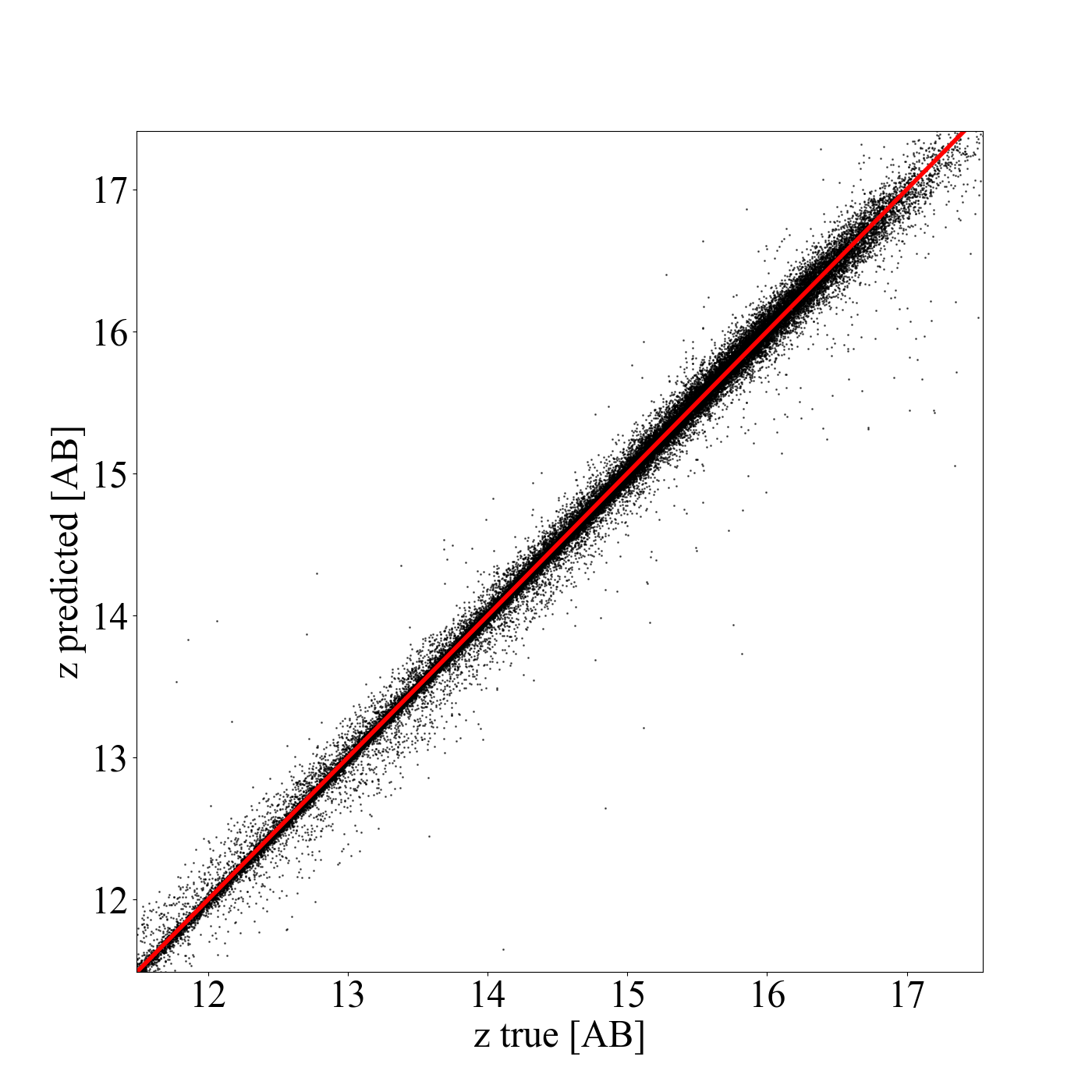}
    \includegraphics[scale=0.17]{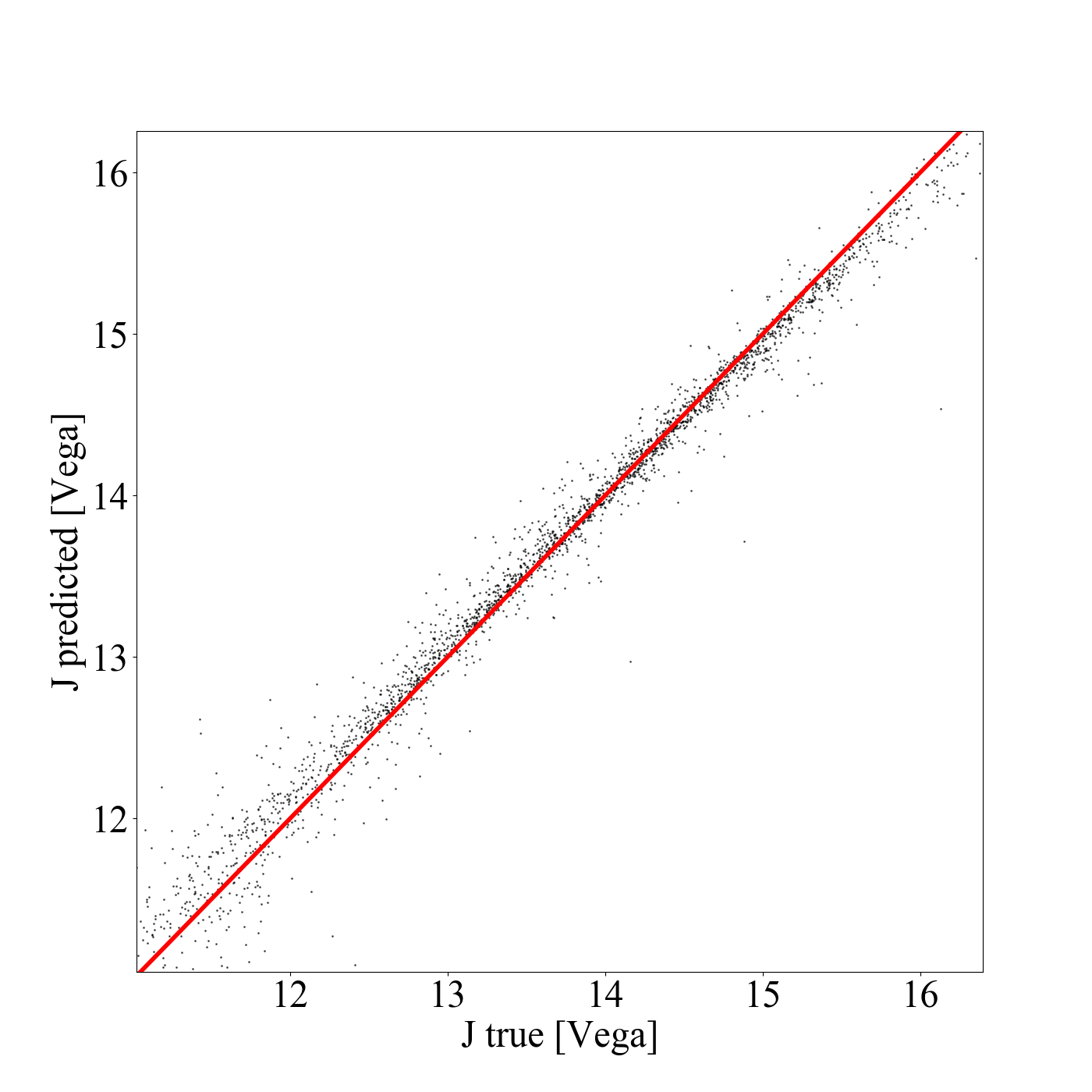}
    \includegraphics[scale=0.17]{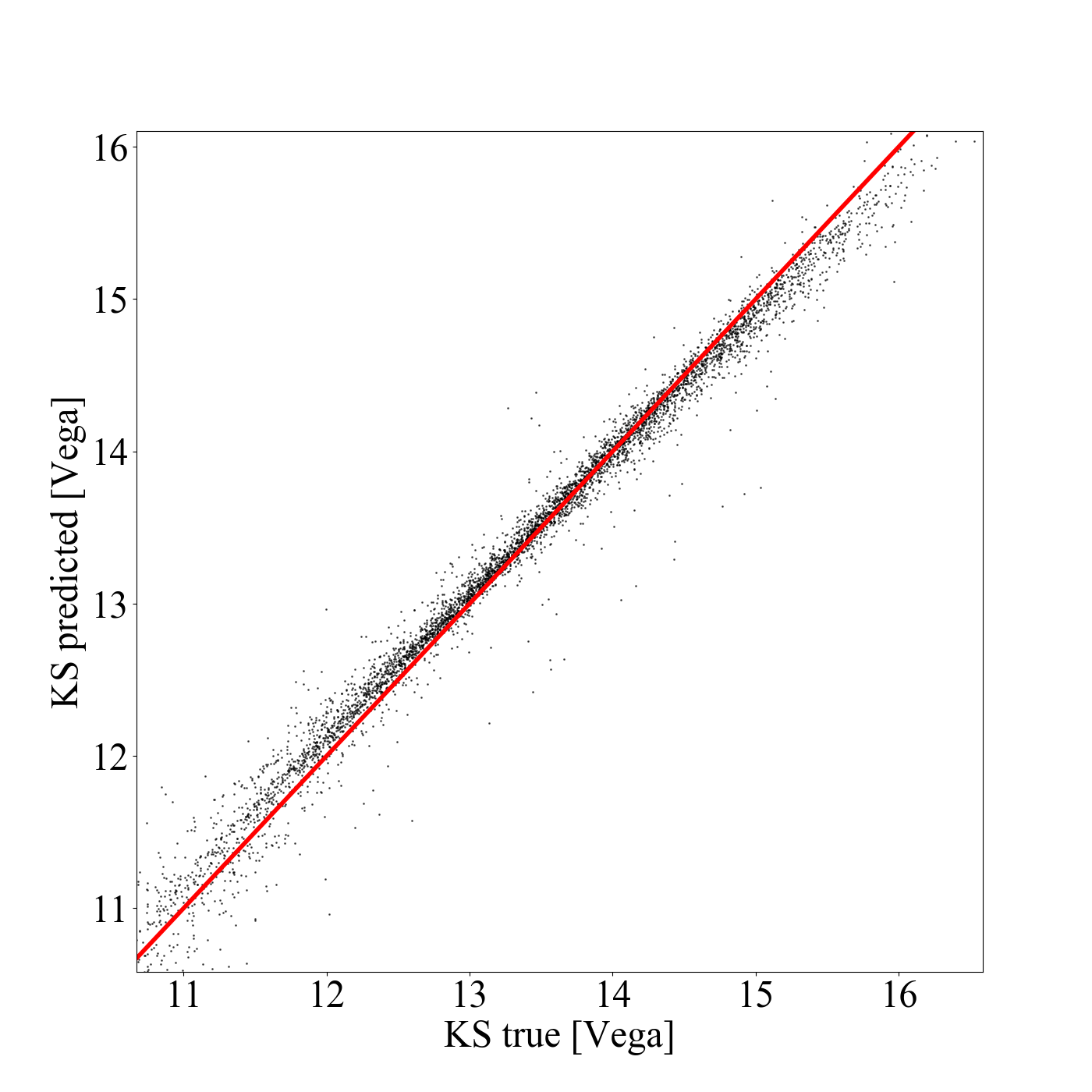}
    \includegraphics[scale=0.17]{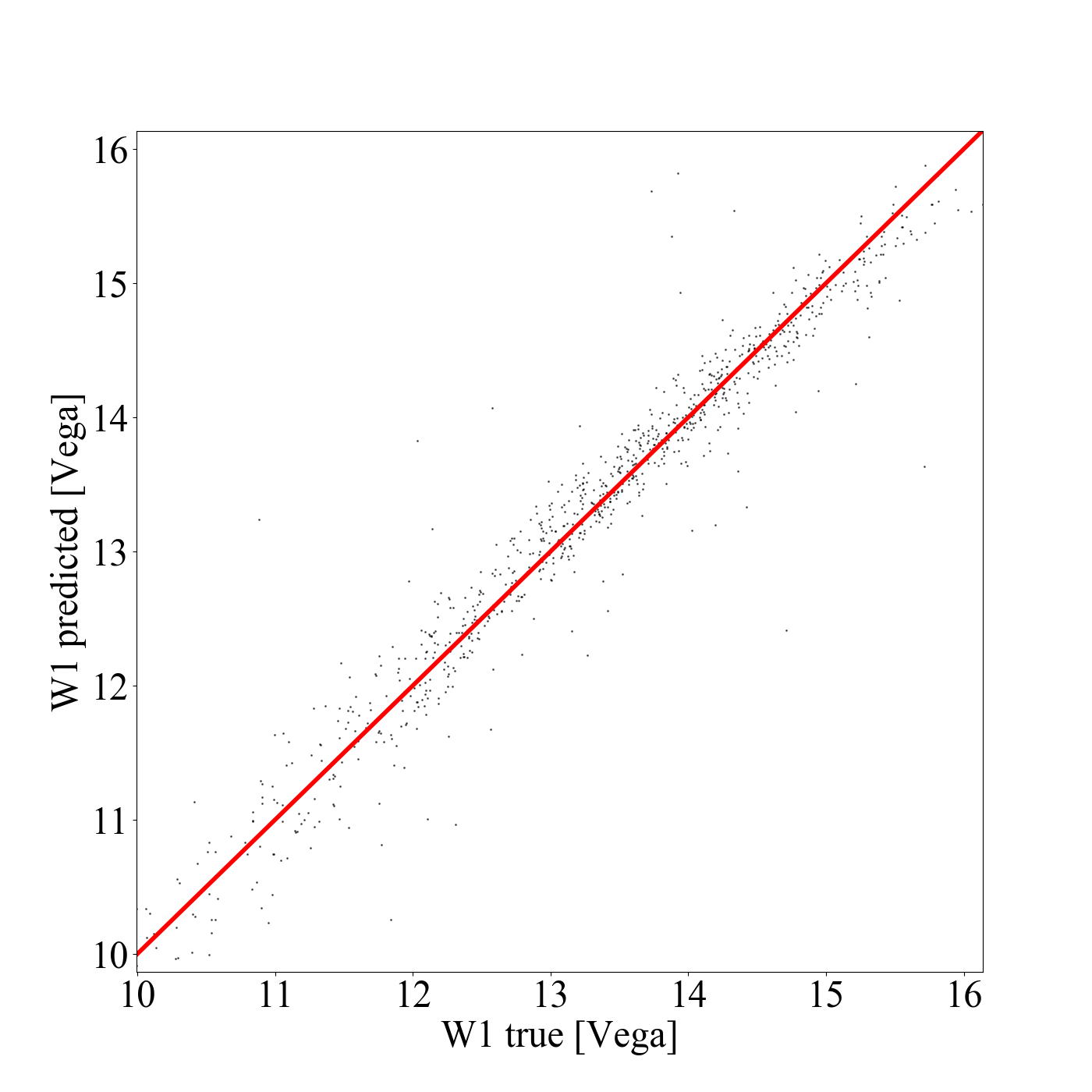}
    \includegraphics[scale=0.17]{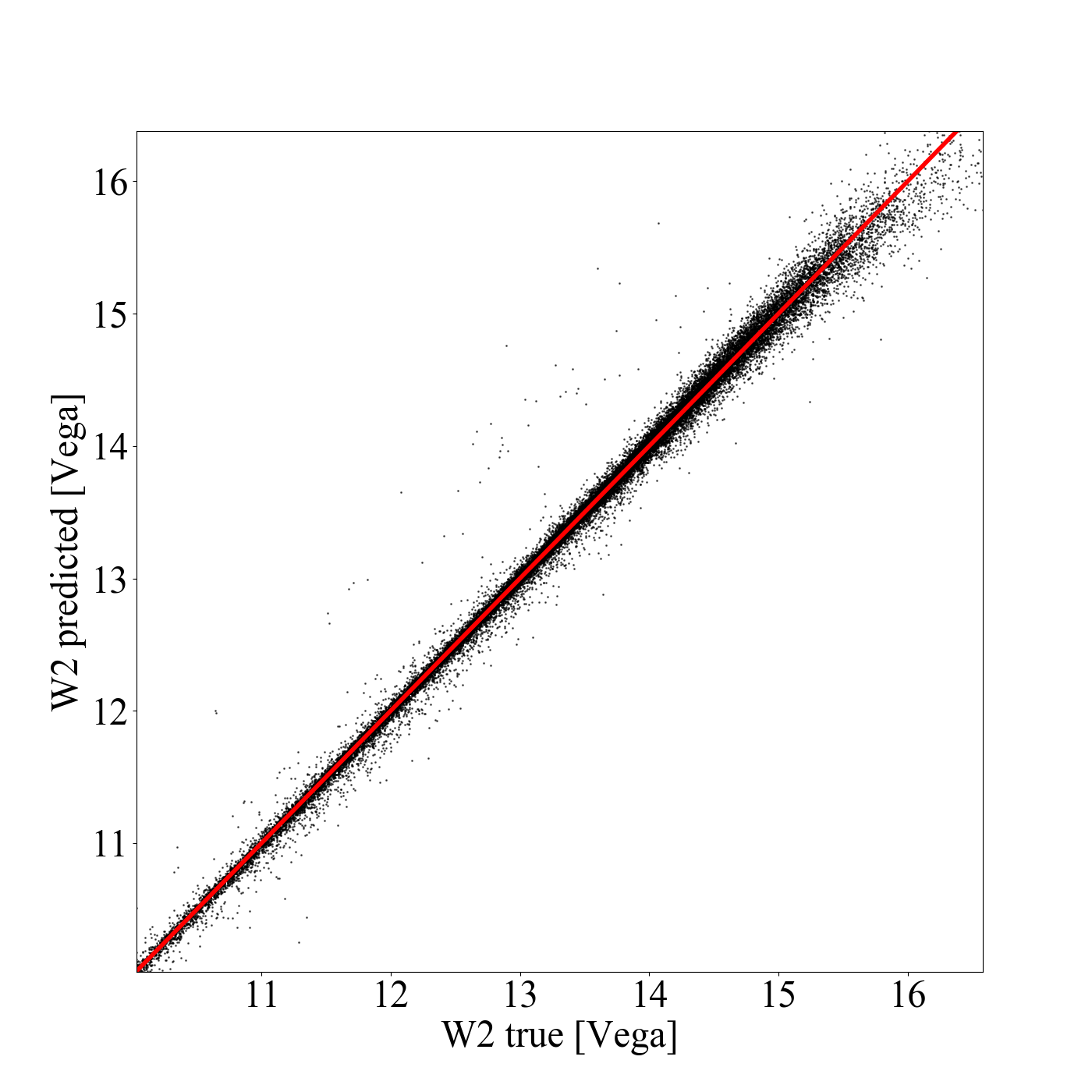}
    \caption{ Results of the magnitude imputation on the training sample for the VEXAS-SMW table. The red line shows the one-to-one correlation. }
    \label{fig:imput_SMW}
\end{figure*}

Finally, we ran the imputation pipeline on the validation objects (the remaining 15\%) with 'hidden' magnitudes. We therefore obtained imputed values which we then compared to the 'true' ones. The results of this test are presented in Figures \ref{fig:imput_DESW}, \ref{fig:imput_PSW} and \ref{fig:imput_SMW} where we plot the 'true' magnitudes versus the imputed ones for the bands on which imputation has been computed for the VEXAS-DESW, VEXAS-PSW and VEXAS-SMW sources, respectively. 

 
The magnitudes are plotted in their original units: AB for the optical and Vega for the infrared.  
We note that we replicated, in scale, the percentage of object with missing magnitudes to be imputed in each band as presented in Table~\ref{tab:num_obj_percentag}. 

Qualitatively, in all the plots, the imputed magnitudes agree well with the 'true' ones, which already demonstrate the validity of the imputation process. For each band and in each table 
we compute the coefficient of determination $R^2$ between true and imputed magnitudes.For DESW and PSW we obtained $0.88 < R^2 < 0.98$ for all magnitudes. For the SMW , almost all imputed magnitudes follows the  one-to-one line with $R^2>0.95$. In particular $g,r,i,z$ magnitudes were recovered with $R^2>0.99$ while the worst scores were obtained for the $u$ and $W1$ bands. The imputation for the $W1$ magnitude shows the worst results in all tables, possibly due to the relatively small fraction of sources with missing $W1$ magnitude in the training sample\footnote{We remind the readers that the frequency of masking magnitudes to create the training sample was taken from Table~\ref{tab:num_obj_percentag}, and a very small percentage was hidden also for $W1$ as a further test.}. Anyways, this would not affect the classification results at all, since in reality we do not perform any imputation on W1, because by definition 100\% of the VEXAS objects have measured W1 magnitudes. We also note that the VEXAS-SMW table consists mostly of \stars (see e.g. Table~\ref{tab:class_numb_prob}), so the obtained scores could not be representative for the other two classes. This holds particularly for quasars:mong the whole SMW training sample for the AE, quasars constitute only a $\approx0.1\%$ ($\approx8\,000$ objects), which is possibly too small a fraction to obtain a satisfactory imputation of some magnitudes for this subclass of sources. 
Hence, in the next section we focus in particular on VEXAS-SMW, showing that indeed the imputation is sub-optimal, but that this does not bias the classification through the ensemble learning described in the main body. 

\subsection{A sub-optimal imputation for VEXAS-SMW}
\label{app:mag_imp_sm}
\begin{figure*}
    \centering
    \includegraphics[scale=0.7]{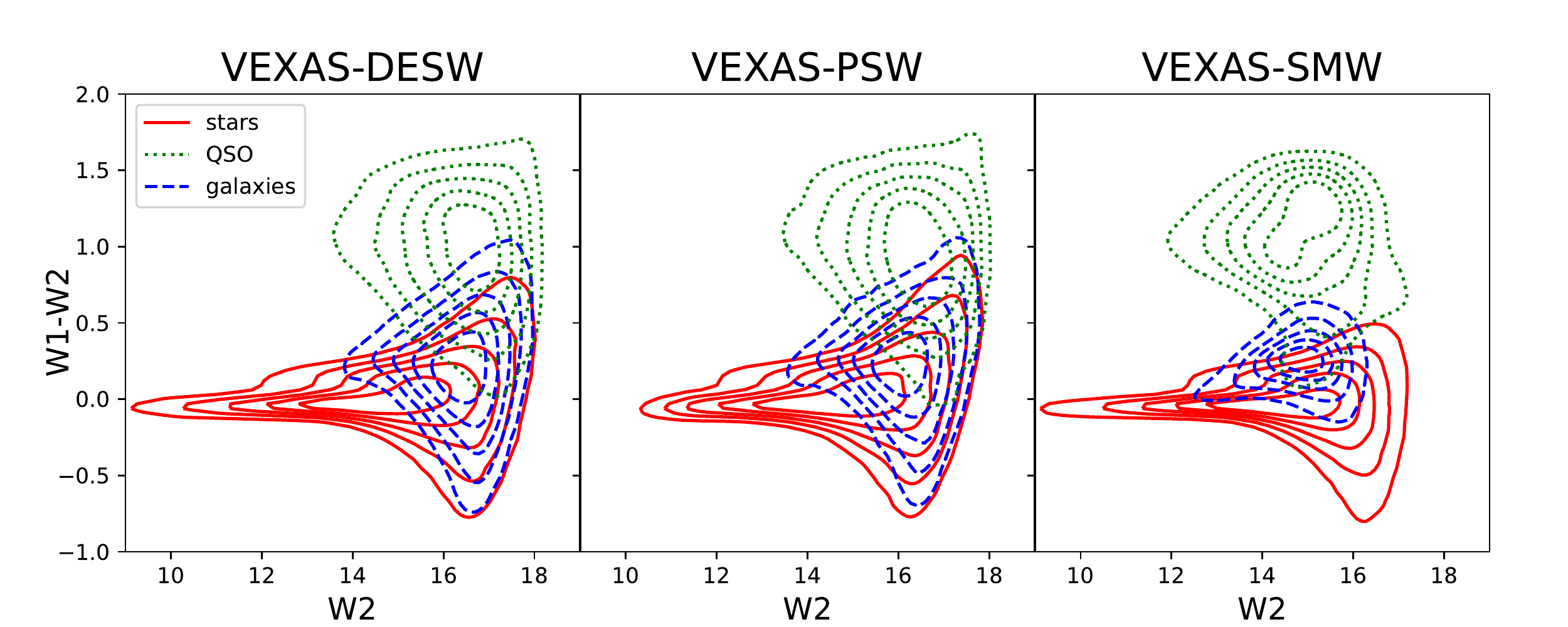}
    \includegraphics[scale=0.7]{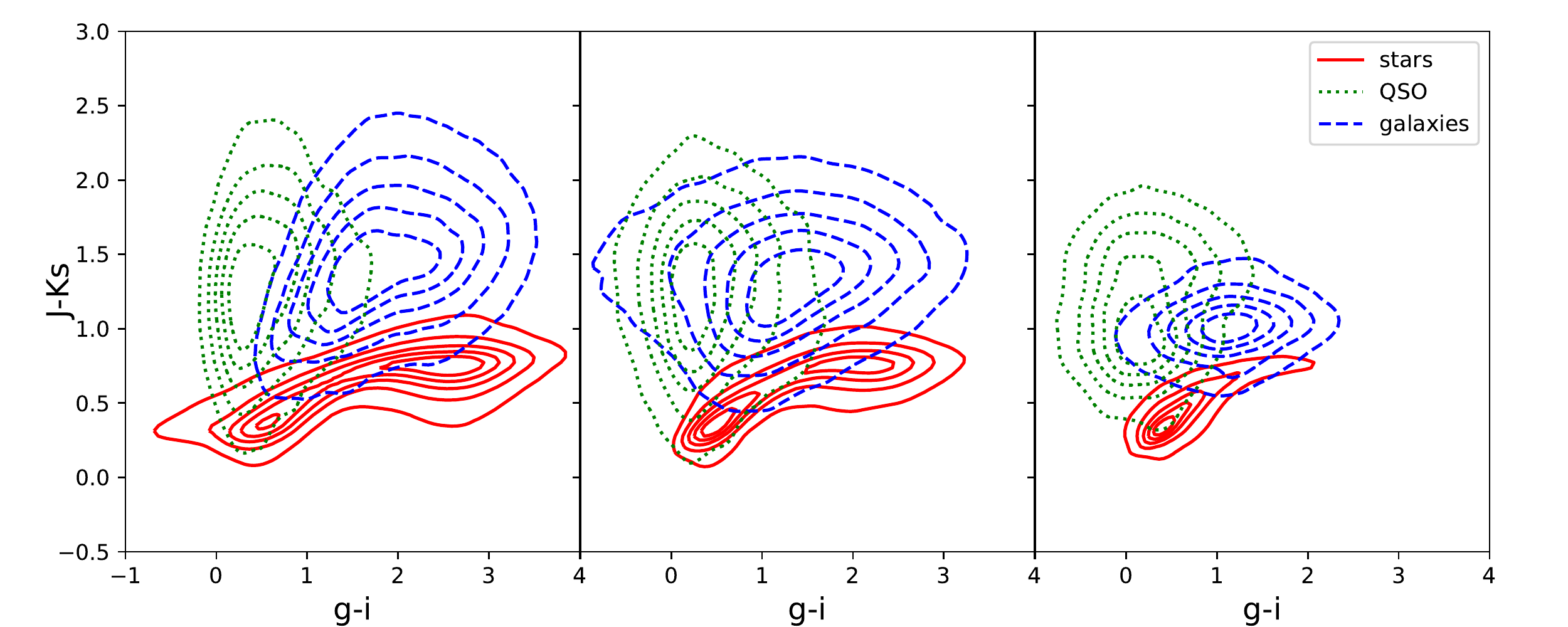}
    \caption{Selected colour-colour and magnitude-colour diagrams of VEXAS high-confidence ($p_{\rm class}=0.7$) \stars (red contours), \qso (green contours) and \galaxy (blue contours), for only objects without missing magnitudes, over the three optical footprints (VEXAS-DESW, left column, VEXAS-PSW, middle column, and VEXAS-SMW  right column), split according to the predicted class. While for the first two tables the situation is unchanged with respect to Figure~\ref{fig:mag-col_plots}, for the VEXAS-SMW the separation obtained without imputation is much more net between the three classes. } 
    \label{fig:mag-col_plots_noIMPUT}
\end{figure*}

In Figure~\ref{fig:mag-col_plots} of the main body, we showed that, while for VEXAS-DESW and VEXAS-PSW a clear (albeit partial) separation between objects belonging to different classes is visible, for the VEXAS-SMW case (right panel) this is not the case. 
To demonstrate that indeed this is caused by the sub-optimal performance of the AE imputation, we plot in Figure~\ref{fig:mag-col_plots_noIMPUT} high confidence objects ($p_{\rm class}=0.7$) for which none of the magnitudes were missing (thus where the imputation was not necessary). 
It is clear that for VEXAS-DESW (left) and VEXAS-PSW (middle) nothing changes, whereas for VEXAS-SMW (right) the confusion between objects classified in different classes is substantially reduced and three different regions are clearly visible. 

Importantly, this sub-optimal imputation does not affect the classification of the objects in the three classes obtained through the ensemble learning (see Fig.~\ref{fig:class_comp_sm}). This is in fact, flexible enough and based on different models, of which only some use imputation. 

\end{appendix}
\end{document}